\documentclass[10pt,twocolumn,showpacs,preprintnumbers,amsmath,amssymb,aps,prb,longbibliography,superscriptaddress,nofootinbib]{revtex4-2}
\usepackage{array,braket,mathtools,siunitx,bbm,enumitem}

\usepackage[unicode=true,
bookmarks=true,bookmarksnumbered=true,bookmarksopen=false,
breaklinks=true,pdfborder={0 0 0},backref=false,colorlinks=true]{hyperref}
\hypersetup{citecolor={blue},urlcolor={magenta}}

\usepackage{cleveref}
\crefname{appendix}{App.}{Apps.}
\crefname{equation}{Eq.}{Eqs.}
\crefname{figure}{Fig.}{Figs.}
\crefname{table}{Tab.}{Tabs.}
\crefname{section}{Sec.}{Secs.}
\crefname{enumi}{Cond.}{Conds.}

\begin{document}
\title{A Hierarchy of Exact Lower Bounds on the Laughlin Spectral Gap}
\author{Michael G. Scheer}
\email{mscheer@fas.harvard.edu}
\affiliation{Department of Physics, Harvard University, Cambridge, Massachusetts 02138, USA}
\author{Eslam Khalaf}
\email{eslam\_khalaf@fas.harvard.edu}
\affiliation{Department of Physics, Harvard University, Cambridge, Massachusetts 02138, USA}
\date{\today}

\begin{abstract}
We develop a bootstrap approach to deriving lower bounds on the spectral gap of the first Haldane pseudopotential Hamiltonian on the sphere with $N_\Phi$ flux quanta for arbitrary $N_\Phi$, including in the thermodynamic limit $N_\Phi \to \infty$. This method can in principle be generalized to pseudopotentials describing other model wavefunctions such as the Moore--Read state or Read--Rezayi sequence. Our approach is based on Knabe's observation that if $H$ is a Hamiltonian with ground state energy zero then the spectral gap is at least $\epsilon > 0$ if and only if $H^2 \succeq \epsilon H$. To show this inequality, we split $H^2$ into two parts, one that can be analytically bounded and another that can be bounded using a finite semidefinite program (SDP). Crucially, the size of this program is determined only by cutoff parameters, and the program depends on $N_\Phi$ only parametrically, allowing us to obtain numerically exact (up to solver precision) lower bounds on the gap. In order to turn these numerical bounds into mathematically rigorous results, we introduce an algorithm that can produce nearly-optimal exact interior feasible points for large rational SDPs even if the affine constraints are highly redundant and ill-conditioned. We additionally provide an algorithm that establishes the rank stability of the SDP in the $N_\Phi \to \infty$ limit, which is needed to certify the bound in the thermodynamic limit. Beyond establishing a Laughlin gap greater than $1/100$ in the thermodynamic limit and revealing a potential method for deriving spectral gap lower bounds for other pseudopotentials, our work provides a general method for turning numerical SDP bounds into rigorous mathematical proofs, which may have wide applicability. In particular, this work demonstrates that bootstrap methods can be used to rigorously resolve major open problems in quantum many-body physics.
\end{abstract}

\maketitle

\section{Introduction}
When a two-dimensional electron gas is subject to a strong perpendicular magnetic field, plateaus with quantized Hall conductance are observed when the filling $\nu$, defined as the number of electrons per flux quantum, takes certain integer \cite{Klitzing1980} or fractional \cite{Tsui1982} values. These experimental observations, called the integer and fractional quantum Hall effects (IQHE and FQHE), have been the subject of intense theoretical interest for over four decades. Thouless, Kohmoto, Nightingale, and den Nijs explained that the IQHE arises from the band topology of Landau levels, all of which are fully filled or fully empty, and therefore the IQHE can be understood in a non-interacting framework \cite{Thouless1982}. On the other hand, the FQHE concerns a system with a partially filled Landau level, so its explanation must take strong electron interactions into account. Laughlin proposed a model wavefunction for fractional quantum Hall states at fillings $\nu = 1/m$ where $m > 1$ is odd \cite{Laughlin1983}. The Laughlin state captures many of the phenomenological properties of the FQHE, including its low-energy excitations which carry fractional charge and anyonic statistics \cite{Laughlin1983,Halperinstatistics}. Haldane \cite{Haldane1983} and Trugman and Kivelson \cite{Trugman1985} constructed idealized parent Hamiltonians (also called pseudopotentials) for which the Laughlin state is an exact zero-energy ground state. Girvin, MacDonald, and Platzman \cite{Girvin1985,Girvin1986} used the single mode approximation to obtain a variational upper bound on the neutral gap for the Laughlin state\footnote{Strictly speaking, they studied the Coulomb interaction but the same analysis can be applied to pseudopotentials.}. However, a lower bound on the neutral gap, the charge gap (incompressibility), or the full spectral gap for the pseudopotential Hamiltonian has been lacking\footnote{See note added in \cref{sec:note-added}.}. We note that pseudopotentials have been defined for a large class of model fractional quantum Hall states at various fillings including the Moore--Read state and the Read--Rezayi sequence \cite{MooreRead,ReadRezayi,SimonPseudopotentials,Bernevig2008a,Bernevig2008b,Bernevig2008c}.

Progress on the Laughlin spectral gap has followed a few lines of thought. First, Refs. \cite{Ortiz2013, Mazaheri2015, Chen2015} constructed a second-quantized representation of the pseudopotentials on various geometries and showed that the Laughlin states and the Laughlin quasihole states exhaust the space of zero modes. Second, Refs. \cite{Nachtergaele2021, Warzel2022, Warzel2023} established a spectral gap in a special class of truncated pseudopotentials relevant to quasi-one-dimensional geometries such as the thin cylinder. Their approach used the observation that the truncated thin cylinder Hamiltonian is effectively a 1D local lattice model. Ref. \cite{Lemm2026} has also proved that the charge gap is lower-bounded by the neutral gap. Intuitively, neutral excitations can be understood as bound states of quasiholes, which are zero-energy states, and quasielectrons, which are finite energy states. Since the interaction between quasiholes and quasielectrons is attractive, we expect the neutral gap to be smaller than the charge gap. Additionally, Refs. \cite{Rougerie2015a,Rougerie2015b,Lieb2018,Lieb2019,Olgiati2020,Thoma2024} derived incompressibility estimates for the Laughlin state. See also Refs. \cite{Rougerie2018,Rougerie2022} for an overview of the Laughlin gap problem.

\subsection{Spectral gap bounds with semidefinite programming}\label{sec:intro-SDP}
In Ref. \cite{Knabe1988}, Knabe observed that the spectral gap of a Hamiltonian $H$ with ground-state energy zero is at least $\epsilon > 0$ if and only if $H^2 \succeq \epsilon H$, where $X \succeq Y$ means that $X - Y$ is positive semidefinite (PSD). He then used this idea to prove a spectral gap for the AKLT model and certain other lattice spin models. More recently, Rai \emph{et al.} \cite{Rai2026} generalized Knabe's approach by formulating a semidefinite program (SDP) whose solution lower bounds the spectral gap in certain lattice spin models. Semidefinite programming or bootstrap methods have been applied to a variety of problems in many-body physics \cite{Barthel2012,Baumgratz2012,Mazziotti2023,Gao2025,Scheer2026,Han2020,Lin2020,Kazakov2023,Navascues2007,Navascues2008,Berenstein2024,Wang2024,Schouten2025,Fawzi2024,Cho2022,Cho2025,Cho2026a,Cho2026b,Paul2026,Gao2026a,Scheer2025,Chadha2026,Onder2026,Cho2026c,Hastings2022}, and Ref. \cite{Rai2026} showed that spectral gap lower bounds are also amenable to this approach.

However, the approach of Ref. \cite{Rai2026} relies crucially on the Hamiltonian being a sum of strictly local PSD lattice operators, and as such it cannot be directly applied to fractional quantum Hall pseudopotentials. To understand this, consider a Hamiltonian $H = \sum_{j=0}^{L-1} h_j$ where $h_j$ is a PSD spin operator supported on sites with distance at most $l_0$ from $j$ in a 1D lattice with $L \geq 2 l_0 + 1$ sites and periodic boundary conditions\footnote{In periodic boundary conditions, the distance $|j-k|$ means $\min(j - k \mod L, k - j \mod L)$.}. We assume that $H$ has ground-state energy zero, which is to say that $H$ is frustration-free. Additionally, we assume that $H$ is translation invariant, so there is a group of unitary translation operators $T_j$ such that $T_j h_k T_j^\dagger = h_{j+k}$. The main difficulty in lower bounding $H^2$ is due to its nonlocality: $H^2 = \sum_{j,k} h_j h_k$ contains terms coupling arbitrarily distant lattice sites. Ref. \cite{Rai2026} solves this problem by using the fact that $[h_j, h_k] = 0$ for $|j-k| > 2l_0$. In this case, one can lower bound
\begin{equation}
H^2 \succeq \sum_j h_{j,l}, \quad h_{j,l} = \frac{1}{2}\sum_{|j-k| \leq 2l} \{h_j, h_k\}
\end{equation}
for any $l \geq l_0$ because the product of two commuting PSD operators is PSD. The authors then formulate a finite SDP which lower bounds the spectral gap for all $L$ sufficiently large. Although they state it quite differently, their program is equivalent to
\begin{align}
&\max_{\epsilon, \{\mathcal{O}_n\}} \epsilon\\
&\text{s.t. } \mathcal{S}\left(h_{0,l} - \epsilon h_0 - \sum_n \mathcal{O}^\dagger_n \mathcal{O}_n\right) = 0\label{eq:H-squared-identity}
\end{align}
where $\mathcal{S}$ is the symmetrization map $\mathcal{S}(X) = \frac{1}{L}\sum_{j=0}^{L-1} T_j X T_j^\dagger$ as in Ref. \cite{Scheer2026}, and the operators $\mathcal{O}_n$ form a finite collection of linear combinations of operators with support only on sites in some interval around $0$, which is independent of $L$. If \cref{eq:H-squared-identity} is satisfied, then $\epsilon$ is a lower bound on the spectral gap for all system sizes $L$ sufficiently large, so any admissible point proves a spectral gap lower bound for $H$ in the thermodynamic limit.

\subsection{Pseudopotential Hamiltonians}
Many fractional quantum Hall wavefunctions such as the Laughlin and Moore--Read states, as well as the Read--Rezayi sequence, satisfy a zero-clustering property and can be realized as an exact zero-energy ground state of a pseudopotential Hamiltonian \cite{Haldane1983,Trugman1985,Bernevig2008a,Bernevig2008b,Bernevig2008c}. In second quantization, these Hamiltonians take the form $\sum_M Q_M^\dagger Q_M$ where $Q_M$ is an annihilation operator for some number of fermions \cite{Ortiz2013,Chen2015,Mazaheri2015}. In this paper, we will focus on the first pseudopotential (sometimes called the $V_1$ Hamiltonian), which is a two-body interaction, meaning that $Q_M$ is a fermion pair annihilation operator. The index $M$ labels some quantum number of the pair: on the sphere or plane this is angular momentum, on a cylinder it is linear momentum along the periodic direction. The Hamiltonian can also be expressed in real-space in terms of projected continuum local operators $c_z$ \cite{Schleith2026,Gao2026b} as $H = \int d\tilde{\mu}(z)\, Q_z^\dagger Q_z$ where $Q_z \propto c_z \partial_z c_z$ (see \cref{sec:discussions} and \cref{sec:coherent} for more detail).

While pseudopotentials share structural similarity to frustration-free local lattice Hamiltonians, they differ in a crucial way: the operators $Q_M$ and $Q_{M'}^\dagger$ generally do not commute and are non-local. More specifically, the operator $Q_M$ is a linear combination of fermion bilinears $c_n c_{M-n}$ with coefficients that decay in $|M - 2n|$ on a scale that diverges with system size. For example, on the cylinder, $M = k_y L_y/(2\pi)$ is an integer labeling the momentum along the periodic direction. $Q_M$ is a linear combination of terms $c_n c_{M-n}$ with coefficients which are Gaussian in $|M - 2n|$ with width proportional to $L_y/\ell_B$ ($\ell_B$ is the magnetic length) which diverges in the 2D limit $L_y \rightarrow \infty$. In fact, the thin cylinder limit $L_y \ll \ell_B$ where such Hamiltonians can be truncated to a local 1D lattice model is the only\footnote{See note added in \cref{sec:note-added}.} case in which the Laughlin gap has been proved, for the truncated Hamiltonian \cite{Nachtergaele2021,Warzel2022,Warzel2023}. A similar consideration applies on the sphere, where $M$ labels angular momentum and the operator $Q_M$ is a linear combination of terms $c_n c_{M-n}$ with coefficients which decay in $|M - 2n|$ over a scale proportional to $\sqrt{N_\Phi}$ for $M$ close to zero \cite{Ortiz2013}.

Alternatively, we can study the interaction in real-space where the operators $Q_z \propto c_z \partial_z c_z$ are labeled by the continuum 2D complex position $z$. Here, we also find that $Q_z$ and $Q_w^\dagger$ do not commute, although the operator norm of their commutator decays as a Gaussian whose width is $\ell_B$. We may then hope to analytically bound the long-distance part and use an SDP to bound the short-distance part. However, the real-space continuum representation has the additional difficulty that the remaining short-distance part is still an integral that cannot be expressed simply in terms of a finite number of operators, which makes it difficult to apply semidefinite programming methods. These issues make a direct application of the method of Ref. \cite{Rai2026} to pseudopotentials challenging.

\subsection{Summary of our contributions}
In this work, we solve the difficulties just described and derive a family of finite SDPs which provide rigorous lower bounds on the spectral gap for the first pseudopotential Hamiltonian. Working on the sphere with $N_\Phi$ flux quanta, a key ingredient in our approach is to use $\text{SU}(2)$ symmetry to combine the continuum real-space representation and the discrete angular momentum one. In particular, while we express our Hamiltonian in terms of angular momentum operators, we maintain explicit spatial symmetrization which enables us to map the angular momentum operators to the simplest ones close to the south pole. This allows us to write $H^2$ as a sum of two terms $H^2 = K_1 + K_2$ which can be roughly understood as describing the short-distance ($K_1$) and long-distance ($K_2$) pieces of $H^2$. We derive an analytical bound $K_2 \succeq -\omega_2 H$ and use a finite SDP to prove $K_1 \succeq \omega_1 H$, so that $H^2 \succeq \epsilon H$ with $\epsilon = \omega_1 - \omega_2$. The size of the SDP for $K_1$ is independent of $N_\Phi$, but the SDP data does depend parametrically on $N_\Phi$. This is an important difference from spectral gap lower bounds in local lattice models, e.g. in Ref. \cite{Rai2026}. In that case, one can find an SDP that does not explicitly depend on the system size and that provides bounds valid for all sufficiently large systems with periodic boundary conditions. Here, the SDP size is determined by a pair of cutoff parameters $r' \leq r \leq N_\Phi$, and the spectral gap bound generally increases with both $r'$ and $r$. As shown in \cref{fig:bounds}, the numerically obtained gap bounds become positive when the cutoffs exceed certain thresholds that generally depend on $N_\Phi$. In the thermodynamic limit, we obtain a numerical lower bound of $0.0693$. In principle, we expect our bound to approach the true spectral gap with sufficient resources. We are careful to normalize everything in such a way that the limit of the SDP data with $N_\Phi \to \infty$ converges to a well defined SDP, results from which are included in \cref{fig:bounds}. Interestingly, the SDP does not depend on filling, so our results lower bound the overall spectral gap of the Hamiltonian. Stated another way, we lower bound the spectral gap for all fillings for which $H$ has a zero mode. These correspond to $N$ electrons with $N \leq N_* = \lfloor \frac{N_\Phi}{3} \rfloor + 1$ corresponding to fillings $0 \leq \nu \leq 1/3$. These results also imply the charge gap at $N_*$ is at least $\epsilon$. To see this, recall the charge gap is defined as $E(N + 1) + E(N - 1) - 2E(N)$, where $E(N)$ is the ground-state energy for $N$ electrons at fixed number of flux quanta $N_\Phi$, and note that $E(N_*) = E(N_* - 1) = 0$ while $E(N_* + 1) \geq \epsilon$ since there are no zero modes for $N > N_*$ \cite{Ortiz2013,Chen2015, Mazaheri2015}.

It is almost always the case in the bootstrap literature that SDPs are solved numerically, not exactly. In order to turn our numerical spectral gap lower bounds into mathematically rigorous results, we develop a novel algorithmic approach. First, we show that a basis transformation renders all SDP data polynomial in $t = 1/N_\Phi$ with rational coefficients. Next, we introduce an algorithm which produces exact rational nearly-optimal lower bounds for a strictly feasible rational SDP using iterative refinement, $p$-adic lifting with large random primes, and rational reconstruction. This allows us to find exact rational certificates for spectral gap lower bounds at arbitrary rational $t$, including $t = 0$. Our algorithm is related to earlier approaches based on rounding and projection of numerical solutions, as in Refs. \cite{Peyrl2008,Kaltofen2012,Naceur2025}, and involves the solution of a subsystem of the affine constraints, as in Ref. \cite{Jansson2008}. However, our method differs in that it produces exact rational certificates for large SDPs with highly redundant and ill-conditioned affine constraints, without relying on special structure.

A lower bound on the SDP at $t = 0$ is not by itself enough to lower bound the limit infimum as $t \to 0$ of the spectral gap, which is the precise meaning of the spectral gap in the thermodynamic limit\footnote{Proving a positive lower bound on the limit infimum of the gap as $t \to 0$ is equivalent to proving a positive lower bound on the gap for all $N_\Phi$ above some finite threshold.}. To prove this, it is sufficient to additionally show that the rank of the affine part of the SDP does not drop in the $t \to 0$ limit\footnote{Strictly speaking, one additionally needs to know that the affine constraints are consistent for all nonzero $t$ in some neighborhood of zero, but in our case this is guaranteed by construction.}. This is similar to Robinson's constraint qualification in the perturbation theory of conic programs \cite{Robinson1975,BonnansShapiro2000} but differs in that we explicitly treat the case of redundant affine constraints. To prove that this rank stability criterion holds, we introduce an algorithm that exactly computes the rank of a matrix of polynomials in $t$ with rational coefficients in some open neighborhood of $t = 0$, based on Taylor series lifting modulo large random primes and rational reconstruction. We note that these algorithms may be useful for turning bootstrap results into rigorous theorems in many other contexts.

Along with this paper, we release on GitHub \cite{LaughlinGapCertificate} a verifier program and an exact rational certificate for the SDP with $r' = 15$, $r = 22$, and $t = 0$. Together with the rank stability results, this establishes a spectral gap greater than $1/100$ in the thermodynamic limit. We note that this lower bound can be improved by increasing $r'$ and $r$ as discussed earlier.

\section{Derivation of the SDP}
We now describe our approach to lower bounding the spectral gap of the first pseudopotential Hamiltonian on the sphere. We review the basic theory of pseudopotentials and the Laughlin state on the sphere using Haldane's formalism in \cref{app:fqhe-sphere}. However, as we will see, the pseudopotentials on the sphere are fully determined by symmetry so this section should be understandable to any reader familiar with second quantization and the representation theory of $\text{SU}(2)$. The starting point is the fact that the lowest Landau level (LLL) on the sphere forms a spin-$s$ unitary irreducible representation (irrep) of $\text{SU}(2)$, where $s = N_\Phi/2$ and $N_\Phi \geq 0$ is the number of flux quanta. We denote the LLL by $W_s$ and take this as the single-particle Hilbert space. By Schur's lemma, any $\text{SU}(2)$ symmetric two-particle interaction can be written in the block diagonal form
\begin{equation}
H^{(2)} = \sum_{k=0}^{2s} V_k P_{2s-k}
\end{equation}
where $P_J$ is the orthogonal projector onto the spin-$J$ irrep $\tilde{W}_J$ defined by the addition of angular momentum formula $W_s \otimes W_s = \oplus_{J=0}^{2s} \tilde{W}_J$. We note that $\tilde{W}_{2s-k}$ is the space of two-particle states with relative angular momentum $k$. For an interaction that decays sufficiently quickly, the coefficients $V_k$ for small $k$ will be dominant.

The second quantized form of this Hamiltonian in the angular momentum basis takes the form
\begin{equation}\label{eq:second-quantized-pseudopotentials}
H = \sum_{k=0}^{2s} V_k \sum_{M=-(2s-k)}^{2s-k} Q^\dagger_{2s-k,M} Q_{2s-k,M}
\end{equation}
where
\begin{equation}\label{eq:define-Q-J-M}
Q_{J,M} = \frac{1}{\sqrt{2}} \sum_{m_1=-s}^s \sum_{m_2=-s}^s C^{J,M}_{s,m_1,s,m_2} c_{m_2} c_{m_1},
\end{equation}
$C^{2s-k,M}_{s,m_1,s,m_2}$ is a Clebsch--Gordan (CG) coefficient, and $c_m$ is a fermion annihilation operator with $\{c_n, c_m\} = 0$ and $\{c_n, c^\dagger_m\} = \delta_{m,n}$. Useful properties of the CG tensor $C^{J,M}_{j_1,m_1,j_2,m_2}$ are reviewed in \cref{app:Clebsch-Gordan}. The term with coefficient $V_k$ in \cref{eq:second-quantized-pseudopotentials} is called the $k$th pseudopotential. The operator $Q_{2s-k,M}$ is a pair annihilation operator where the $z$-component of the pair angular momentum is $M$.

Fermion antisymmetry and the exchange relation for the CG tensor imply that $Q_{2s-k,M} = 0$ for all even $k$. The dominant interaction is therefore the first pseudopotential. Going forward, we consider only this term. That is to say, we take $2s \geq 1$, $V_1 = 1$, and $V_k = 0$ for all $k \neq 1$. For notational simplicity, we denote $Q_{2s-1,M}$ by $Q_M$ so that the Hamiltonian simplifies to
\begin{equation}\label{eq:H-Q-M}
H = \sum_{M=-(2s-1)}^{2s-1} Q^\dagger_M Q_M.
\end{equation}
This Hamiltonian is a sum of positive semidefinite (PSD) operators. Additionally, it has at least one zero-energy ground state for each electron number $N \leq \lfloor\frac{N_\Phi}{3}\rfloor + 1$, corresponding to fillings $0 \leq \nu \leq 1/3$ \cite{Laughlin1983,Trugman1985,Ortiz2013,Chen2015}. These zero-energy states are the Laughlin state with possibly some number of quasiholes.

\subsection{Symmetrized gap identity}
As mentioned in \cref{sec:intro-SDP}, the spectral gap of $H$ is at least $\epsilon > 0$ if and only if
\begin{equation}\label{eq:gap-inequality}
H^2 \succeq  \epsilon H.
\end{equation}
We now begin working towards a proof of \cref{eq:gap-inequality} by deriving a useful expression for $H^2$.

For notational simplicity, from here on we will leave limits off of sums and leave parameter ranges implicit when they can be inferred from context. We denote by $U$ the unitary representation of $\text{SU}(2)$ on the many-body Hilbert space. This is generated by angular momentum operators $\mathcal{L}_x, \mathcal{L}_y, \mathcal{L}_z$ where
\begin{align}
\mathcal{L}_z &= \sum_m m c^\dagger_m c_m\\
\mathcal{L}_\pm &= \sum_m \sqrt{s(s+1) - m(m \pm 1)} c^\dagger_{m \pm 1} c_m
\end{align}
and $\mathcal{L}_x = (\mathcal{L}_- + \mathcal{L}_+)/2$, $\mathcal{L}_y = i(\mathcal{L}_- - \mathcal{L}_+)/2$. The number operator is
\begin{equation}
\mathcal{N} = \sum_m c^\dagger_m c_m.
\end{equation}
We use $D(J,g)$ to denote the matrix representation for the unitary spin-$J$ irrep of $\text{SU}(2)$ using the conventions in \cref{app:Clebsch-Gordan}. For example, the $c^\dagger_m$ operators transform under $D(s,g)$, by which we mean
\begin{equation}
U(g) c^\dagger_m U^\dagger(g) = \sum_n D_{n,m}(s,g) c^\dagger_n
\end{equation}
for all $g \in \text{SU}(2)$. By \cref{eq:define-Q-J-M} and the definition of the CG coefficients, the $Q^\dagger_M$ operators transform under $D(2s-1,g)$.

Following Ref. \cite{Scheer2026}, we define the symmetry averaging map\footnote{We could include the number $\text{U}(1)$ and complex conjugation symmetries in the symmetry averaging map as well, following Ref. \cite{Scheer2026}. However, since these symmetries are so simple it is more straightforward in this case to handle them separately.}
\begin{equation}
\mathcal{S}(X) = \int_{g\in\text{SU}(2)} U(g) X U^\dagger(g) \, d\mu
\end{equation}
where $\mu$ is the left-invariant probability measure on $\text{SU}(2)$ provided by Haar's theorem. Since $H$ is symmetric under $\text{SU}(2)$, we have $\mathcal{S}(H) = H$ and $\mathcal{S}(H^2) = H^2$.

We begin by using $[Q_{M_0}, Q_{M_1}] = 0$ to write
\begin{equation}\label{eq:H-squared-initial}
H^2 = \sum_M \mathcal{S}(Q^\dagger_M H Q_M) + \sum_{M_0,M_1} \mathcal{S}(Q^\dagger_{M_0}[Q_{M_0}, Q^\dagger_{M_1}] Q_{M_1}).
\end{equation}
The second term can be simplified to
\begin{equation}\label{eq:H-squared-second-term-initial}
\begin{split}
&\sum_{M_0,M_1} \mathcal{S}(Q^\dagger_{M_0}[Q_{M_0}, Q^\dagger_{M_1}] Q_{M_1}) = -H\\
&+2\sum_{l,M_0,M_1} C^{2s-1,M_0}_{s,l,s,M_0-l} C^{2s-1,M_1}_{s,l,s,M_1-l} \mathcal{S}(Q^\dagger_{M_0} c_{M_0-l} c^\dagger_{M_1-l} Q_{M_1}).
\end{split}
\end{equation}
This expression is written in terms of the trion operator $Q^\dagger_M c_{M-l}$ which creates two particles and one hole for which the total $z$-component of angular momentum is $l$.

We now seek to simplify $\mathcal{S}(Q^\dagger_{M_0} c_{M_0-l} c^\dagger_{M_1-l} Q_{M_1})$. To do so, we will recombine the $Q^\dagger$ and $c$ operators into combinations that transform according to specific angular momentum representations. Note that while the $c_m$ operators transform under the conjugate of $D(s,g)$, the operators $c^-_m = (-1)^{s-m} c_{-m}$ transform under $D(s,g)$, which means that $Q^\dagger c^-$ operators transform as creation operators of three particles rather than two-particles and a hole. We therefore define operators $R^\dagger_{J,j}$ for each $|s-1| \leq J \leq 3s-1$ and $-J \leq j \leq J$ which transform under $D(J, g)$ by
\begin{align}
R^\dagger_{J,j} &= \sum_{M,m} \tilde C^{J,j}_{2s-1,M,s,m} Q^\dagger_M c_m \label{eq:define-R}\\
\tilde{C}^{J,j}_{2s-1,M,s,m} &= (-1)^{s-m} C^{J,j}_{2s-1,M,s,-m}.\label{eq:define-tilde-C}
\end{align}
This relation can be inverted to give
\begin{equation}\label{eq:Q-c-minus-R}
Q^\dagger_M c_m = \sum_{J,j} \tilde C^{J,j}_{2s-1,M,s,m} R^\dagger_{J,j}.
\end{equation}
It is straightforward to show using Schur's lemma that
\begin{equation}\label{eq:R-symmetrization}
\mathcal{S}(R^\dagger_{J,j} R_{K,k}) = \delta_{J,K} \delta_{j,k} \mathcal{S}(R^\dagger_{J,j_0} R_{J,j_0})
\end{equation}
for any $j_0$. Combining \cref{eq:define-R,eq:Q-c-minus-R,eq:R-symmetrization} produces
\begin{equation}\label{eq:Q-c-symmetrization}
\begin{split}
&\mathcal{S}(Q^\dagger_{M_0} c_{M_0-l} c^\dagger_{M_1-l} Q_{M_1}) = \\
&\sum_J \tilde{C}^{J, l}_{2s-1,M_0,s,M_0-l} \tilde{C}^{J, l}_{2s-1,M_1,s,M_1-l} \mathcal{S}(R^\dagger_{J,j_0} R_{J,j_0}).
\end{split}
\end{equation}
Combining \cref{eq:H-squared-second-term-initial,eq:Q-c-symmetrization} with the identity
\begin{equation}\label{eq:CG-sum-identity}
\sum_l \left(\sum_M C^{2s-1,M}_{s,l,s,M-l} \tilde{C}^{J, l}_{2s-1,M,s,M-l}\right)^2 = \delta_{J,s}(4s-1)
\end{equation}
which is proved in \cref{app:CG-summation-identity}, we obtain
\begin{equation}\label{eq:H-squared-second-term-intermediate}
\begin{split}
&\sum_{M_0,M_1} \mathcal{S}(Q^\dagger_{M_0}[Q_{M_0}, Q^\dagger_{M_1}] Q_{M_1}) =\\
&-H +2  (4s-1) \mathcal{S}(R^\dagger_{s,s_0} R_{s,s_0})
\end{split}
\end{equation}
for any $s_0$. Schur's lemma again implies\footnote{We note that \cref{eq:H-symmetrized,eq:Q-H-Q-symmetrized} imply that $\mathcal{S}(Q^\dagger_M Q_M)$ and $\mathcal{S}(Q^\dagger_M H Q_M)$ are independent of $M$.}
\begin{align}
H &= (4s-1)\mathcal{S}(Q^\dagger_M Q_M)\label{eq:H-symmetrized}\\
\sum_{M'} Q^\dagger_{M'} H Q_{M'} &= (4s-1)\mathcal{S}(Q^\dagger_M H Q_M)\label{eq:Q-H-Q-symmetrized}
\end{align}
for any $M$. Finally, we combine \cref{eq:H-squared-initial,eq:H-squared-second-term-intermediate,eq:H-symmetrized,eq:Q-H-Q-symmetrized} to obtain
\begin{equation}\label{eq:H-squared-symmetrized}
H^2 = (4s-1)\mathcal{S}(Q^\dagger_M H Q_M + 2 R^\dagger_{s,s_0} R_{s,s_0} - Q^\dagger_M Q_M)
\end{equation}
for any $M$ and $s_0$. Importantly, this identity only contains trion operators $R$ of spin $s$, and the scale factor $4s-1$ is common to all terms.

Using \cref{eq:H-symmetrized,eq:H-squared-symmetrized}, we see that \cref{eq:gap-inequality} is equivalent to
\begin{equation}\label{eq:symmetrized-gap-inequality}
\mathcal{S}(Q^\dagger_M H Q_M + 2 R^\dagger_{s,s_0} R_{s,s_0} - (1 + \epsilon)Q^\dagger_M Q_M) \succeq 0
\end{equation}
for any $M$ and $s_0$. As a result, if \cref{eq:symmetrized-gap-inequality} holds then the spectral gap of $H$ is at least $\epsilon$. This equation notably does not depend on filling and can therefore be used to establish the spectral gap for all fillings $\nu \leq 1/3$.

\subsection{South pole expansion}
As explained in \cref{app:fqhe-sphere}, the coordinates of the spherical LLL near the south pole converge in the thermodynamic limit to holomorphic coordinates for the planar LLL. This motivates us to relabel operators according to their angular momentum relative to its minimal possible value. Specifically, we define
\begin{equation}\label{eq:define-tilde-c}
\tilde{c}_m = c_{-s+m}
\end{equation}
for $0 \leq m \leq 2s$ and
\begin{equation}\label{eq:define-tilde-Q}
\begin{split}
\tilde{Q}_M &= Q_{-2s+M}\\
&= \frac{1}{\sqrt{2}} \sum_m C^{2s-1,-2s+M}_{s,-s+m,s,-s+ M-m} \tilde{c}_{M-m} \tilde{c}_m
\end{split}
\end{equation}
for $1 \leq M \leq 4s-1$. The number operator is
\begin{equation}
\mathcal{N} = \sum_m \tilde{c}^\dagger_m \tilde{c}_m
\end{equation}
and we have $\mathcal{L}_z = -s \mathcal{N} + \mathcal{J}_z$ where
\begin{equation}
\mathcal{J}_z = \sum_m m \tilde{c}^\dagger_m \tilde{c}_m.
\end{equation}
Additionally, $\mathcal{L}_\pm = \sqrt{2s} \mathcal{J}_{\pm}$ where
\begin{align}
\mathcal{J}_+ &= \frac{1}{\sqrt{2s}}\sum_m \sqrt{(2s-m) (m+1)} \tilde{c}^\dagger_{m+1} \tilde{c}_m\\
\mathcal{J}_- &= \frac{1}{\sqrt{2s}} \sum_m \sqrt{(2s+1-m) m} \tilde{c}^\dagger_{m-1} \tilde{c}_m.
\end{align}
We have included the factors of $\sqrt{2s}$ so that the coefficients of $\mathcal{J}_\pm$ converge in the $s \to \infty$ limit\footnote{It is conceptually clarifying to see how the symmetries of the plane emerge from those of the sphere. In the $s \to \infty$ limit, we have $\mathcal{J}_+ \to \sum_m \sqrt{m+1} \tilde{c}^\dagger_{m+1} \tilde{c}_m$ and $\mathcal{J}_- \to \sum_m \sqrt{m} \tilde{c}^\dagger_{m-1} \tilde{c}_m$. We recognize these expressions as raising and lowering operators for a bosonic mode, namely the magnetic translations of the plane. Additionally, $\mathcal{J}_z$ can be interpreted as the generator of rotations of the plane. In the presence of both number and angular momentum symmetries, the $s$-dependent shift between $\mathcal{L}_z$ and $\mathcal{J}_z$ will not cause difficulty.}.

A crucial step in our approach is to split the various terms in $H^2$ in \cref{eq:H-squared-symmetrized} into a south pole piece that is a finite sum even as $s \to \infty$ and a remainder piece that we will bound analytically. Specifically, for some threshold value $1 \leq r \leq 2s$ and integers $1 \leq M_0, M_1 \leq r$, we will define operators $\Omega_1$ and $\Omega_2$ such that
\begin{equation}\label{eq:Omega-1-Omega-2}
\Omega_1 + \Omega_2 = \tilde{Q}^\dagger_{M_0} H \tilde{Q}_{M_0} + 2 R^\dagger_{s,-s} R_{s,-s} - \tilde{Q}^\dagger_{M_1} \tilde{Q}_{M_1}
\end{equation}
and such that $\Omega_1$ is a polynomial of $\tilde{c}^\dagger_m$ and $\tilde{c}_m$ with $0 \leq m \leq r$.

Writing $R_{s,-s}^\dagger$ in terms of the south pole operators, we find
\begin{equation}
R^\dagger_{s,-s} = \sum_M \tilde{C}^{s,-s}_{2s-1,-2s+M,s,-s+M} \tilde{Q}^\dagger_M \tilde{c}_M.
\end{equation}
which combines particle pair and hole creation operators at the same angular momentum relative to the south pole. We partition $R^\dagger_{s,-s}$ into a sum of two operators
\begin{align}
\tilde{R}^\dagger_1 &= \sum_{M=1}^r \tilde{C}^{s,-s}_{2s-1,-2s+M,s,-s+M} \tilde{Q}^\dagger_M \tilde{c}_M\label{eq:define-tilde-R-1}\\
\tilde{R}^\dagger_2 &= \sum_{M=r+1}^{2s} \tilde{C}^{s,-s}_{2s-1,-2s+M,s,-s+M} \tilde{Q}^\dagger_M \tilde{c}_M.\label{eq:define-tilde-R-2}
\end{align}
For some threshold parameter $1 \leq r' \leq r$ which we will discuss more later, we now define
\begin{align}
\Omega_1 &= \sum_{N=1}^{r'} \tilde{Q}^\dagger_{M_0} \tilde{Q}^\dagger_N \tilde{Q}_N \tilde{Q}_{M_0} + 2 \tilde{R}^\dagger_1 \tilde{R}_1 - \tilde{Q}^\dagger_{M_1} \tilde{Q}_{M_1}\label{eq:Omega-1}\\
\Omega_2 &= \sum_{N=r'+1}^{4s-1} \tilde{Q}^\dagger_{M_0} \tilde{Q}^\dagger_N \tilde{Q}_N \tilde{Q}_{M_0} + 2\tilde{R}^\dagger_2 \tilde{R}_2\label{eq:Omega-2-PSD}\\
&+ 2\tilde{R}^\dagger_1 \tilde{R}_2 + 2\tilde{R}^\dagger_2 \tilde{R}_1\label{eq:Omega-2-non-PSD}
\end{align}
and note that \cref{eq:Omega-1-Omega-2} is satisfied.

The next step is to derive an analytical bound on $\mathcal{S}(\Omega_2)$ that is proportional to the Hamiltonian. First, we note that the first two terms in $\Omega_2$ are PSD so that $\mathcal{S}(\Omega_2) \succeq 2\mathcal{S}(\tilde{R}^\dagger_1 \tilde{R}_2 + \tilde{R}^\dagger_2 \tilde{R}_1)$. Second, we use the operator inequality $ x(X^\dagger Y + Y^\dagger X) \succeq -|x|(X^\dagger X + Y^\dagger Y)$, with $x$ real, for each term in $\tilde R_1^\dagger \tilde R_2 + \tilde R_2^\dagger \tilde R_1$. Finally, we use $-\mathcal S(\tilde Q_M^\dagger \tilde c_M \tilde c_M^\dagger \tilde Q_M) = -\mathcal S(\tilde Q_M^\dagger  \tilde Q_M) + \mathcal S(\tilde Q_M^\dagger \tilde c_M^\dagger \tilde c_M \tilde Q_M)$ which implies $-\mathcal S(\tilde Q_M^\dagger \tilde c_M \tilde c_M^\dagger \tilde Q_M) \succeq -\mathcal S(\tilde Q_M^\dagger  \tilde Q_M)$. Leaving further details to \cref{app:analytical-bound}, this yields the analytical bound
\begin{equation}\label{eq:bound-equation-Omega-2}
\mathcal{S}(\Omega_2) \succeq -\omega_2 \mathcal{S}(\tilde{Q}^\dagger_M \tilde{Q}_M)
\end{equation}
for any $M$, where
\begin{align}
\omega_2 &= 4 \chi_r (\chi_{2s} - \chi_r)\label{eq:define-omega_2}\\
\chi_M &= \sum_{N=1}^M |C^{s,-s}_{2s-1,-2s+N,s,s-N}|.\label{eq:define-chi_M}
\end{align}
Our goal is now to use an SDP to prove the additional bound
\begin{equation}\label{eq:bound-equation-Omega-1}
\mathcal{S}(\Omega_1) \succeq \omega_1 \mathcal{S}(\tilde{Q}^\dagger_M \tilde{Q}_M)
\end{equation}
which only includes operators $\tilde{c}_m$ with $m \leq r$. Then \cref{eq:symmetrized-gap-inequality} follows with $\epsilon = \omega_1 - \omega_2$. As a result, if \cref{eq:bound-equation-Omega-1} holds then the spectral gap of $H$ is at least $\omega_1 - \omega_2$.

\subsection{Semidefinite program}\label{sec:SDP-numerical}
Suppose we are able to derive an equation of the form
\begin{equation}\label{eq:Omega-1-bootstrap}
\begin{split}
\Omega_1 &= \omega_1 \tilde{Q}^\dagger_{M_1} \tilde{Q}_{M_1} + \sum_j \mathcal{O}^\dagger_j \mathcal{O}_j\\
&+ \frac{1}{2} ([\mathcal{J}_+, \tilde{\mathcal{O}}] + [\tilde{\mathcal{O}}^\dagger, \mathcal{J}_-])
\end{split}
\end{equation}
for some operators $\mathcal{O}_j, \tilde{\mathcal{O}}$. Clearly, $\mathcal{S}(\mathcal{O}^\dagger \mathcal{O}) \succeq 0$ for any operator $\mathcal{O}$ and $\mathcal{S}([\mathcal{J}_\pm, \tilde{\mathcal{O}}]) = 0$ for any operator $\tilde{\mathcal{O}}$. As a result, \cref{eq:bound-equation-Omega-1} follows from \cref{eq:Omega-1-bootstrap} upon the application of $\mathcal{S}$. We now explain how \cref{eq:Omega-1-bootstrap} can be derived using a semidefinite program.

We call an operator of the form
\begin{equation}\label{eq:define-tilde-c-monomials}
\tilde{c}^\dagger_{m_1} \cdots \tilde{c}^\dagger_{m_j} \tilde{c}_{n_k} \cdots \tilde{c}_{n_1}
\end{equation}
for $m_1 < \cdots < m_j$ and $n_1 < \cdots < n_k$ a $\tilde{c}$-monomial. The index ordering convention ensures that the set of $\tilde{c}$-monomials is closed under adjoints. We additionally say that a linear combination of $\tilde{c}$-monomials is a $\tilde{c}$-polynomial. Let $\mathcal{P}_\lambda$ for $1 \leq \lambda \leq \Lambda$ and $\tilde{\mathcal{P}}$ be sets of $\tilde{c}$-polynomials. We choose a set $\mathcal{Q}$ of operators of the form $(q^\dagger + q)/2$ for $\tilde{c}$-monomials $q$ which forms a basis for the span of
\begin{equation}
\begin{split}
&\bigcup_\lambda \left\{p^\dagger_1 p_2 + p_2^\dagger p_1 \Big| p_1, p_2 \in \mathcal{P}_\lambda\right\}\\
&\bigcup \left\{[\mathcal{J}_+, \tilde{p}] + [\tilde{p}^\dagger, \mathcal{J}_-] \Big| \tilde{p} \in \tilde{\mathcal{P}}\right\}.
\end{split}
\end{equation}
This allows us to define the symmetric structure constant matrices $\Gamma^{\lambda,q}$ by
\begin{equation}\label{eq:define-Gamma-lambda-q}
\frac{1}{2}(p^\dagger_1 p_2 + p_2^\dagger p_1) = \sum_{q \in \mathcal{Q}} \Gamma^{\lambda,q}_{p_1,p_2} q \text{ for } p_1, p_2 \in \mathcal{P}_\lambda
\end{equation}
and the structure constant vectors $\tilde{\Gamma}^q$ by
\begin{equation}\label{eq:define-tilde-Gamma-q}
\frac{1}{2}([\mathcal{J}_+, \tilde{p}] + [\tilde{p}^\dagger, \mathcal{J}_-]) = \sum_{q \in \mathcal{Q}} \tilde{\Gamma}^q_{\tilde{p}} q \text{ for } \tilde{p} \in \tilde{\mathcal{P}}.
\end{equation}

For each $\lambda$, let $Z_\lambda$ be a real PSD matrix of dimension $|\mathcal{P}_\lambda|$. We can decompose $Z_\lambda$ in the form
\begin{equation}
Z_\lambda = \sum_{k=1}^{\text{rank}(Z_\lambda)} z^{(\lambda,k)} (z^{(\lambda,k)})^T
\end{equation}
for some real vectors $z^{(\lambda,k)}$. Now let
\begin{equation}
\mathcal{O}_{\lambda,k} = \sum_{p \in \mathcal{P}_\lambda} z^{(\lambda,k)}_p p.
\end{equation}
\cref{eq:define-Gamma-lambda-q} implies
\begin{equation}
\mathcal{O}^\dagger_{\lambda,k} \mathcal{O}_{\lambda,k} = \sum_q (z^{(\lambda,k)})^T \Gamma^{\lambda,q} z^{(\lambda,k)} q
\end{equation}
so that
\begin{equation}\label{eq:sum-O-dag-O}
\sum_{\lambda,k} \mathcal{O}^\dagger_{\lambda,k} \mathcal{O}_{\lambda,k} = \sum_{\lambda,q} \text{tr}(\Gamma^{\lambda,q} Z_\lambda) q.
\end{equation}
Similarly, let $\tilde{\mathcal{O}} = \sum_{\tilde{p} \in \tilde{\mathcal{P}}} z_{\tilde{p}} \tilde{p}$ for some real vector $z$ with length $|\tilde{\mathcal{P}}|$. \cref{eq:define-tilde-Gamma-q} implies
\begin{equation}\label{eq:J-tilde-O-commutators}
\frac{1}{2} ([\mathcal{J}_+, \tilde{\mathcal{O}}] + [\tilde{\mathcal{O}}^\dagger, \mathcal{J}_-]) = \sum_q (z^T \tilde{\Gamma}^q) q.
\end{equation}

We now assume that $\mathcal{Q}$ is large enough so that $\Omega_1$ and $\tilde{Q}^\dagger_{M_1} \tilde{Q}_{M_1}$ lie in its span. We can then write
\begin{align}
\Omega_1 &= \sum_q X_q q\label{eq:H-squared-decomposition}\\
\tilde{Q}^\dagger_{M_1} \tilde{Q}_{M_1} &= \sum_q Y_q q\label{eq:normalization-decomposition}
\end{align}
for some real coefficients $X_q$ and $Y_q$.

Finally, we define the spectrahedron
\begin{equation}\label{eq:spectrahedron}
\begin{split}
&K = \left\{(\omega_1,z,Z_1, \dots, Z_\Lambda) \Bigg | \begin{aligned}
&(1) \text{ holds for all } \lambda,\\
&(2) \text{ holds for all } q
\end{aligned}\right\}\\
&(1):\quad Z_\lambda \succeq 0\\
&(2):\quad X_q = \omega_1 Y_q + \sum_\lambda \text{tr}(\Gamma^{\lambda,q} Z_\lambda) + (\tilde{\Gamma}^q)^T z.
\end{split}
\end{equation}
\cref{eq:sum-O-dag-O,eq:J-tilde-O-commutators,eq:H-squared-decomposition,eq:normalization-decomposition} imply that each element of $K$ yields an equation of the form in \cref{eq:Omega-1-bootstrap}. The convex optimization problem
\begin{equation}\label{eq:SDP}
\max_{(\omega_1,z,Z_1, \dots, Z_\Lambda) \in K} \omega_1
\end{equation}
is an SDP\footnote{Strictly speaking, in all SDPs in this paper ``max" should really be ``sup" meaning supremum, as it is not important if the supremum value is actually attained. However, we will write ``max" for simplicity.}.

We now specify the sets $\mathcal{P}_\lambda$, $\tilde{\mathcal{P}}$, and $\mathcal{Q}$ along with the integers $M_0$ and $M_1$ which all together define the spectrahedron in \cref{eq:spectrahedron} and SDP in \cref{eq:SDP}. In terms of our two threshold parameters $1 \leq r' \leq r \leq 2s$, we define the set
\begin{equation}\label{eq:define-P}
\begin{split}
\mathcal{P} = &\{\tilde{Q}_M | 1 \leq M \leq r'\}\\
&\bigcup\left\{\tilde{c}^\dagger_m \tilde{c}_n \tilde{Q}_M \Bigg| \begin{aligned}
&0 \leq m, n \leq r', 1 \leq M \leq r'\\
&1 \leq  -m + n + M \leq r'
\end{aligned}\right\}\\
&\bigcup\left\{\tilde{c}_m \tilde{c}_n \tilde{Q}_M \Bigg | \begin{aligned}
&0 \leq m < n \leq r', 1 \leq M \leq r'\\
&m + n + M \leq 2 r' + 4
\end{aligned}\right\}\\
&\bigcup\{\tilde{c}^\dagger_M \tilde{Q}_M | 1 \leq M \leq r\}\\
&\bigcup\{\tilde{c}_M \tilde{Q}_M | 1 \leq M \leq r\}.
\end{split}
\end{equation}
For any $\tilde{c}$-polynomial $p$ we can assign the particle number $n_0$ and angular momentum $j_0$ quantum numbers by
\begin{align}
[\mathcal{N}, p] &= n_0 p\\
[\mathcal{J}_z, p] &= j_0 p.
\end{align}
Suppose that $\Lambda$ is the number of distinct quantum number tuples $(n_0,j_0)$ which occur for $p \in \mathcal{P}$. If we enumerate these distinct quantum number tuples by a label $1 \leq \lambda \leq \Lambda$ we can define
\begin{equation}
\mathcal{P}_\lambda = \{p \in \mathcal{P} | p \text{ has quantum number tuple } \lambda\}.
\end{equation}

Next, we define $\tilde{\mathcal{Q}}$ to be the smallest set of $\tilde{c}$-monomials which spans
\begin{equation}
\bigcup_\lambda \{p_1^\dagger p_2 | p_1, p_2 \in \mathcal{P}_\lambda\}
\end{equation}
and then
\begin{equation}
\mathcal{Q} = \{(q^\dagger + q)/2 | q \in \tilde{\mathcal{Q}}\}.
\end{equation}
We define $\tilde{\mathcal{P}}_{v,d}$ to be the set of $\tilde{c}$-monomials $p$ of the form in \cref{eq:define-tilde-c-monomials} for which $m_j \leq v$, $n_k \leq v$, $j \leq d$, $k \leq d$, the quantum number tuple is $(n_0, j_0) = (0, -1)$, and
\begin{equation}
[\mathcal{J}_+, p] + [p^\dagger, \mathcal{J}_-] \in \text{Span}(\mathcal{Q}).
\end{equation}
We then define
\begin{equation}
\tilde{\mathcal{P}} = \tilde{\mathcal{P}}_{r', 4} \bigcup \tilde{\mathcal{P}}_{r, 3}.
\end{equation}

Finally, we choose $M_0 = r'$ and $M_1 = 1$. Having defined $\mathcal{P}_\lambda$, $\tilde{\mathcal{P}}$, $\mathcal{Q}$, $M_0$, and $M_1$ we have specified a spectrahedron and SDP for each choice of thresholds $r'$ and $r$.

\subsection{Numerical results}
\begin{figure}
\centering
\includegraphics[scale=0.55]{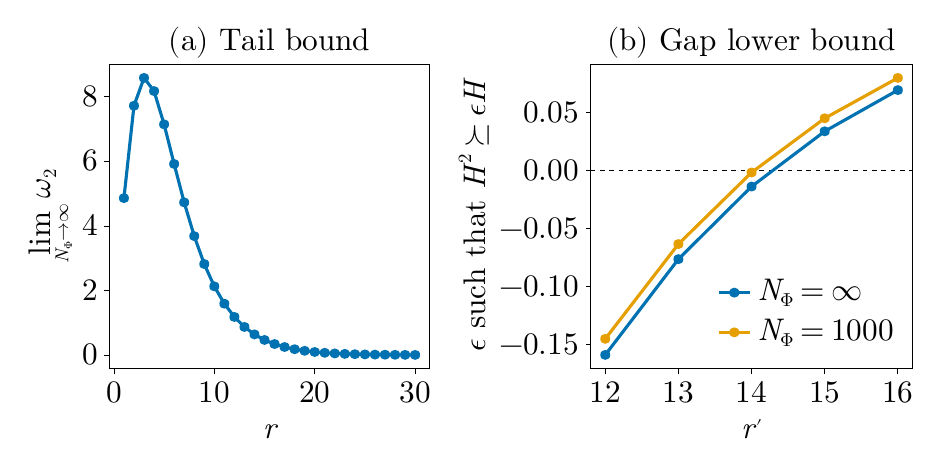}
\caption{(a) The tail bound $\omega_2$ in the thermodynamic limit $N_\Phi \to \infty$. We numerically compute these values using the upper bound in \cref{eq:tilde-omega_2-bounded} with $N_0 = 200$. Strictly speaking, this is an upper bound on $\lim_{N_\Phi \to \infty} \omega_2$ but the error due to the finite $N_0$ value is far less than that intrinsic to floating point arithmetic. (b) Spectral gap lower bound for $N_\Phi = \infty$ and $N_\Phi = 1000$ with $r = 25$. For $N_\Phi = \infty$, the last two data points are $0.0338$ for $r'=15$ and $0.0693$ for $r'=16$ (rounded down). For $N_\Phi = 1000$ the last two data points are $0.0450$ for $r'=15$ and $0.0797$ for $r'=16$ (rounded down).}
\label{fig:bounds}
\end{figure}

Our SDP produces a numerical lower bound on the gap that depends explicitly on $N_\Phi$ and the cutoffs $r'$ and $r$. The tail bound $\omega_2$ depends on $r$ but not on $r'$ (see \cref{fig:bounds}(a)). The SDP size depends on both $r'$ and $r$, but it grows much faster with $r'$ than with $r$. For example, the size of the set $\mathcal{P}$ in \cref{eq:define-P} grows cubically with $r'$ but only linearly with $r$. As a result, it is reasonable to choose a relatively large $r$ such that the tail contribution $\omega_2$ is small and study the SDP bound as a function of $r'$. For numerical results, we use the MOSEK solver\footnote{For convenience, we solve the SDP described in \cref{sec:SDP-rescaled} rather than the one described in \cref{sec:SDP-numerical}. However, these SDPs differ only by rescalings, exact relations, and a constant offset of their optimal objective values, so the results are mathematically equivalent. Additionally, we rescale the SDP data to reduce its dynamic range before solving SDPs since MOSEK uses $64$-bit floating point arithmetic.}\cite{mosek}.

In \cref{fig:bounds}(b), we choose $r = 25$ for which $\lim_{N_\Phi \to \infty} \omega_2 = 0.0184$ (rounded up) and plot the gap lower bound $\epsilon = \omega_1 - \omega_2$ as a function of $r'$ for $N_\Phi = 1000$ and $N_\Phi = \infty$. Both bounds become positive for $r' \geq 15$ and increase with increasing $r'$. For $r' = 16$ (the largest value reported), the numerical gap lower bound is $0.0693$ (rounded down) for $N_\Phi = \infty$ and $0.0797$ (rounded down) for $N_\Phi = 1000$. We observe that our lower bound decreases as $N_\Phi$ increases, which is in line with the gap evolution observed in exact diagonalization studies \cite{Haldane1985}.

\section{Exact certification and the thermodynamic limit}\label{sec:exact-certification}
Now that we have seen numerically that a positive spectral gap lower bound is possible, we would like to make this bound mathematically rigorous. The first difficulty is that the SDP data contains square roots and depends on $s$ in a non-trivial way. However, we will see that it is possible to render the $s$ dependence of all SDP data to be polynomial in
\begin{equation}\label{eq:define-t}
t = \frac{1}{2s} = \frac{1}{N_\Phi} \leq 1
\end{equation}
with rational coefficients by simply rescaling the $\tilde{c}$ operators. This basis transformation will give us much greater algorithmic and analytical control.

We will use the term $t$-polynomial to refer to a polynomial in $t$ with rational coefficients. First, we define
\begin{equation}\label{eq:define-f}
f_m = \sqrt{\xi_m} \tilde{c}_m
\end{equation}
where $\xi_m$ is the $t$-polynomial
\begin{equation}
\xi_m = m! \prod_{n=0}^{m-1} (1-nt)\label{eq:define-xi_m}.
\end{equation}
We call an operator of the form
\begin{equation}\label{eq:define-f-monomials}
f^\dagger_{m_1} \cdots f^\dagger_{m_j} f_{n_k} \cdots f_{n_1}
\end{equation}
for $m_1 < \cdots < m_j$ and $n_1 < \cdots < n_k$ an $f$-monomial. As in the case of $\tilde{c}$-monomials, the index ordering convention here ensures that the set of $f$-monomials is closed under adjoints. We refer to an operator which is a linear combination of $f$-monomials with coefficients that are $t$-polynomials (coefficients that are rational) as an $ft$-polynomial ($f$-polynomial). The set of $ft$-polynomials is closed under adjoints.

The $f_m$ operators satisfy
\begin{align}
\{f_m, f_n\} &= 0\label{eq:f-anticommutator-0}\\
\{f_m, f^\dagger_n\} &= \delta_{m,n} \xi_m.\label{eq:f-anticommutator-delta}
\end{align}
\cref{eq:define-xi_m,eq:f-anticommutator-0,eq:f-anticommutator-delta} can be used to write any product $p_1 p_2$ as an $ft$-polynomial where $p_1$ and $p_2$ are $f$-monomials. Consequently, the set of $ft$-polynomials is closed under products.

Next, we note that $\mathcal{J}_+$ can be written in the form
\begin{equation}
\mathcal{J}_+ = \sum_m \frac{1}{\xi_m} f^\dagger_{m+1} f_m
\end{equation}
from which we find\footnote{\cref{eq:J-fm,eq:J-fm-dagger} are understood to hold for all $m$ for which all terms are defined. The edge cases are $[\mathcal{J}_+, f_0] = [\mathcal{J}_+, f^\dagger_{2s}] = 0$.}
\begin{align}
[\mathcal{J}_+, f_m] &= -m (1-(m-1)t) f_{m-1}\label{eq:J-fm}\\
[\mathcal{J}_+, f^\dagger_m] &= f^\dagger_{m+1}.\label{eq:J-fm-dagger}
\end{align}
These relations can be used to write $[\mathcal{J}_+, p]$ and $[\mathcal{J}_-, p]$ as $ft$-polynomials for any $f$-monomial $p$. Consequently, the set of $ft$-polynomials is closed under commutators with $\mathcal{J}_\pm$.

We show in \cref{app:tilde-Q-f_m} that
\begin{align}
\tilde{Q}_M &= q_M \sqrt{\frac{(M-1)!}{\eta_M}}\label{eq:Q-in-q}\\
q_M &= \frac{1}{2}\sum_m \frac{2m-M}{m! (M-m)!} f_{M-m} f_m\label{eq:define-q_M}\\
\eta_M &= \prod_{l=2}^M (2 - lt).\label{eq:define-eta_M}
\end{align}
Additionally, we show in \cref{app:tilde-R_1-f_m} that
\begin{equation}
\mathcal{S}(2\tilde{R}^\dagger_1 \tilde{R}_1) = (2-\delta) \mathcal{S}(Q^\dagger_M Q_M) - \mathcal{S}(W)
\end{equation}
for any $M$, where\footnote{The way $\delta$ is defined here reflects the fact that $\lim_{s\to\infty} \delta$ is a positive function of $r$ which vanishes in the $r \to \infty$ limit. Specifically, we prove in \cref{app:omega-2-delta-thermodynamic} that $\lim_{s\to\infty} \delta = (r+2)/2^r$.}
\begin{align}
\label{eq:define-delta}
\delta &= 2 - \frac{1+t}{2-t} \sum_{M=1}^r \frac{\xi_M}{(M-1)!\eta_M}\\
W &= \frac{1+t}{2-t} \sum_{M,N=1}^r \frac{1}{\eta_M\eta_N} q^\dagger_M f^\dagger_N f_M q_N.\label{eq:define-W}
\end{align}
This means
\begin{equation}
\mathcal{S}(\Omega_1) = \mathcal{S}(\Omega'_1) - \delta\mathcal{S}(\tilde{Q}^\dagger_M \tilde{Q}_M)
\end{equation}
for any $M$, where\footnote{It is more computationally convenient to work with $\Omega'_1$ than $\Omega_1$ because all operators in $\Omega'_1$ are normal ordered.}
\begin{equation}\label{eq:define-Omega-prime-1}
\Omega'_1 = \sum_{N=1}^{r'} \tilde{Q}^\dagger_{M_0} \tilde{Q}^\dagger_N \tilde{Q}_N \tilde{Q}_{M_0} - W + \tilde{Q}^\dagger_{M_1} \tilde{Q}_{M_1}.
\end{equation}

Note that $\tilde{Q}$ operators enter $\Omega'_1$ only in pairs $\tilde{Q}^\dagger_M$ and $\tilde{Q}_M$. By \cref{eq:Q-in-q}, these pairs provide denominators $\eta_M$ which depend polynomially on $t$. Additionally, $W$ has factors of $2-t$ and $\eta_M$ in the denominator. In order to cancel these polynomials, we define the $t$-polynomial
\begin{equation}
\tilde{\eta} = \frac{1}{2^{2r-1}} (2-t)\eta_r^2
\end{equation}
and note that $\tilde{\eta} > 0$ and additionally
\begin{equation}
\lim_{t \to 0} \tilde{\eta} = 1.
\end{equation}
By \cref{eq:Q-in-q,eq:define-q_M,eq:define-eta_M,eq:define-W,eq:define-Omega-prime-1}, we see that $\tilde{\eta}\Omega'_1$ and $\tilde{\eta} \tilde{Q}^\dagger_{M_1} \tilde{Q}_{M_1}$ are $ft$-polynomials. Furthermore, if we are able to show
\begin{equation}\label{eq:bound-equation-Omega-prime-1}
\tilde{\eta}\mathcal{S}(\Omega'_1 - \mu_1 \tilde{Q}^\dagger_{M_1} \tilde{Q}_{M_1}) \succeq 0
\end{equation}
then \cref{eq:bound-equation-Omega-1} follows with $\omega_1 = \mu_1 - \delta$. As a result, if \cref{eq:bound-equation-Omega-prime-1} holds then the spectral gap of $H$ is at least $\mu_1 - \delta - \omega_2$.

\subsection{Rescaled semidefinite program}\label{sec:SDP-rescaled}
We now describe the modifications to the SDP in \cref{sec:SDP-numerical} which are needed for exact certification. Instead of \cref{eq:Omega-1-bootstrap}, we aim to prove
\begin{equation}\label{eq:Omega-prime-1-bootstrap}
\tilde{\eta}(\Omega'_1 - \mu_1 \tilde{Q}^\dagger_{M_1} \tilde{Q}_{M_1}) = \sum_j \mathcal{O}^\dagger_j \mathcal{O}_j + \frac{1}{2} ([\mathcal{J}_+, \tilde{\mathcal{O}}] + [\tilde{\mathcal{O}}^\dagger, \mathcal{J}_-])
\end{equation}
for some operators $\mathcal{O}_j, \tilde{\mathcal{O}}$, noting that \cref{eq:bound-equation-Omega-prime-1} follows from \cref{eq:Omega-prime-1-bootstrap} upon the application of $\mathcal{S}$. We now proceed as in \cref{sec:SDP-numerical} except that each $\tilde{Q}_M$ is replaced by $q_M$, each $\tilde{c}_m$ is replaced by $f_m$, $\mathcal{P}_\lambda$ and $\tilde{\mathcal{P}}$ are sets of $ft$-polynomials, $\mathcal{Q}$ is a set of $f$-polynomials, and $X$ and $Y$ are $t$-polynomials defined by
\begin{align}
\tilde{\eta}\Omega'_1 &= \sum_q X_q q\label{eq:H-squared-decomposition-t}\\
\tilde{\eta} \tilde{Q}^\dagger_{M_1} \tilde{Q}_{M_1} &= \sum_q Y_q q.\label{eq:normalization-decomposition-t}
\end{align}
Importantly, this means that all SDP data, by which we mean $\Gamma^{\lambda,q}_{p_1,p_2}$, $\tilde{\Gamma}^q_{\tilde{p}}$, $X_q$, and $Y_q$ are $t$-polynomials. The spectrahedron is
\begin{equation}\label{eq:spectrahedron-t}
\begin{split}
&K = \left\{(\mu_1,z,Z_1, \dots, Z_\Lambda) \Bigg | \begin{aligned}
&(1) \text{ holds for all } \lambda,\\
&(2) \text{ holds for all } q
\end{aligned}\right\}\\
&(1):\quad Z_\lambda \succeq 0\\
&(2):\quad X_q = \mu_1 Y_q + \sum_\lambda \text{tr}(\Gamma^{\lambda,q} Z_\lambda) + (\tilde{\Gamma}^q)^T z.
\end{split}
\end{equation}
and the SDP is
\begin{equation}\label{eq:SDP-t}
\max_{(\mu_1,z,Z_1, \dots, Z_\Lambda) \in K} \mu_1
\end{equation}

\subsection{Thermodynamic limit}
We now consider the thermodynamic limit $N_\Phi \to \infty$ or equivalently $t \to 0$ of the spectral gap lower bound $\mu_1 - \delta - \omega_2$. Recall that $\delta$ and $\omega_2$ are given by explicit expressions in \cref{eq:define-delta,eq:define-omega_2}. We explain in \cref{app:omega-2-delta-thermodynamic} how we can exactly compute $\tilde{\delta} = \lim_{N_\Phi \to \infty} \delta$ and how we can compute a rigorous upper bound to $\tilde{\omega}_2 = \lim_{N_\Phi \to \infty} \omega_2$ using interval arithmetic. The main challenge is to rigorously certify a lower bound for $\tilde{\mu}_1 = \liminf_{N_\Phi \to \infty} \mu_1$.

We begin by defining a vectorization map which can be applied to elements of $K$. For an $n\times n$ symmetric matrix $Z$, we define the lower triangular vectorization of $Z$ to be the length $n(n+1)/2$ vector
\begin{equation}
(Z_{1,1}, Z_{2,1}, \dots, Z_{n,1}, Z_{2,2}, \dots, Z_{n,2}, \dots, Z_{n,n})
\end{equation}
formed by stacking the lower triangular parts of each column. For $u = (\mu_1, z, Z_1, \dots, Z_\Lambda)$, we define $\text{vec}(u)$ to be the concatenation of $\mu_1$, $z$, and the lower triangular vectorizations of $Z_1, \dots, Z_\Lambda$. The map $\text{vec}$ is linear and invertible and can be applied to elements of $K$.

We define the cone $C$ to consist of all vectors $v$ such that $\text{vec}^{-1}(v) = (\mu_1, z, Z_1, \dots, Z_\Lambda)$ and each $Z_\lambda$ is PSD. Additionally, we define a matrix $A$ of $t$-polynomials and a vector $b$ of $t$-polynomials such that $u$ satisfies condition (2) in \cref{eq:spectrahedron-t} if and only if $A \text{vec}(u) = b$. $A$ is formed from elements of $Y$, $\Gamma$, and $\tilde{\Gamma}$ while $b$ is formed from $X$. We see that $\text{vec}$ maps $K$ to
\begin{equation}\label{eq:mathcal-K}
\mathcal{K} = \{v \in C | A v = b\}
\end{equation}
so the SDP can be written
\begin{equation}\label{eq:sdp-mathcal-K}
\max_{v \in \mathcal{K}} c^T v
\end{equation}
where $c$ is the vector with $1$ in the first component and $0$ everywhere else.

We now include the argument $t$ for $\mathcal{K}$, $A$, and $b$ to emphasize that these objects depend on $t$. We have
\begin{equation}\label{eq:mathcal-K-t}
\mathcal{K}(t) = \{v \in C | A(t) v = b(t)\}
\end{equation}
and we define the optimal objective values
\begin{equation}
d(t) = \max_{v \in \mathcal{K}(t)} c^T v
\end{equation}
for each $t$.

Although $t$ physically must be an inverse integer, since the elements of $A(t)$ and $b(t)$ are $t$-polynomials we now extend $t$ to take any real value. We note that, by construction, the vector $b(t)$ is in the image of $A(t)$ for all nonzero $t$ in some neighborhood of zero\footnote{To see this, it is easiest to start with the SDP in \cref{eq:spectrahedron,eq:SDP} with the given choices for $\mathcal{P}_\lambda$, $\tilde{\mathcal{P}}$, $\mathcal{Q}$, $M_0$, and $M_1$. Here, the statement that the objective lies in the image of the equality constraints follows directly from the definition of $\Omega_1$. The transformation to the SDP given in \cref{eq:spectrahedron-t,eq:SDP-t} involves rescaling, exact operator identities, removal of terms proportional to the normalization, and a symmetrization identity which can be accounted for by commutators with $\mathcal{J}_+$. This implies that $b(t) \in \text{im}(A(t))$ whenever $1/t$ is a sufficiently large positive integer. Since a nonzero polynomial cannot have infinitely many zeros, it follows that $b(t) \in \text{im}(A(t))$ for all but finitely many values $t$.}. We would like to say that $\lim_{t \to 0} d(t) = d(0)$ so that $d(0)$ can be used as $\mu_1$ in the spectral gap lower bound. However, this type of convergence does not always hold for one-parameter families of SDPs for the following reasons. First, we do not know that $\mathcal{K}(t)$ is always nonempty so it could be that some or all $d(t)$ values are not well defined. Second, even if $d(t)$ is well defined for all sufficiently small $t$, it could be that $\lim_{t\to 0} d(t)$ does not converge or even that $\liminf_{t\to 0} d(t) < d(0)$.

There is, however, a simple criterion under which we can conclude that $d(t)$ is well defined for all sufficiently small $t$ and $\liminf_{t\to 0} d(t) \geq d(0)$, which is precisely the result needed for the spectral gap lower bound. To explain this, let $\text{int}(C) \subset C$ denote the interior of $C$, by which we mean the set of vectors $v$ such that $\text{vec}^{-1}(v) = (\mu_1, z, Z_1, \dots, Z_\Lambda)$ and each $Z_\lambda$ is strictly positive definite. Now suppose that the conditions
\begin{equation}
\begin{split}
&(1):\quad \mathcal{K}(0) \cap \text{int}(C) \neq \{\}\\
&(2):\quad \lim_{t\to 0} \text{rank}(A(t)) = \text{rank}(A(0))
\end{split}
\end{equation}
are satisfied. The essential idea is that condition $(1)$ means that for $t = 0$, the SDP has an interior feasible point, and condition $(2)$ means that any such point can be perturbed to find an interior feasible point for any sufficiently small $t$. More formally, condition $(1)$ implies that there exists an interior feasible point $v_0 \in \mathcal{K}(0) \cap \text{int}(C)$ for the SDP with $t = 0$, which in particular means $A(0) v_0 = b(0)$. If we take $v(t) = v_0 + t v_1$ for some $v_1$, then the equation $A(t) v(t) = b(t)$ is equivalent to $A(t) v_1 = r(t)$ where $r(t) = (b(t) - A(t) v_0)/t$ is a vector of $t$-polynomials. In particular, $\lim_{t\to 0} r(t)$ is well defined. Since $b(t)$ is in the image of $A(t)$ for all $t$ in some neighborhood of zero, we see that the same is true of $r(t)$ as well. This allows us to solve for $v_1$ as $v_1 = A^+(t) r(t)$ where $A^+(t)$ is the Moore--Penrose inverse of $A(t)$. Condition $(2)$ means that the smallest positive singular value of $A(t)$ does not vanish as $t \to 0$, and therefore $\lim_{t \to 0} ||A^+(t)||$ is well defined. We then have $\lim_{t \to 0} ||v(t) - v_0|| = \lim_{t \to 0} t ||A^+(t) r(t)|| = 0$. Since a positive definite matrix remains positive definite under all sufficiently small perturbations, it follows that $v(t) \in \mathcal{K}(t) \cap \text{int}(C)$ for all $t$ sufficiently small. This implies that for all $t$ sufficiently small, $d(t)$ is well defined and $d(t) \geq c^T v(t)$ so that $\liminf_{t\to 0} d(t) \geq c^T v_0$. Since this holds for all $v_0 \in \mathcal{K}(0) \cap \text{int}(C)$ and interior feasible points can be chosen with objective value arbitrarily close to $d(0)$, we conclude $\liminf_{t\to 0} d(t) \geq d(0)$, as desired.

In order to rigorously certify our SDP results, we will find an interior feasible point $v_0 \in \mathcal{K}(0) \cap \text{int}(C)$ and prove that condition $(2)$ is satisfied. We will then have $\liminf_{t\to 0} d(t) \geq c^T v_0$ so that $c^T v_0$ is a rigorously certified lower bound to $\tilde{\mu}_1$.

\subsection{Exact feasible interior points}\label{sec:exact-feasible-interior-points}
We now consider the general problem of finding an interior feasible point for an SDP of the form in \cref{eq:mathcal-K,eq:sdp-mathcal-K} with nearly optimal objective value. We take the cone $C$ to be a product of free scalar variables\footnote{Previously we had one scalar variable and a vector of scalar variables. Here, for simplicity we absorb all scalar variables into one vector $z$.} $z$ and vectorized PSD variables $Z_1, \dots, Z_\Lambda$. The generalization of this method to SDPs including nonnegative scalar variables is straightforward, so we will not consider them here. Additionally, going forward we will identify a point $v \in C$ with its inverse vectorization $\text{vec}^{-1}(v) = (z, Z_1, \dots, Z_\Lambda)$.

We denote by $\text{int}(C)$ the elements of $C$ for which all PSD variables are strictly positive definite. We assume that the matrix $A$ and vector $b$ are rational and $b$ is in the image of $A$. For any linear system of equations $A' v' = b'$, we say that the residual error of $v'$ is $||b' - A' v'||_\infty$ where $||w||_\infty = \max_j |w_j|$ is the infinity norm for a vector $w$. We assume that one has access to an ``oracle" program which can produce approximate primal-dual pairs for an SDP of the type considered here, where the primal and dual positivity conditions are exactly satisfied, where the residual error for the dual solution is at most a fixed value $\epsilon_0 < 1$, and such that the primal-dual pair is nearly optimal\footnote{By ``nearly optimal" we mean that the Karush--Kuhn--Tucker equations are nearly satisfied. However, we leave the notion of ``nearly" vague, since our goal here is to use this method to solve a particular problem, not to produce a formal algorithmic analysis. In practice, the solution accuracy from MOSEK is sufficient for our purposes.}. In practice, we use the MOSEK\footnote{We rescale the SDP data to reduce its dynamic range before solving SDPs since MOSEK uses $64$-bit floating point arithmetic. Additionally, MOSEK may return solutions which violate positivity constraints a very small amount, which is not an issue because we start with a large PSD buffer in \cref{subsec:starting-point}.} numerical SDP solver \cite{mosek}.

Suppose we start with some $v_0 \in \text{int}(C)$ which has nearly optimal objective value. We would like to find $v_1$ such that $v = v_0 + v_1$ is an interior feasible point. We note that $Av = b$ is equivalent to $A v_1 = r_0$ where $r_0 = b - A v_0$ is the initial residual. Since $b$ is in the image of $A$, $r_0$ is as well, so a mathematically valid solution would be to take $v_1 = A^+ r_0$ where $A^+$ is the Moore--Penrose inverse of $A$. Indeed, if $||r_0||$ is small, then $||v_1||$ will also be small, and if $v_0$ is sufficiently interior to $C$, then $v$ will be our desired solution. Although this idea works for small problems, exactly computing $A^+ r_0$ for a large rational matrix $A$ and vector $r_0$ is extremely difficult due to the phenomenon of ``coefficient growth" \cite{vonzurGathen2013}. That is to say, even if the numerators and denominators of $A^+ r_0$ are small, during intermediate steps of such an exact calculation, one often encounters rational numbers with numerators and denominators with an extremely high number of bits, which slows down the calculation and can use incredible amounts of memory. Similar approaches based on LDLT factorization also suffer from this problem. In our experience, methods that involve a large amount of rational arithmetic are not practical for solving SDPs of the size and complexity encountered here.

In our solution, instead of attempting to compute $A^+r_0$, we begin by finding a square maximal rank submatrix of $A$. Without loss of generality, we reorder the rows and columns so that this is the top left block $A_{0,0}$ of $A$. We then write $A v_1 = r_0$ in the block form
\begin{equation}\label{eq:A-v-r-block}
\begin{pmatrix}
A_{0,0} & A_{0,1}\\
A_{1,0} & A_{1,1}
\end{pmatrix}
\begin{pmatrix}
v_{1,0} \\ v_{1,1}
\end{pmatrix} =
\begin{pmatrix}
r_{0,0}\\ r_{0,1}
\end{pmatrix}.
\end{equation}
One solution of this equation is $v_{1,0} = A_{0,0}^{-1} r_{0,0}$ and $v_{1,1} = 0$\footnote{Since $\text{rank}(A_{0,0}) = \text{rank}(A)$ and $r_0$ is in the image of $A$, the equations $A_{1,0} v_{1,0} + A_{1,1} v_{1,1} = r_{0,1}$ are redundant in the presence of $A_{0,0} v_{1,0} + A_{0,1} v_{1,1} = r_{0,0}$. This is why we are able to set $v_{1,1} = 0$ and find a solution for the full system of equations.}. Although this solution may have larger norm than $A^+ r_0$, it will be far easier to compute because it only involves solving the linear system $A_{0,0} v_{1,0} = r_{0,0}$ instead of the factorization needed for the Moore--Penrose inverse. On the other hand, if we can make $||r_0||$ small enough, it will still produce an exact feasible point $v$.

With this motivation, we break the problem into the following primitive pieces.
\begin{enumerate}
\item Find a maximal rank square submatrix $A_{0,0}$ of $A$.
\item Find a starting interior point $w_0$ which is close to an optimal solution but can have nonzero residual and has PSD buffer at least $\gamma$ for some chosen $\gamma > 0$, by which we mean that the eigenvalues of the PSD matrices in $w_0$ are all at least $\gamma$.
\item Use an iterative algorithm involving the oracle to find a sequence of points $w_k$ for $k \geq 1$ such that the residual error decreases exponentially with $k$ (unless the residual error becomes zero) while the objective value decreases by no more than a specified value.
\item Taking $v_0 = w_k$ for some $k$, solve $A_{0,0} v_{1,0} = r_{0,0}$ for $v_{1,0}$. Defining $v_{1,1} = 0$, we then obtain $v_1$ so that $v = v_0 + v_1$ satisfies $A v = b$.
\item Prove that $v \in \text{int}(C)$.
\end{enumerate}
We now describe our solutions to these sub-problems, and then finish by providing the logical control flow of the overall algorithm.

\subsubsection{Maximal rank square submatrix}\label{subsec:maximal-rank}
Since $A$ is a rational matrix, if $p$ is any prime which does not divide any denominators of elements of $A$, we can reduce $A$ modulo $p$. We can then perform Gaussian elimination modulo $p$ which produces the row and column indices $I_R$ and $I_C$ for a maximal rank square submatrix $A[I_R, I_C]$ modulo $p$. Importantly, there is no coefficient growth in calculations modulo $p$. Since $A[I_R, I_C]$ is invertible modulo $p$, it must also be invertible over the rationals. We perform this calculation using several large randomly chosen primes and compute the maximal size of the resulting submatrices. With high probability, any choice among these with the highest observed modular rank will indeed be a maximal rank square submatrix over the rationals. Of course, there is a small probability of underestimating the true rank of $A$, but this does not create a correctness problem because we will exactly verify our linear solution $A v_1 = r_0$.

We finish by mentioning two important implementation details. First, since we use a $64$-bit computer architecture, it is convenient to work with randomly chosen $31$-bit primes. This allows us to perform modular arithmetic operations directly with $64$-bit integers, which is very efficient. We use this idea for all arithmetic modulo primes in this paper. Second, before reducing modulo primes, we use the COLAMD algorithm \cite{Davis2004} to find row and column permutations which are designed to reduce fill-in during Gaussian elimination.

\subsubsection{Starting interior point}\label{subsec:starting-point}
We begin by solving a shifted SDP to find an interior point with PSD buffer at least $\gamma > 0$. By a shifted SDP, we mean that we replace the PSD variables $Z_\lambda$ with $\tilde{Z}_\lambda + \gamma I$ and use $\tilde{Z}_\lambda$ as new PSD variables. If we solve this SDP with the oracle using variables $z$ and $\tilde{Z}_\lambda$ and then transform back to variables $z$ and $Z_\lambda$, we obtain a starting point $w_0$ with PSD buffer $\gamma$. This is important because at the end we will need to prove that $v \in \text{int}(C)$ and we will need nonnegligible PSD buffer in this proof.

\subsubsection{SDP refinement}\label{subsec:sdp-refinement}
We describe here an iterative algorithm which starts at an interior point $w_0$ and uses the oracle to produce a sequence of points $w_k$ for $k \geq 1$ with residual error exponentially decreasing with $k$ (unless the residual error becomes zero). Additionally, the objective value for $w_k$ is lower bounded by $c^T w_k \geq c^T w_0 - B_0$ where $B_0 > 0$ is the total loss budget, which is an input to the algorithm. We note that our algorithm is quite similar to the SDP refinement method of Appendix A in Ref. \cite{Mohammadisiahroudi2025}. The main difference is that their goal is to reduce the complementarity gap and feasibility errors simultaneously, while we focus only on reducing the residual error with bounded objective loss and preserved interiority.

Going forward, when we say that we solve an SDP, we mean that we apply the oracle. Consider step $k \geq 0$, at which we begin with $w_k$ and remaining loss budget $B_k > 0$. Let $r_k = b - A w_k$ and let $\alpha_k = ||r_k||_\infty$. If $\alpha_k = 0$, then we have found an exact feasible point, so we can end the algorithm. Otherwise, suppose $\alpha_k > 0$. We define $w_{k+1} = w_k + \alpha_k u_k$ for some $u_k$. The equation $A w_{k+1} = b$ is equivalent to $A u_k = r_k/\alpha_k$, and notably $||r_k/\alpha_k||_\infty = 1$. For some fixed value $\gamma_{\text{max}} > 0$, we define $\gamma_k = \min(\gamma_{\text{max}}, B_k/\alpha_k)$ and we define an optimization problem
\begin{equation}\label{eq:correction-SDP}
\begin{split}
&\min_{\beta_k \geq 0, u_k} \beta_k\\
&\text{s.t. } \beta_k w_k + u_k \in C,\\
&\quad A u_k = r_k/\alpha_k,\\
&\quad c^T u_k \geq -\gamma_k.
\end{split}
\end{equation}
The Lagrange dual of this optimization problem can be written as an SDP of the form in \cref{eq:mathcal-K,eq:sdp-mathcal-K}, and therefore it can be solved by the oracle after dualization. If $(\beta_k, u_k)$ is the obtained solution, then $w_{k+1} = (1 - \alpha_k \beta_k)w_k + \alpha_k (\beta_k w_k + u_k)$. Since $w_k \in \text{int}(C)$ and $\beta_k w_k + u_k \in C$, we see that $w_{k+1} \in \text{int}(C)$ as long as $\alpha_k \beta_k < 1$. Furthermore, the residual for $w_{k+1}$ is $r_{k+1} = b - A w_{k+1} = \alpha_k(r_k/\alpha_k - A u_k)$. Recalling that the oracle has dual residual accuracy $\epsilon_0 < 1$, we find $\alpha_{k+1} = ||r_{k+1}||_\infty \leq \alpha_k \epsilon_0$. Finally, let $B_{k+1} = B_k + c^T(w_{k+1} - w_k)$ be the remaining loss budget and note that $c^T w_{k+1} \geq c^T w_k - \alpha_k \gamma_k \geq c^T w_k - B_k$. If $B_{k+1} \leq 0$ then the algorithm has failed because we exhausted the loss budget. Otherwise we continue to the next iteration.

Inductively, we see that as long as $\alpha_l \beta_l < 1$ and $B_{l+1} > 0$ for all $0 \leq l < k$, then $w_k$ is an interior point with residual error $\alpha_k \leq \alpha_0 \epsilon_0^k$ and objective value $c^T w_k \geq c^T w_0 - B_0$. In practice, we find the values of $\alpha_k \beta_k$ to be very small, which is consistent with the minimization of $\beta_k$ in \cref{eq:correction-SDP} and the exponential decay of $\alpha_k$. We include the upper bound $\gamma_{\text{max}}$ on $\gamma_k$ for numerical stability, because $\alpha_k$ decays exponentially.

\subsubsection{Linear system solution}\label{subsec:linear-system}
We now consider the problem of solving a linear equation $R x = y$ where $R$ is an invertible square rational matrix and $y$ is a rational vector. We will apply this algorithm to the linear equation $A_{0,0} v_{1,0} = r_{0,0}$.

Rather than working directly over the rationals, where coefficient growth can be a problem, we apply the $p$-adic lifting and rational reconstruction methods of Refs. \cite{Dixon1982,Wang1982}. $p$-adic lifting allows one to iteratively construct solutions of $R x \equiv y \mod p^n$ for a prime $p$ and increasing integers $n \geq 1$. Crucially, the only linear solves involved are modulo $p$ with a fixed left hand side $R \mod p$, which are very efficient especially because we can factor this matrix once and use this factorization for all solves. Rather than using a single prime, we use several concurrent processes with one prime per process. We periodically combine the results across all primes using the Chinese remainder theorem to obtain a solution of $R x \equiv y \mod \mathcal{M}$ where $\mathcal{M} = \prod_j p_j^n$ for some $n$ is the combined modulus. We then use the extended Euclidean algorithm to reconstruct $x$ as a rational vector where each numerator has magnitude at most $\lfloor \sqrt{(\mathcal{M}-1)/2}\rfloor$. We then check if our reconstructed $x$ satisfies $R x \equiv y \mod q$ for another prime $q$, which allows us to rapidly reject incorrect reconstructions. Only if this test modulo $q$ passes do we check if $Rx = y$ over the rationals. After any failed reconstruction or verification, we return to concurrent $p$-adic lifting, and the algorithm only terminates after a successful exact check $R x = y$.

\subsubsection{Proof of $v \in \text{int}(C)$}\label{subsec:positivity-certification}
We recall that $\text{vec}^{-1}(v) = (z, Z_1, \dots, Z_\Lambda)$ and $v \in \text{int}(C)$ is equivalent to saying that all $Z_\lambda$ are strictly positive definite. We now describe a simple method to certify that a rational matrix $Z$ is strictly positive definite. We first numerically diagonalize $Z$, yielding numerical eigenvalues and eigenvectors. We represent these approximate eigenvalues and eigenvectors as exact rational quantities, which gives us $Z \approx V D V^T$ where $V$ is a rational matrix and $D$ is a diagonal matrix with rational diagonal entries $d_1, \dots, d_n$. We define $d_0 = \min(d_1, \dots, d_n)$ and if $d_0 \leq 0$ the certification fails. Assuming $d_0 > 0$, we define
\begin{align}
\sigma &= ||Z - V D V^T||_\infty\label{eq:define-bound-sigma}\\
\tau &= ||I - V V^T||_\infty \label{eq:define-bound-tau}
\end{align}
where the matrix infinity norm is $||M||_\infty = \max_j \sum_k |M_{j,k}|$. We use this norm rather than the Frobenius or operator norm because this way $\sigma$ and $\tau$ are rational and can be computed exactly. Additionally, if $M$ is symmetric then $||M||_\infty$ upper bounds the operator norm $||M||$.

Since $D \succeq d_0 I$, we have $V D V^T \succeq d_0 V V^T$. Additionally, \cref{eq:define-bound-sigma} implies $Z - V D V^T \succeq -\sigma I$ and \cref{eq:define-bound-tau} implies $V V^T \succeq (1-\tau) I$. Putting these relations together gives
\begin{equation}
Z \succeq (d_0(1-\tau) - \sigma)I
\end{equation}
so that if $d_0(1-\tau) - \sigma > 0$ then $Z$ is strictly positive definite. In practice, if $Z$ is positive definite and $d_0$ is not too small, this method can reliably certify the strict positivity of $Z$.

\subsubsection{Overall algorithm}
We are now ready to put all of these pieces together into an algorithm that can produce an exact feasible interior point for a rational SDP of the form in \cref{eq:mathcal-K,eq:sdp-mathcal-K}. A simple version of the algorithm proceeds as follows. We begin by identifying a maximal rank square submatrix (with high probability) as described in \cref{subsec:maximal-rank}. Without loss of generality, we reorder the rows and columns of $A$ so that the maximal rank square submatrix is the top left block $A_{0,0}$, as in \cref{eq:A-v-r-block}. Next, we construct the starting point $w_0$ as described in \cref{subsec:starting-point}. We then enter a loop, for which each iteration begins with a point $w_k$. Initially this is $w_0$, but later it is the output of SDP refinement as explained in \cref{subsec:sdp-refinement}. We construct a potential solution $v = v_0 + v_1$ by taking $v_0 = w_k$ and solving \cref{eq:A-v-r-block} for $v_1$ with $v_{1,1} = 0$, using the method of \cref{subsec:linear-system}. If $c^T v \geq c^T w_0 - B_0$ where $B_0 > 0$ is the total objective loss budget, then we attempt exact certification of $v$ as described in \cref{subsec:positivity-certification}. If certification succeeds, we exit the loop. In all other cases, we perform some number of SDP refinement steps as described in \cref{subsec:sdp-refinement} and go back to the top of the loop. After a successful certification, we also check that $A v = b$ since we only know that $A_{0,0}$ has maximal rank with high probability. If we find $Av \neq b$, then we need to start over and choose different primes (although we can keep the same $w_0$). However, this is an exceedingly unlikely occurrence.

In practice, we have a more complex heuristic which enables us to avoid doing many reconstruction and certification steps. This only affects the overall runtime of the algorithm and not its correctness or underlying principles.

\subsection{Rank stability}\label{sec:rank-stability}
Let $A(t)$ be a matrix of polynomials of $t$ with rational coefficients. We now describe a general algorithm which is capable of proving that $A(t)$ is rank-stable, by which we mean
\begin{equation}\label{eq:rank-stability}
\lim_{t\to 0} \text{rank}(A(t)) = \text{rank}(A(0)).
\end{equation}
First, continuity implies that $\text{rank}(A(0))$ is a lower bound on $\text{rank}(A(t))$ for all $t$ sufficiently small so that $\lim_{t\to 0} \text{rank}(A(t)) \geq \text{rank}(A(0))$. Additionally, we can use the algorithm described in \cref{subsec:maximal-rank} to identify a maximal rank submatrix of $A(0)$ with high probability. Without loss of generality, we reorder the rows and columns of $A(t)$ so that the maximal rank submatrix of $A(0)$ is $A_{0,0}(0)$, where $A(t)$ is written in the block form
\begin{equation}
A(t) = \begin{pmatrix}
A_{0,0}(t) & A_{0,1}(t)\\
A_{1,0}(t) & A_{1,1}(t)
\end{pmatrix}.
\end{equation}
Although $A_{0,0}(0)$ is only maximal rank with high probability, it is certainly true that $\text{rank}(A_{0,0}(0)) \leq \text{rank}(A(0))$. Our goal now is to show that each row of $\begin{pmatrix} A_{1,0}(t) & A_{1,1}(t) \end{pmatrix}$ can be written as a linear combination of rows of $\begin{pmatrix} A_{0,0}(t) & A_{0,1}(t) \end{pmatrix}$ where the coefficients are rational functions of $t$ with rational coefficients which are regular at $t = 0$. If we can find these linear relations, it follows that $A_{0,0}(t)$ is a maximal rank submatrix of $A(t)$ for all $t$ in some open neighborhood of $0$, and so \cref{eq:rank-stability} is proved.

We have reduced the problem to solving all equations of the form
\begin{equation}
\begin{pmatrix}
A_{0,0}^T(t) \\ A_{0,1}^T(t)
\end{pmatrix} v(t) = \begin{pmatrix}
w_0(t) \\ w_1(t)
\end{pmatrix}
\end{equation}
where the right hand side is a column vector formed by transposing a row of $\begin{pmatrix} A_{1,0}(t) & A_{1,1}(t) \end{pmatrix}$ and $v(t)$ is a vector of rational functions of $t$ with rational coefficients which are regular at $t = 0$. Since $A_{0,0}(t)$ is invertible for all $t$ sufficiently small, $v(t)$ is uniquely determined by solving $A^T_{0,0}(t) v(t) = w_0(t)$. Once we have determined $v(t)$, we simply check $A^T_{0,1}(t) v(t) = w_1(t)$.

Now we explain how we solve $A^T_{0,0}(t) v(t) = w_0(t)$. We choose a prime $p$ which does not divide any denominators in $A_{0,0}(t)$ or $w_0(t)$ and for which $A_{0,0}(0)$ is invertible modulo $p$, and reduce this equation modulo $p$. We then solve the reduced equation modulo $t^n$ for increasing powers $n$ using Taylor series lifting. This algorithm is exactly analogous to $p$-adic lifting under the replacement of the ring of integers and the prime by the ring $\mathbb{F}_p[t]$ of polynomials in $t$ with coefficients in the field $\mathbb{F}_p$ and the irreducible polynomial $t$, respectively. After some number of iterations, we have evaluated $v(t)$ modulo $p$ and $t^n$ for some $n$. Let $m$ be the number of components of $v(t)$. We are interested in determining modulo $p$ a numerator vector $N(t)$ and denominator $D(t)$ with $D(0) = 1$ which solve the equation $N(t) \equiv v(t) D(t) \mod{p, t^n}$. Suppose that we require $\text{deg}(N) \leq d$ and $\text{deg}(D) \leq d$ for some integer $d$. Comparing coefficients of degrees from $d+1$ to $n-1$ gives us $m(n-1-d)$ equations in the $d$ unknown coefficients of $D(t)$. We choose $d = \lfloor m(n-1)/(m+1)\rfloor$, which is the largest value such that $m(n-1-d) \geq d$, and we then solve these equations for $D(t)$ mod $p$. After that, we probabilistically validate our solution by computing $N(t_0)$ modulo $p$ for several random values of $t_0$ by truncating $v(t) D(t)$ to degree at most $d$ and then substituting $t = t_0$. If for several randomly chosen values $t_0$ we find $A_{0,0}^T(t_0) N(t_0) \equiv D(t_0) w_0(t_0) \mod p$ then with high probability we have reconstructed $D(t) \mod{p}$ and we stop Taylor series lifting. If the test fails, we continue Taylor series lifting and attempt the reconstruction again after some number of iterations.

We perform the above algorithm for several primes $p = p_1, p_2, \dots$ and collect the resulting denominators. We apply the Chinese remainder theorem and rational reconstruction to all denominator polynomials of maximal degree to produce a proposed denominator polynomial $D(t)$ with rational coefficients. By construction, we will have $D(0) = 1$. We then reduce $D(t)$ modulo a held out prime $q$, and check $A^T_{0,0}(t_0) N(t_0) \equiv D(t_0) w_0(t_0) \mod q$ for several values of $t_0$ using the truncated method described above. If this test passes, then with high probability we have constructed the correct denominator $D(t)$ over the rationals. If not, we repeat reconstruction using additional primes until a candidate denominator polynomial $D(t)$ passes this test. At this point, we embark upon the expensive step of computing $N(t)$ exactly by solving $A^T_{0,0}(t)N(t) = D(t)w_0(t)$ coefficient by coefficient using the method of \cref{subsec:linear-system} and checking that this equality holds to all orders. If that succeeds, we check $A^T_{0,1}(t) N(t) = D(t) w_1(t)$. If this also succeeds, then we have proved the desired linear relation with $v(t) = N(t)/D(t)$ where $D(0) = 1$.

In practice, our implementation of this algorithm is highly parallel and optimized. However, the above explanation captures the key ideas and demonstrates why the resulting certification is fully rigorous.

\subsection{Exact results}\label{sec:exact-results}
\begin{figure}
	\centering
	\includegraphics[scale=0.55]{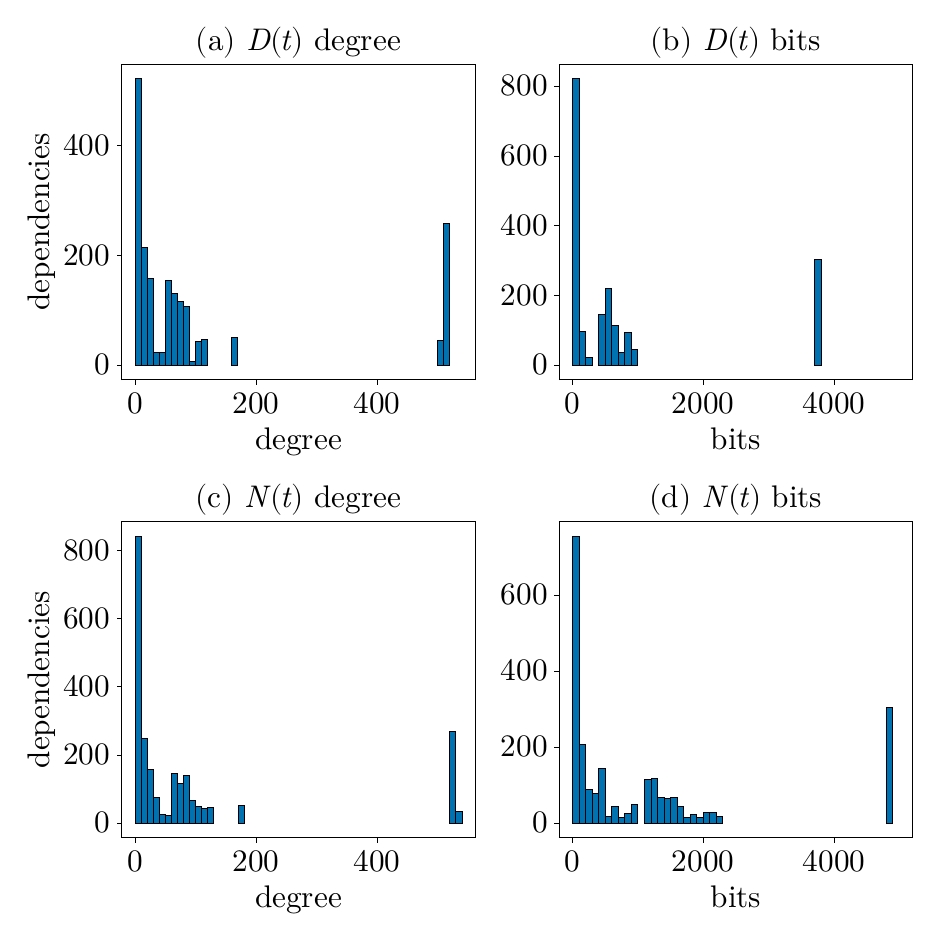}
	\caption{Rank stability results for the $r'=15$, $r=22$ SDP. Data is shown for $D(t)$ in the $1908$ cases in which the degree is positive. Data is shown for $N(t)$ in the $2338$ cases in which the degree is positive. (a) and (c) show the degree for $D(t)$ and $N(t)$ while (b) and (d) show the number of bits for $D(t)$ and $N(t)$. See \cref{sec:exact-results} for more details.}
	\label{fig:rank-stability}
\end{figure}
We apply the methods of \cref{sec:exact-certification} to the SDP in \cref{eq:spectrahedron-t,eq:SDP-t} with $r' = 15$ and $r = 22$. Using the notation of \cref{eq:mathcal-K-t}, the matrix $A(t)$ has dimension $65742 \times 357662$. Using the rank stability algorithm of \cref{sec:rank-stability} we find that for all $t$ in some neighborhood of zero, $A(t)$ has rank exactly $57639$, meaning that there are $8103$ redundant rows. Along with this paper, we release on GitHub \cite{LaughlinGapCertificate} a verifier program and an exact rational certificate for this SDP at $t = 0$ as described in \cref{sec:exact-feasible-interior-points}. The released data also includes the row and column indices of the maximal rank square submatrix used in the rank stability algorithm along with a prime which can be used to check its invertibility.

We now describe the results of the rank stability algorithm. First, we consider the denominator polynomial $D(t) = \sum_{n=0}^d D_n t^n$ for rational coefficients $D_n$ with $D_0 = 1$. The degree of $D(t)$ is $d$ and we define the number of bits of $D(t)$ to be the maximum number of binary digits required to represent any numerator or denominator among the coefficients $D_n$, with all common factors removed. In $6195$ of the dependencies, the denominator polynomial $D(t)$ has degree zero. \cref{fig:rank-stability}(a) and (b) show the degree and number of bits for $D(t)$ in the remaining $1908$ cases. We see that most dependencies have a relatively small degree and number of bits, but there is also a minority of cases with much higher degree and number of bits. The maximum degree is $517$ and the maximum number of bits is $3770$.

Next, we consider the numerator vector of polynomials $N(t)$. We define the degree and number of bits for $N(t)$ to be the maximum of these quantities, as defined above for $D(t)$, across the components of $N(t)$. We observe that the degree and number of bits for $N(t)$ are at least those for $D(t)$ in all cases. In $5765$ of the dependencies, the degree of $N(t)$ is zero, and in these cases the corresponding $D(t)$ also has degree zero. \cref{fig:rank-stability}(c) and (d) show the degree and number of bits for $N(t)$ in the remaining $2338$ cases. The observed pattern for the $D(t)$ histograms is also present for the $N(t)$ histograms. The maximum degree is $532$ and the maximum number of bits is $4882$.

\section{Discussion}
We close by discussing extensions and generalizations of the gap bounding approach introduced here.

\subsection{Generalization to other model states}
The generalization of our approach to other model fractional quantum Hall states is conceptually straightforward \cite{MooreRead,ReadRezayi,SimonPseudopotentials,Bernevig2008b,Bernevig2008c,Bernevig2008a}. For instance, the Moore--Read state is a zero mode for a $3$-body pseudopotential. In second quantization this can be written in a form similar to \cref{eq:H-Q-M} with $Q \sim ccc$ being a $3$-body annihilation operator. The form of the Hamiltonian is fully fixed by $\text{SU}(2)$ symmetry as in the case of the Laughlin state discussed above. The $H^2$ identity in \cref{eq:H-squared-symmetrized} can be generalized in a similar way and the south pole expansion, the splitting into $\Omega_1$ and $\Omega_2$, and the analytical bound proceed very similarly. Three main differences make the problem more computationally challenging: the decay of the error terms from the analytical bound is expected to be slower so a larger cutoff is needed to establish a gap, the gap itself is expected to be smaller, and the operator set needs to be larger since the term $Q^\dagger H Q$ now becomes a $6$-body operator (it contains $12$ creation/annihilation operators). Generalization to bosonic states also appears straightforward.

\subsection{Generalization to other geometries and beyond uniform magnetic field}\label{sec:discussions}
A question that is relevant to fractional quantum Hall states realized in flat Chern bands such as those observed recently in moir\'e platforms \cite{WuMacdonald,xu2023observation,cai2023signatures,park2023observation,zeng2023thermodynamic} is whether the same gap bounds can be established for inhomogeneous magnetic fields. A recent body of work has shown that under certain idealized conditions, the flat bands in these systems can be mapped to the LLL of an Aharonov--Casher Hamiltonian \cite{aharonov1979ground} in an inhomogeneous periodic magnetic field \cite{ledwithFractionalChernInsulator2020a,Wang2021exact,ledwith_vortexability_2023,TMDAharonocCasher}. Because the magnetic field is periodic, such problems are defined on the torus rather than the sphere. In the uniform case, the gap obtained on the torus should agree with that obtained on the sphere in the thermodynamic limit. Recent work \cite{Wang2021exact,ledwith_vortexability_2023,TMDAharonocCasher} has shown that pseudopotential arguments generalize to this situation leading to exact zero modes that generalize their uniform field counterparts.

A convenient way to define the problem in this case is to consider an LLL spanned by single-particle orbitals of the form $\psi_\alpha(\bar z,z) = \phi_\alpha(z) e^{-K(\bar z,z)}$ where $\phi$ is holomorphic in $z = x + i y$ and $K(\bar z,z)$ is a K\"ahler potential defined via $B(\bar z,z) = \Delta_g K(\bar z,z)$ where $\Delta_g = \frac{4}{\sqrt{g}}\partial_z \partial_{\bar z}$ is the Laplace--Beltrami operator. For instance, a uniform field on the plane gives $K = \frac{z \bar z}{4 \ell_B^2}$ corresponding to $B = 4\partial_z \partial_{\bar z} K = \frac{1}{\ell_B^2}$. On the sphere with radius $1$ we have $K = s \ln (1 + |z|^2)$ and $\sqrt{g} = 4/(1+|z|^2)^2$ so that $B = s = 1/\ell_B^2$. We can then define the coherent state operator $c_z = \sum_\alpha \phi_\alpha(z) c_\alpha$ which satisfies
\begin{equation}
    \{c_z, c_w^\dagger \} = \sum_\alpha \phi_\alpha(z) \bar \phi_\alpha(w) = P(z,\bar w)
\end{equation}
where $P(z, \bar w)$ is the holomorphic LLL kernel. The LLL also comes with an integration measure $d\mu(z) \propto e^{-2K} \sqrt{g} d^2 z$ defined such that the wavefunctions $\phi_\alpha(z)$ are orthonormal. On the infinite plane with uniform field, $P(z, \bar w)$ is the Bergman reproducing kernel with the Bargmann--Fock measure given by $d\mu(z) \propto e^{-\frac{1}{2\ell_B^2}|z|^2} d^2 z$. The Hamiltonian is defined as $\int d\tilde \mu(z) Q_z^\dagger Q_z$ in terms of the operators $Q_z \propto c_z \partial_z c_z$ which annihilate the Laughlin state. The measure $\tilde \mu(z)$ is generally not the same as $\mu(z)$. It is fixed by symmetry ($\text{SU}(2)$ or magnetic translation) for uniform fields. For example, on the sphere we have $d\tilde \mu(z) = d\mu_{2s-1}(z)$. In \cref{sec:coherent}, we show how this reproduces the expressions studied here when limiting to the case of uniform field on the sphere.

One important ingredient in our proof was the use of ${\rm SU}(2)$ symmetry which enables us to write the symmetrized form of $H^2$ in \cref{eq:H-squared-symmetrized} and map everything to the south pole where splitting $H^2$ into two pieces is possible. The generalization of ${\rm SU}(2)$ symmetry to the torus with uniform magnetic field is the magnetic translation group. However, it is unclear how to generalize \cref{eq:H-squared-symmetrized} to the non-uniform magnetic fields relevant to Chern bands. Furthermore, there is no clear analog of the south pole mapping on the torus. One possibility is to define $d_n \propto  \partial_z^n c_z|_{z=0}$. This can be defined on any geometry and even for non-uniform field. On the sphere, $d_n$ coincides with the angular momentum basis measured relative to the south pole $\tilde c_n$. On other geometries such as the torus, these operators will not have simple anticommutation relations since $\{d_m, d_n^\dagger\} \propto \partial_z^m \partial_{\bar w}^n P(z, \bar w)|_{z=w=0}$ which is only diagonal when the kernel is rotationally symmetric. Nevertheless, these operators are expected to provide a similar notion of the ``south pole" for large enough systems. Using these operators to generalize our results to the torus case is a promising direction that we leave to future work.

\subsection{Exact certification of semidefinite programs and their limits}
Beyond the specific case of the Laughlin gap or more generally pseudopotentials, we have presented in \cref{sec:exact-feasible-interior-points} an algorithm which uses a numerical SDP solver to produce a mathematically rigorous lower bound on the optimal objective value of a strictly feasible rational SDP. The output of the algorithm is an exact rational interior feasible point, which acts as an efficiently verifiable certificate. We have released such a certificate along with an accompanying verifier for the $r' = 15$, $r = 22$, $t = 0$ SDP on GitHub \cite{LaughlinGapCertificate}. In contrast to earlier approaches to rigorous certification of SDP bounds \cite{Peyrl2008,Jansson2008,Kaltofen2012,Naceur2025}, our method produces exact rational certificates for large SDPs with highly redundant and ill-conditioned affine constraints, without relying on special structure.

In addition to this, we have provided in \cref{sec:rank-stability} an algorithm that certifies the rank stability of a one-parameter family of SDPs in which all data are polynomials of a variable $t$ with rational coefficients. In combination with the first algorithm, this allows us to certify a lower bound on the limit infimum of the SDP objective value in the limit $t \to 0$ (or more generally $t \to t_0$ for some rational constant $t_0$). Aside from the use-case here in which $t$ is proportional to the inverse system size so that $t \to 0$ is the thermodynamic limit, this method may also be applicable to perturbative response functions or continuum limits.

\section{Note added}\label{sec:note-added}
We obtained a positive numerical lower bound on the Laughlin spectral gap using the $t = 1/N_\Phi = 0$ SDP on June 30th, 2026. We obtained the corresponding exact rational certificate on August 13th, 2026. During the final calculations to establish the rank stability of the SDP in the $t \to 0$ limit, OpenAI publicly released a collection of AI-generated manuscripts on October 6th, 2026 \cite{OpenAI2026MathRelease}. Among them is a manuscript \cite{OpenAI2026LaughlinGap} proving a positive gap for the first pseudopotential Hamiltonian on the sphere for all $N_\Phi$ sufficiently large. Their approach shares several ingredients with ours: Knabe's inequality, $\text{SU}(2)$ averaging on the sphere, a positivity inequality using operators supported on finitely many orbitals, and an analytical tail bound. On the other hand, some of the important differences between the approaches are as follows. Our positivity inequality is determined by an SDP hierarchy and an exact SDP certification algorithm, whereas Ref. \cite{OpenAI2026LaughlinGap} uses a fixed set of operators to obtain a sum-of-squares certificate but does not describe how it is obtained. Our approach can provide bounds for arbitrary $N_\Phi$ including the $N_\Phi \to \infty$ limit, whereas Ref. \cite{OpenAI2026LaughlinGap} only provides bounds on the sphere for $N_\Phi$ sufficiently large. Additionally, we justify the $N_\Phi \to \infty$ limit by certifying the rank stability of the SDP, whereas Ref. \cite{OpenAI2026LaughlinGap} uses a different approach. We note also that Ref. \cite{Gao2026} has used an SDP calculation to significantly improve the gap bound using the approach of Ref. \cite{OpenAI2026LaughlinGap}.

\begin{acknowledgements}
MGS is supported by a postdoctoral fellowship from the Julian Schwinger Foundation. Research reported in this publication was supported by an award from the Harvard FAS Dean's Competitive Fund for Promising Scholarship. We acknowledge discussions and previous collaborations with Nisarg Chadha, Ashvin Vishwanath, Patrick Ledwith, Jonah Herzog-Arbeitman, and Francesco Cioni. We acknowledge the use of AI tools ChatGPT 5.5, 5.6 Sol, and 5.6 Astra, and Claude Fable 5.1 and Opus 5.5. We used these tools to search the literature for useful identities and known results, for assistance with algorithms and coding, and for assistance with deriving or verifying some of the calculations. The AI tools did not contribute in a substantial way to the main ideas presented here or to the writing of this article. The computations in this paper were run on the FASRC Cannon cluster supported by the FAS Division of Science Research Computing Group at Harvard University.
\end{acknowledgements}

\bibliography{bibliography}

\begin{thebibliography}{87}%
\makeatletter
\providecommand \@ifxundefined [1]{%
 \@ifx{#1\undefined}
}%
\providecommand \@ifnum [1]{%
 \ifnum #1\expandafter \@firstoftwo
 \else \expandafter \@secondoftwo
 \fi
}%
\providecommand \@ifx [1]{%
 \ifx #1\expandafter \@firstoftwo
 \else \expandafter \@secondoftwo
 \fi
}%
\providecommand \natexlab [1]{#1}%
\providecommand \enquote  [1]{``#1''}%
\providecommand \bibnamefont  [1]{#1}%
\providecommand \bibfnamefont [1]{#1}%
\providecommand \citenamefont [1]{#1}%
\providecommand \href@noop [0]{\@secondoftwo}%
\providecommand \href [0]{\begingroup \@sanitize@url \@href}%
\providecommand \@href[1]{\@@startlink{#1}\@@href}%
\providecommand \@@href[1]{\endgroup#1\@@endlink}%
\providecommand \@sanitize@url [0]{\catcode `\\12\catcode `\$12\catcode
  `\&12\catcode `\#12\catcode `\^12\catcode `\_12\catcode `\%12\relax}%
\providecommand \@@startlink[1]{}%
\providecommand \@@endlink[0]{}%
\providecommand \url  [0]{\begingroup\@sanitize@url \@url }%
\providecommand \@url [1]{\endgroup\@href {#1}{\urlprefix }}%
\providecommand \urlprefix  [0]{URL }%
\providecommand \Eprint [0]{\href }%
\providecommand \doibase [0]{https://doi.org/}%
\providecommand \selectlanguage [0]{\@gobble}%
\providecommand \bibinfo  [0]{\@secondoftwo}%
\providecommand \bibfield  [0]{\@secondoftwo}%
\providecommand \translation [1]{[#1]}%
\providecommand \BibitemOpen [0]{}%
\providecommand \bibitemStop [0]{}%
\providecommand \bibitemNoStop [0]{.\EOS\space}%
\providecommand \EOS [0]{\spacefactor3000\relax}%
\providecommand \BibitemShut  [1]{\csname bibitem#1\endcsname}%
\let\auto@bib@innerbib\@empty
\bibitem [{\citenamefont {Klitzing}\ \emph {et~al.}(1980)\citenamefont
  {Klitzing}, \citenamefont {Dorda},\ and\ \citenamefont
  {Pepper}}]{Klitzing1980}%
  \BibitemOpen
  \bibfield  {author} {\bibinfo {author} {\bibfnamefont {K.~v.}\ \bibnamefont
  {Klitzing}}, \bibinfo {author} {\bibfnamefont {G.}~\bibnamefont {Dorda}},\
  and\ \bibinfo {author} {\bibfnamefont {M.}~\bibnamefont {Pepper}},\
  }\bibfield  {title} {\bibinfo {title} {New method for high-accuracy
  determination of the fine-structure constant based on quantized {H}all
  resistance},\ }\href {https://doi.org/10.1103/PhysRevLett.45.494} {\bibfield
  {journal} {\bibinfo  {journal} {Phys. Rev. Lett.}\ }\textbf {\bibinfo
  {volume} {45}},\ \bibinfo {pages} {494} (\bibinfo {year} {1980})}\BibitemShut
  {NoStop}%
\bibitem [{\citenamefont {Tsui}\ \emph {et~al.}(1982)\citenamefont {Tsui},
  \citenamefont {Stormer},\ and\ \citenamefont {Gossard}}]{Tsui1982}%
  \BibitemOpen
  \bibfield  {author} {\bibinfo {author} {\bibfnamefont {D.~C.}\ \bibnamefont
  {Tsui}}, \bibinfo {author} {\bibfnamefont {H.~L.}\ \bibnamefont {Stormer}},\
  and\ \bibinfo {author} {\bibfnamefont {A.~C.}\ \bibnamefont {Gossard}},\
  }\bibfield  {title} {\bibinfo {title} {Two-dimensional magnetotransport in
  the extreme quantum limit},\ }\href
  {https://doi.org/10.1103/PhysRevLett.48.1559} {\bibfield  {journal} {\bibinfo
   {journal} {Phys. Rev. Lett.}\ }\textbf {\bibinfo {volume} {48}},\ \bibinfo
  {pages} {1559} (\bibinfo {year} {1982})}\BibitemShut {NoStop}%
\bibitem [{\citenamefont {Thouless}\ \emph {et~al.}(1982)\citenamefont
  {Thouless}, \citenamefont {Kohmoto}, \citenamefont {Nightingale},\ and\
  \citenamefont {den Nijs}}]{Thouless1982}%
  \BibitemOpen
  \bibfield  {author} {\bibinfo {author} {\bibfnamefont {D.~J.}\ \bibnamefont
  {Thouless}}, \bibinfo {author} {\bibfnamefont {M.}~\bibnamefont {Kohmoto}},
  \bibinfo {author} {\bibfnamefont {M.~P.}\ \bibnamefont {Nightingale}},\ and\
  \bibinfo {author} {\bibfnamefont {M.}~\bibnamefont {den Nijs}},\ }\bibfield
  {title} {\bibinfo {title} {Quantized {H}all conductance in a two-dimensional
  periodic potential},\ }\href {https://doi.org/10.1103/PhysRevLett.49.405}
  {\bibfield  {journal} {\bibinfo  {journal} {Phys. Rev. Lett.}\ }\textbf
  {\bibinfo {volume} {49}},\ \bibinfo {pages} {405} (\bibinfo {year}
  {1982})}\BibitemShut {NoStop}%
\bibitem [{\citenamefont {Laughlin}(1983)}]{Laughlin1983}%
  \BibitemOpen
  \bibfield  {author} {\bibinfo {author} {\bibfnamefont {R.~B.}\ \bibnamefont
  {Laughlin}},\ }\bibfield  {title} {\bibinfo {title} {Anomalous quantum {H}all
  effect: An incompressible quantum fluid with fractionally charged
  excitations},\ }\href {https://doi.org/10.1103/PhysRevLett.50.1395}
  {\bibfield  {journal} {\bibinfo  {journal} {Phys. Rev. Lett.}\ }\textbf
  {\bibinfo {volume} {50}},\ \bibinfo {pages} {1395} (\bibinfo {year}
  {1983})}\BibitemShut {NoStop}%
\bibitem [{\citenamefont {Halperin}(1984)}]{Halperinstatistics}%
  \BibitemOpen
  \bibfield  {author} {\bibinfo {author} {\bibfnamefont {B.~I.}\ \bibnamefont
  {Halperin}},\ }\bibfield  {title} {\bibinfo {title} {Statistics of
  quasiparticles and the hierarchy of fractional quantized {H}all states},\
  }\href@noop {} {\bibfield  {journal} {\bibinfo  {journal} {Physical Review
  Letters}\ }\textbf {\bibinfo {volume} {52}},\ \bibinfo {pages} {1583}
  (\bibinfo {year} {1984})}\BibitemShut {NoStop}%
\bibitem [{\citenamefont {Haldane}(1983)}]{Haldane1983}%
  \BibitemOpen
  \bibfield  {author} {\bibinfo {author} {\bibfnamefont {F.~D.~M.}\
  \bibnamefont {Haldane}},\ }\bibfield  {title} {\bibinfo {title} {Fractional
  quantization of the {H}all effect: A hierarchy of incompressible quantum
  fluid states},\ }\href {https://doi.org/10.1103/PhysRevLett.51.605}
  {\bibfield  {journal} {\bibinfo  {journal} {Phys. Rev. Lett.}\ }\textbf
  {\bibinfo {volume} {51}},\ \bibinfo {pages} {605} (\bibinfo {year}
  {1983})}\BibitemShut {NoStop}%
\bibitem [{\citenamefont {Trugman}\ and\ \citenamefont
  {Kivelson}(1985)}]{Trugman1985}%
  \BibitemOpen
  \bibfield  {author} {\bibinfo {author} {\bibfnamefont {S.~A.}\ \bibnamefont
  {Trugman}}\ and\ \bibinfo {author} {\bibfnamefont {S.}~\bibnamefont
  {Kivelson}},\ }\bibfield  {title} {\bibinfo {title} {Exact results for the
  fractional quantum {H}all effect with general interactions},\ }\href
  {https://doi.org/10.1103/PhysRevB.31.5280} {\bibfield  {journal} {\bibinfo
  {journal} {Phys. Rev. B}\ }\textbf {\bibinfo {volume} {31}},\ \bibinfo
  {pages} {5280} (\bibinfo {year} {1985})}\BibitemShut {NoStop}%
\bibitem [{\citenamefont {Girvin}\ \emph {et~al.}(1985)\citenamefont {Girvin},
  \citenamefont {MacDonald},\ and\ \citenamefont {Platzman}}]{Girvin1985}%
  \BibitemOpen
  \bibfield  {author} {\bibinfo {author} {\bibfnamefont {S.~M.}\ \bibnamefont
  {Girvin}}, \bibinfo {author} {\bibfnamefont {A.~H.}\ \bibnamefont
  {MacDonald}},\ and\ \bibinfo {author} {\bibfnamefont {P.~M.}\ \bibnamefont
  {Platzman}},\ }\bibfield  {title} {\bibinfo {title} {Collective-excitation
  gap in the fractional quantum {H}all effect},\ }\href
  {https://doi.org/10.1103/PhysRevLett.54.581} {\bibfield  {journal} {\bibinfo
  {journal} {Phys. Rev. Lett.}\ }\textbf {\bibinfo {volume} {54}},\ \bibinfo
  {pages} {581} (\bibinfo {year} {1985})}\BibitemShut {NoStop}%
\bibitem [{\citenamefont {Girvin}\ \emph {et~al.}(1986)\citenamefont {Girvin},
  \citenamefont {MacDonald},\ and\ \citenamefont {Platzman}}]{Girvin1986}%
  \BibitemOpen
  \bibfield  {author} {\bibinfo {author} {\bibfnamefont {S.~M.}\ \bibnamefont
  {Girvin}}, \bibinfo {author} {\bibfnamefont {A.~H.}\ \bibnamefont
  {MacDonald}},\ and\ \bibinfo {author} {\bibfnamefont {P.~M.}\ \bibnamefont
  {Platzman}},\ }\bibfield  {title} {\bibinfo {title} {Magneto-roton theory of
  collective excitations in the fractional quantum {H}all effect},\ }\href
  {https://doi.org/10.1103/PhysRevB.33.2481} {\bibfield  {journal} {\bibinfo
  {journal} {Phys. Rev. B}\ }\textbf {\bibinfo {volume} {33}},\ \bibinfo
  {pages} {2481} (\bibinfo {year} {1986})}\BibitemShut {NoStop}%
\bibitem [{\citenamefont {Moore}\ and\ \citenamefont {Read}(1991)}]{MooreRead}%
  \BibitemOpen
  \bibfield  {author} {\bibinfo {author} {\bibfnamefont {G.}~\bibnamefont
  {Moore}}\ and\ \bibinfo {author} {\bibfnamefont {N.}~\bibnamefont {Read}},\
  }\bibfield  {title} {\bibinfo {title} {Nonabelions in the fractional quantum
  {H}all effect},\ }\href@noop {} {\bibfield  {journal} {\bibinfo  {journal}
  {Nuclear Physics B}\ }\textbf {\bibinfo {volume} {360}},\ \bibinfo {pages}
  {362} (\bibinfo {year} {1991})}\BibitemShut {NoStop}%
\bibitem [{\citenamefont {Read}\ and\ \citenamefont
  {Rezayi}(1999)}]{ReadRezayi}%
  \BibitemOpen
  \bibfield  {author} {\bibinfo {author} {\bibfnamefont {N.}~\bibnamefont
  {Read}}\ and\ \bibinfo {author} {\bibfnamefont {E.}~\bibnamefont {Rezayi}},\
  }\bibfield  {title} {\bibinfo {title} {Beyond paired quantum {H}all states:
  Parafermions and incompressible states in the first excited {L}andau level},\
  }\href@noop {} {\bibfield  {journal} {\bibinfo  {journal} {Physical Review
  B}\ }\textbf {\bibinfo {volume} {59}},\ \bibinfo {pages} {8084} (\bibinfo
  {year} {1999})}\BibitemShut {NoStop}%
\bibitem [{\citenamefont {Simon}\ \emph {et~al.}(2007)\citenamefont {Simon},
  \citenamefont {Rezayi},\ and\ \citenamefont
  {Cooper}}]{SimonPseudopotentials}%
  \BibitemOpen
  \bibfield  {author} {\bibinfo {author} {\bibfnamefont {S.~H.}\ \bibnamefont
  {Simon}}, \bibinfo {author} {\bibfnamefont {E.}~\bibnamefont {Rezayi}},\ and\
  \bibinfo {author} {\bibfnamefont {N.~R.}\ \bibnamefont {Cooper}},\ }\bibfield
   {title} {\bibinfo {title} {Pseudopotentials for multiparticle interactions
  in the quantum {H}all regime},\ }\href@noop {} {\bibfield  {journal}
  {\bibinfo  {journal} {Physical Review B—Condensed Matter and Materials
  Physics}\ }\textbf {\bibinfo {volume} {75}},\ \bibinfo {pages} {195306}
  (\bibinfo {year} {2007})}\BibitemShut {NoStop}%
\bibitem [{\citenamefont {Bernevig}\ and\ \citenamefont
  {Haldane}(2008{\natexlab{a}})}]{Bernevig2008a}%
  \BibitemOpen
  \bibfield  {author} {\bibinfo {author} {\bibfnamefont {B.~A.}\ \bibnamefont
  {Bernevig}}\ and\ \bibinfo {author} {\bibfnamefont {F.~D.~M.}\ \bibnamefont
  {Haldane}},\ }\bibfield  {title} {\bibinfo {title} {Generalized clustering
  conditions of {Jack} polynomials at negative {Jack} parameter $\alpha$},\
  }\href {https://doi.org/10.1103/PhysRevB.77.184502} {\bibfield  {journal}
  {\bibinfo  {journal} {Physical Review B}\ }\textbf {\bibinfo {volume} {77}},\
  \bibinfo {pages} {184502} (\bibinfo {year} {2008}{\natexlab{a}})}\BibitemShut
  {NoStop}%
\bibitem [{\citenamefont {Bernevig}\ and\ \citenamefont
  {Haldane}(2008{\natexlab{b}})}]{Bernevig2008b}%
  \BibitemOpen
  \bibfield  {author} {\bibinfo {author} {\bibfnamefont {B.~A.}\ \bibnamefont
  {Bernevig}}\ and\ \bibinfo {author} {\bibfnamefont {F.~D.~M.}\ \bibnamefont
  {Haldane}},\ }\bibfield  {title} {\bibinfo {title} {Model fractional quantum
  {H}all states and {J}ack polynomials},\ }\href
  {https://doi.org/10.1103/PhysRevLett.100.246802} {\bibfield  {journal}
  {\bibinfo  {journal} {Physical Review Letters}\ }\textbf {\bibinfo {volume}
  {100}},\ \bibinfo {pages} {246802} (\bibinfo {year}
  {2008}{\natexlab{b}})}\BibitemShut {NoStop}%
\bibitem [{\citenamefont {Bernevig}\ and\ \citenamefont
  {Haldane}(2008{\natexlab{c}})}]{Bernevig2008c}%
  \BibitemOpen
  \bibfield  {author} {\bibinfo {author} {\bibfnamefont {B.~A.}\ \bibnamefont
  {Bernevig}}\ and\ \bibinfo {author} {\bibfnamefont {F.~D.~M.}\ \bibnamefont
  {Haldane}},\ }\bibfield  {title} {\bibinfo {title} {Properties of non-abelian
  fractional quantum {H}all states at filling $\nu = k / r$},\ }\href
  {https://doi.org/10.1103/PhysRevLett.101.246806} {\bibfield  {journal}
  {\bibinfo  {journal} {Physical Review Letters}\ }\textbf {\bibinfo {volume}
  {101}},\ \bibinfo {pages} {246806} (\bibinfo {year}
  {2008}{\natexlab{c}})}\BibitemShut {NoStop}%
\bibitem [{\citenamefont {Ortiz}\ \emph {et~al.}(2013)\citenamefont {Ortiz},
  \citenamefont {Nussinov}, \citenamefont {Dukelsky},\ and\ \citenamefont
  {Seidel}}]{Ortiz2013}%
  \BibitemOpen
  \bibfield  {author} {\bibinfo {author} {\bibfnamefont {G.}~\bibnamefont
  {Ortiz}}, \bibinfo {author} {\bibfnamefont {Z.}~\bibnamefont {Nussinov}},
  \bibinfo {author} {\bibfnamefont {J.}~\bibnamefont {Dukelsky}},\ and\
  \bibinfo {author} {\bibfnamefont {A.}~\bibnamefont {Seidel}},\ }\bibfield
  {title} {\bibinfo {title} {Repulsive interactions in quantum {H}all systems
  as a pairing problem},\ }\href {https://doi.org/10.1103/PhysRevB.88.165303}
  {\bibfield  {journal} {\bibinfo  {journal} {Phys. Rev. B}\ }\textbf {\bibinfo
  {volume} {88}},\ \bibinfo {pages} {165303} (\bibinfo {year}
  {2013})}\BibitemShut {NoStop}%
\bibitem [{\citenamefont {Mazaheri}\ \emph {et~al.}(2015)\citenamefont
  {Mazaheri}, \citenamefont {Ortiz}, \citenamefont {Nussinov},\ and\
  \citenamefont {Seidel}}]{Mazaheri2015}%
  \BibitemOpen
  \bibfield  {author} {\bibinfo {author} {\bibfnamefont {T.}~\bibnamefont
  {Mazaheri}}, \bibinfo {author} {\bibfnamefont {G.}~\bibnamefont {Ortiz}},
  \bibinfo {author} {\bibfnamefont {Z.}~\bibnamefont {Nussinov}},\ and\
  \bibinfo {author} {\bibfnamefont {A.}~\bibnamefont {Seidel}},\ }\bibfield
  {title} {\bibinfo {title} {Zero modes, bosonization, and topological quantum
  order: The {L}aughlin state in second quantization},\ }\href
  {https://doi.org/10.1103/PhysRevB.91.085115} {\bibfield  {journal} {\bibinfo
  {journal} {Phys. Rev. B}\ }\textbf {\bibinfo {volume} {91}},\ \bibinfo
  {pages} {085115} (\bibinfo {year} {2015})}\BibitemShut {NoStop}%
\bibitem [{\citenamefont {Chen}\ and\ \citenamefont {Seidel}(2015)}]{Chen2015}%
  \BibitemOpen
  \bibfield  {author} {\bibinfo {author} {\bibfnamefont {L.}~\bibnamefont
  {Chen}}\ and\ \bibinfo {author} {\bibfnamefont {A.}~\bibnamefont {Seidel}},\
  }\bibfield  {title} {\bibinfo {title} {Algebraic approach to the study of
  zero modes of {H}aldane pseudopotentials},\ }\href@noop {} {\bibfield
  {journal} {\bibinfo  {journal} {Physical Review B}\ }\textbf {\bibinfo
  {volume} {91}},\ \bibinfo {pages} {085103} (\bibinfo {year}
  {2015})}\BibitemShut {NoStop}%
\bibitem [{\citenamefont {Nachtergaele}\ \emph {et~al.}(2021)\citenamefont
  {Nachtergaele}, \citenamefont {Warzel},\ and\ \citenamefont
  {Young}}]{Nachtergaele2021}%
  \BibitemOpen
  \bibfield  {author} {\bibinfo {author} {\bibfnamefont {B.}~\bibnamefont
  {Nachtergaele}}, \bibinfo {author} {\bibfnamefont {S.}~\bibnamefont
  {Warzel}},\ and\ \bibinfo {author} {\bibfnamefont {A.}~\bibnamefont
  {Young}},\ }\bibfield  {title} {\bibinfo {title} {Spectral gaps and
  incompressibility in a $\nu=1/3$ fractional quantum {H}all system},\ }\href
  {https://doi.org/10.1007/s00220-021-03997-0} {\bibfield  {journal} {\bibinfo
  {journal} {Communications in Mathematical Physics}\ }\textbf {\bibinfo
  {volume} {383}},\ \bibinfo {pages} {1093} (\bibinfo {year}
  {2021})}\BibitemShut {NoStop}%
\bibitem [{\citenamefont {Warzel}\ and\ \citenamefont
  {Young}(2022)}]{Warzel2022}%
  \BibitemOpen
  \bibfield  {author} {\bibinfo {author} {\bibfnamefont {S.}~\bibnamefont
  {Warzel}}\ and\ \bibinfo {author} {\bibfnamefont {A.}~\bibnamefont {Young}},\
  }\bibfield  {title} {\bibinfo {title} {The spectral gap of a fractional
  quantum {H}all system on a thin torus},\ }\href
  {https://doi.org/10.1063/5.0084677} {\bibfield  {journal} {\bibinfo
  {journal} {Journal of Mathematical Physics}\ }\textbf {\bibinfo {volume}
  {63}},\ \bibinfo {pages} {041901} (\bibinfo {year} {2022})}\BibitemShut
  {NoStop}%
\bibitem [{\citenamefont {Warzel}\ and\ \citenamefont
  {Young}(2023)}]{Warzel2023}%
  \BibitemOpen
  \bibfield  {author} {\bibinfo {author} {\bibfnamefont {S.}~\bibnamefont
  {Warzel}}\ and\ \bibinfo {author} {\bibfnamefont {A.}~\bibnamefont {Young}},\
  }\bibfield  {title} {\bibinfo {title} {A bulk spectral gap in the presence of
  edge states for a truncated pseudopotential},\ }\href
  {https://doi.org/10.1007/s00023-022-01210-z} {\bibfield  {journal} {\bibinfo
  {journal} {Annales Henri Poincar{\'e}}\ }\textbf {\bibinfo {volume} {24}},\
  \bibinfo {pages} {133} (\bibinfo {year} {2023})}\BibitemShut {NoStop}%
\bibitem [{\citenamefont {Lemm}\ \emph {et~al.}(2026)\citenamefont {Lemm},
  \citenamefont {Nachtergaele}, \citenamefont {Warzel},\ and\ \citenamefont
  {Young}}]{Lemm2026}%
  \BibitemOpen
  \bibfield  {author} {\bibinfo {author} {\bibfnamefont {M.}~\bibnamefont
  {Lemm}}, \bibinfo {author} {\bibfnamefont {B.}~\bibnamefont {Nachtergaele}},
  \bibinfo {author} {\bibfnamefont {S.}~\bibnamefont {Warzel}},\ and\ \bibinfo
  {author} {\bibfnamefont {A.}~\bibnamefont {Young}},\ }\bibfield  {title}
  {\bibinfo {title} {Recursive spectral relations and the charge vs neutral gap
  in fractional quantum {H}all systems},\ }\href
  {https://doi.org/10.1063/5.0292975} {\bibfield  {journal} {\bibinfo
  {journal} {Journal of Mathematical Physics}\ }\textbf {\bibinfo {volume}
  {67}},\ \bibinfo {pages} {011901} (\bibinfo {year} {2026})}\BibitemShut
  {NoStop}%
\bibitem [{\citenamefont {Rougerie}\ and\ \citenamefont
  {Yngvason}(2015{\natexlab{a}})}]{Rougerie2015a}%
  \BibitemOpen
  \bibfield  {author} {\bibinfo {author} {\bibfnamefont {N.}~\bibnamefont
  {Rougerie}}\ and\ \bibinfo {author} {\bibfnamefont {J.}~\bibnamefont
  {Yngvason}},\ }\bibfield  {title} {\bibinfo {title} {Incompressibility
  estimates for the {L}aughlin phase},\ }\href
  {https://doi.org/10.1007/s00220-014-2232-5} {\bibfield  {journal} {\bibinfo
  {journal} {Communications in Mathematical Physics}\ }\textbf {\bibinfo
  {volume} {336}},\ \bibinfo {pages} {1109} (\bibinfo {year}
  {2015}{\natexlab{a}})}\BibitemShut {NoStop}%
\bibitem [{\citenamefont {Rougerie}\ and\ \citenamefont
  {Yngvason}(2015{\natexlab{b}})}]{Rougerie2015b}%
  \BibitemOpen
  \bibfield  {author} {\bibinfo {author} {\bibfnamefont {N.}~\bibnamefont
  {Rougerie}}\ and\ \bibinfo {author} {\bibfnamefont {J.}~\bibnamefont
  {Yngvason}},\ }\bibfield  {title} {\bibinfo {title} {Incompressibility
  estimates for the {L}aughlin phase, part {II}},\ }\href
  {https://doi.org/10.1007/s00220-015-2400-2} {\bibfield  {journal} {\bibinfo
  {journal} {Communications in Mathematical Physics}\ }\textbf {\bibinfo
  {volume} {339}},\ \bibinfo {pages} {263} (\bibinfo {year}
  {2015}{\natexlab{b}})}\BibitemShut {NoStop}%
\bibitem [{\citenamefont {Lieb}\ \emph {et~al.}(2018)\citenamefont {Lieb},
  \citenamefont {Rougerie},\ and\ \citenamefont {Yngvason}}]{Lieb2018}%
  \BibitemOpen
  \bibfield  {author} {\bibinfo {author} {\bibfnamefont {E.~H.}\ \bibnamefont
  {Lieb}}, \bibinfo {author} {\bibfnamefont {N.}~\bibnamefont {Rougerie}},\
  and\ \bibinfo {author} {\bibfnamefont {J.}~\bibnamefont {Yngvason}},\
  }\bibfield  {title} {\bibinfo {title} {Rigidity of the {L}aughlin liquid},\
  }\href {https://doi.org/10.1007/s10955-018-2082-1} {\bibfield  {journal}
  {\bibinfo  {journal} {Journal of Statistical Physics}\ }\textbf {\bibinfo
  {volume} {172}},\ \bibinfo {pages} {544} (\bibinfo {year}
  {2018})}\BibitemShut {NoStop}%
\bibitem [{\citenamefont {Lieb}\ \emph {et~al.}(2019)\citenamefont {Lieb},
  \citenamefont {Rougerie},\ and\ \citenamefont {Yngvason}}]{Lieb2019}%
  \BibitemOpen
  \bibfield  {author} {\bibinfo {author} {\bibfnamefont {E.~H.}\ \bibnamefont
  {Lieb}}, \bibinfo {author} {\bibfnamefont {N.}~\bibnamefont {Rougerie}},\
  and\ \bibinfo {author} {\bibfnamefont {J.}~\bibnamefont {Yngvason}},\
  }\bibfield  {title} {\bibinfo {title} {Local incompressibility estimates for
  the {L}aughlin phase},\ }\href@noop {} {\bibfield  {journal} {\bibinfo
  {journal} {Communications in Mathematical Physics}\ }\textbf {\bibinfo
  {volume} {365}},\ \bibinfo {pages} {431} (\bibinfo {year}
  {2019})}\BibitemShut {NoStop}%
\bibitem [{\citenamefont {Olgiati}\ and\ \citenamefont
  {Rougerie}(2020)}]{Olgiati2020}%
  \BibitemOpen
  \bibfield  {author} {\bibinfo {author} {\bibfnamefont {A.}~\bibnamefont
  {Olgiati}}\ and\ \bibinfo {author} {\bibfnamefont {N.}~\bibnamefont
  {Rougerie}},\ }\bibfield  {title} {\bibinfo {title} {Stability of the
  {L}aughlin phase against long-range interactions},\ }\href@noop {} {\bibfield
   {journal} {\bibinfo  {journal} {Archive for Rational Mechanics and
  Analysis}\ }\textbf {\bibinfo {volume} {237}},\ \bibinfo {pages} {1475}
  (\bibinfo {year} {2020})}\BibitemShut {NoStop}%
\bibitem [{\citenamefont {Thoma}(2024)}]{Thoma2024}%
  \BibitemOpen
  \bibfield  {author} {\bibinfo {author} {\bibfnamefont {E.}~\bibnamefont
  {Thoma}},\ }\bibfield  {title} {\bibinfo {title} {Overcrowding and separation
  estimates for the {C}oulomb gas},\ }\href {https://doi.org/10.1002/cpa.22188}
  {\bibfield  {journal} {\bibinfo  {journal} {Communications on Pure and
  Applied Mathematics}\ }\textbf {\bibinfo {volume} {77}},\ \bibinfo {pages}
  {3227} (\bibinfo {year} {2024})},\ \Eprint
  {https://arxiv.org/abs/https://onlinelibrary.wiley.com/doi/pdf/10.1002/cpa.22188}
  {https://onlinelibrary.wiley.com/doi/pdf/10.1002/cpa.22188} \BibitemShut
  {NoStop}%
\bibitem [{\citenamefont {Rougerie}(2019)}]{Rougerie2018}%
  \BibitemOpen
  \bibfield  {author} {\bibinfo {author} {\bibfnamefont {N.}~\bibnamefont
  {Rougerie}},\ }\bibfield  {title} {\bibinfo {title} {On the {L}aughlin
  function and its perturbations},\ }in\ \href
  {https://doi.org/10.5802/slsedp.131} {\emph {\bibinfo {booktitle}
  {S{\'e}minaire Laurent Schwartz — EDP et applications}}},\ Vol.\ \bibinfo
  {volume} {2018-2019}\ (\bibinfo {year} {2018-2019})\ pp.\ \bibinfo {pages}
  {Exposé no. 2, 17 p.}\BibitemShut {Stop}%
\bibitem [{\citenamefont {Rougerie}(2022)}]{Rougerie2022}%
  \BibitemOpen
  \bibfield  {author} {\bibinfo {author} {\bibfnamefont {N.}~\bibnamefont
  {Rougerie}},\ }\bibfield  {title} {\bibinfo {title} {{On the stability of
  {L}aughlin's fractional quantum {H}all phase}},\ }in\ \href@noop {} {\emph
  {\bibinfo {booktitle} {Encyclopedia of condensed matter physics, 2nd
  edition}}}\ (\bibinfo  {publisher} {Elsevier},\ \bibinfo {year}
  {2022})\BibitemShut {NoStop}%
\bibitem [{\citenamefont {Knabe}(1988)}]{Knabe1988}%
  \BibitemOpen
  \bibfield  {author} {\bibinfo {author} {\bibfnamefont {S.}~\bibnamefont
  {Knabe}},\ }\bibfield  {title} {\bibinfo {title} {Energy gaps and elementary
  excitations for certain {VBS}-quantum antiferromagnets},\ }\href@noop {}
  {\bibfield  {journal} {\bibinfo  {journal} {Journal of statistical physics}\
  }\textbf {\bibinfo {volume} {52}},\ \bibinfo {pages} {627} (\bibinfo {year}
  {1988})}\BibitemShut {NoStop}%
\bibitem [{\citenamefont {Rai}\ \emph {et~al.}(2026)\citenamefont {Rai},
  \citenamefont {Kull}, \citenamefont {Emonts}, \citenamefont {Tura},
  \citenamefont {Schuch},\ and\ \citenamefont {Baccari}}]{Rai2026}%
  \BibitemOpen
  \bibfield  {author} {\bibinfo {author} {\bibfnamefont {K.~S.}\ \bibnamefont
  {Rai}}, \bibinfo {author} {\bibfnamefont {I.}~\bibnamefont {Kull}}, \bibinfo
  {author} {\bibfnamefont {P.}~\bibnamefont {Emonts}}, \bibinfo {author}
  {\bibfnamefont {J.}~\bibnamefont {Tura}}, \bibinfo {author} {\bibfnamefont
  {N.}~\bibnamefont {Schuch}},\ and\ \bibinfo {author} {\bibfnamefont
  {F.}~\bibnamefont {Baccari}},\ }\bibfield  {title} {\bibinfo {title} {A
  hierarchy of spectral gap certificates for frustration-free spin systems},\
  }\href {https://doi.org/10.22331/q-2026-04-13-2065} {\bibfield  {journal}
  {\bibinfo  {journal} {Quantum}\ }\textbf {\bibinfo {volume} {10}},\ \bibinfo
  {pages} {2065} (\bibinfo {year} {2026})}\BibitemShut {NoStop}%
\bibitem [{\citenamefont {Barthel}\ and\ \citenamefont
  {H\"ubener}(2012)}]{Barthel2012}%
  \BibitemOpen
  \bibfield  {author} {\bibinfo {author} {\bibfnamefont {T.}~\bibnamefont
  {Barthel}}\ and\ \bibinfo {author} {\bibfnamefont {R.}~\bibnamefont
  {H\"ubener}},\ }\bibfield  {title} {\bibinfo {title} {Solving
  condensed-matter ground-state problems by semidefinite relaxations},\ }\href
  {https://doi.org/10.1103/PhysRevLett.108.200404} {\bibfield  {journal}
  {\bibinfo  {journal} {Phys. Rev. Lett.}\ }\textbf {\bibinfo {volume} {108}},\
  \bibinfo {pages} {200404} (\bibinfo {year} {2012})}\BibitemShut {NoStop}%
\bibitem [{\citenamefont {Baumgratz}\ and\ \citenamefont
  {Plenio}(2012)}]{Baumgratz2012}%
  \BibitemOpen
  \bibfield  {author} {\bibinfo {author} {\bibfnamefont {T.}~\bibnamefont
  {Baumgratz}}\ and\ \bibinfo {author} {\bibfnamefont {M.~B.}\ \bibnamefont
  {Plenio}},\ }\bibfield  {title} {\bibinfo {title} {Lower bounds for ground
  states of condensed matter systems},\ }\href
  {https://doi.org/10.1088/1367-2630/14/2/023027} {\bibfield  {journal}
  {\bibinfo  {journal} {New Journal of Physics}\ }\textbf {\bibinfo {volume}
  {14}},\ \bibinfo {pages} {023027} (\bibinfo {year} {2012})}\BibitemShut
  {NoStop}%
\bibitem [{\citenamefont {Mazziotti}(2023)}]{Mazziotti2023}%
  \BibitemOpen
  \bibfield  {author} {\bibinfo {author} {\bibfnamefont {D.~A.}\ \bibnamefont
  {Mazziotti}},\ }\bibfield  {title} {\bibinfo {title} {Quantum many-body
  theory from a solution of the {$N$}-representability problem},\ }\href
  {https://doi.org/10.1103/PhysRevLett.130.153001} {\bibfield  {journal}
  {\bibinfo  {journal} {Phys. Rev. Lett.}\ }\textbf {\bibinfo {volume} {130}},\
  \bibinfo {pages} {153001} (\bibinfo {year} {2023})}\BibitemShut {NoStop}%
\bibitem [{\citenamefont {Gao}\ \emph {et~al.}(2025)\citenamefont {Gao},
  \citenamefont {Lanzetta}, \citenamefont {Ledwith}, \citenamefont {Wang},\
  and\ \citenamefont {Khalaf}}]{Gao2025}%
  \BibitemOpen
  \bibfield  {author} {\bibinfo {author} {\bibfnamefont {Q.}~\bibnamefont
  {Gao}}, \bibinfo {author} {\bibfnamefont {R.~A.}\ \bibnamefont {Lanzetta}},
  \bibinfo {author} {\bibfnamefont {P.}~\bibnamefont {Ledwith}}, \bibinfo
  {author} {\bibfnamefont {J.}~\bibnamefont {Wang}},\ and\ \bibinfo {author}
  {\bibfnamefont {E.}~\bibnamefont {Khalaf}},\ }\bibfield  {title} {\bibinfo
  {title} {Bootstrapping the quantum {H}all problem},\ }\href
  {https://doi.org/10.1103/csnn-vjhn} {\bibfield  {journal} {\bibinfo
  {journal} {Phys. Rev. X}\ }\textbf {\bibinfo {volume} {15}},\ \bibinfo
  {pages} {031034} (\bibinfo {year} {2025})}\BibitemShut {NoStop}%
\bibitem [{\citenamefont {Scheer}(2026)}]{Scheer2026}%
  \BibitemOpen
  \bibfield  {author} {\bibinfo {author} {\bibfnamefont {M.~G.}\ \bibnamefont
  {Scheer}},\ }\bibfield  {title} {\bibinfo {title} {Bootstrapping quantum
  {H}amiltonians with symmetry},\ }\href {https://doi.org/10.1103/1757-fx9c}
  {\bibfield  {journal} {\bibinfo  {journal} {Phys. Rev. Lett.}\ }\textbf
  {\bibinfo {volume} {136}},\ \bibinfo {pages} {246501} (\bibinfo {year}
  {2026})}\BibitemShut {NoStop}%
\bibitem [{\citenamefont {Han}\ \emph {et~al.}(2020)\citenamefont {Han},
  \citenamefont {Hartnoll},\ and\ \citenamefont {Kruthoff}}]{Han2020}%
  \BibitemOpen
  \bibfield  {author} {\bibinfo {author} {\bibfnamefont {X.}~\bibnamefont
  {Han}}, \bibinfo {author} {\bibfnamefont {S.~A.}\ \bibnamefont {Hartnoll}},\
  and\ \bibinfo {author} {\bibfnamefont {J.}~\bibnamefont {Kruthoff}},\
  }\bibfield  {title} {\bibinfo {title} {Bootstrapping matrix quantum
  mechanics},\ }\bibfield  {journal} {\bibinfo  {journal} {Physical Review
  Letters}\ }\textbf {\bibinfo {volume} {125}},\ \href
  {https://doi.org/10.1103/physrevlett.125.041601}
  {10.1103/physrevlett.125.041601} (\bibinfo {year} {2020})\BibitemShut
  {NoStop}%
\bibitem [{\citenamefont {Lin}(2020)}]{Lin2020}%
  \BibitemOpen
  \bibfield  {author} {\bibinfo {author} {\bibfnamefont {H.~W.}\ \bibnamefont
  {Lin}},\ }\bibfield  {title} {\bibinfo {title} {Bootstraps to strings:
  solving random matrix models with positivity},\ }\bibfield  {journal}
  {\bibinfo  {journal} {Journal of High Energy Physics}\ }\textbf {\bibinfo
  {volume} {2020}},\ \href {https://doi.org/10.1007/jhep06(2020)090}
  {10.1007/jhep06(2020)090} (\bibinfo {year} {2020})\BibitemShut {NoStop}%
\bibitem [{\citenamefont {Kazakov}\ and\ \citenamefont
  {Zheng}(2023)}]{Kazakov2023}%
  \BibitemOpen
  \bibfield  {author} {\bibinfo {author} {\bibfnamefont {V.}~\bibnamefont
  {Kazakov}}\ and\ \bibinfo {author} {\bibfnamefont {Z.}~\bibnamefont
  {Zheng}},\ }\bibfield  {title} {\bibinfo {title} {Bootstrap for lattice
  {Y}ang-{M}ills theory},\ }\bibfield  {journal} {\bibinfo  {journal} {Physical
  Review D}\ }\textbf {\bibinfo {volume} {107}},\ \href
  {https://doi.org/10.1103/physrevd.107.l051501} {10.1103/physrevd.107.l051501}
  (\bibinfo {year} {2023})\BibitemShut {NoStop}%
\bibitem [{\citenamefont {Navascués}\ \emph {et~al.}(2007)\citenamefont
  {Navascués}, \citenamefont {Pironio},\ and\ \citenamefont
  {Acín}}]{Navascues2007}%
  \BibitemOpen
  \bibfield  {author} {\bibinfo {author} {\bibfnamefont {M.}~\bibnamefont
  {Navascués}}, \bibinfo {author} {\bibfnamefont {S.}~\bibnamefont
  {Pironio}},\ and\ \bibinfo {author} {\bibfnamefont {A.}~\bibnamefont
  {Acín}},\ }\bibfield  {title} {\bibinfo {title} {Bounding the set of quantum
  correlations},\ }\bibfield  {journal} {\bibinfo  {journal} {Physical Review
  Letters}\ }\textbf {\bibinfo {volume} {98}},\ \href
  {https://doi.org/10.1103/physrevlett.98.010401}
  {10.1103/physrevlett.98.010401} (\bibinfo {year} {2007})\BibitemShut
  {NoStop}%
\bibitem [{\citenamefont {Navascués}\ \emph {et~al.}(2008)\citenamefont
  {Navascués}, \citenamefont {Pironio},\ and\ \citenamefont
  {Acín}}]{Navascues2008}%
  \BibitemOpen
  \bibfield  {author} {\bibinfo {author} {\bibfnamefont {M.}~\bibnamefont
  {Navascués}}, \bibinfo {author} {\bibfnamefont {S.}~\bibnamefont
  {Pironio}},\ and\ \bibinfo {author} {\bibfnamefont {A.}~\bibnamefont
  {Acín}},\ }\bibfield  {title} {\bibinfo {title} {A convergent hierarchy of
  semidefinite programs characterizing the set of quantum correlations},\
  }\href {https://doi.org/10.1088/1367-2630/10/7/073013} {\bibfield  {journal}
  {\bibinfo  {journal} {New Journal of Physics}\ }\textbf {\bibinfo {volume}
  {10}},\ \bibinfo {pages} {073013} (\bibinfo {year} {2008})}\BibitemShut
  {NoStop}%
\bibitem [{\citenamefont {Berenstein}\ \emph {et~al.}(2024)\citenamefont
  {Berenstein}, \citenamefont {Hulsey},\ and\ \citenamefont
  {Lloyd}}]{Berenstein2024}%
  \BibitemOpen
  \bibfield  {author} {\bibinfo {author} {\bibfnamefont {D.}~\bibnamefont
  {Berenstein}}, \bibinfo {author} {\bibfnamefont {G.}~\bibnamefont {Hulsey}},\
  and\ \bibinfo {author} {\bibfnamefont {P.~N.~T.}\ \bibnamefont {Lloyd}},\
  }\href {https://arxiv.org/abs/2406.17844} {\bibinfo {title} {Numerical
  exploration of the bootstrap in spin chain models}} (\bibinfo {year}
  {2024}),\ \Eprint {https://arxiv.org/abs/2406.17844} {arXiv:2406.17844
  [hep-th]} \BibitemShut {NoStop}%
\bibitem [{\citenamefont {Wang}\ \emph {et~al.}(2024)\citenamefont {Wang},
  \citenamefont {Surace}, \citenamefont {Fr\'erot}, \citenamefont {Legat},
  \citenamefont {Renou}, \citenamefont {Magron},\ and\ \citenamefont
  {Ac\'{\i}n}}]{Wang2024}%
  \BibitemOpen
  \bibfield  {author} {\bibinfo {author} {\bibfnamefont {J.}~\bibnamefont
  {Wang}}, \bibinfo {author} {\bibfnamefont {J.}~\bibnamefont {Surace}},
  \bibinfo {author} {\bibfnamefont {I.}~\bibnamefont {Fr\'erot}}, \bibinfo
  {author} {\bibfnamefont {B.}~\bibnamefont {Legat}}, \bibinfo {author}
  {\bibfnamefont {M.-O.}\ \bibnamefont {Renou}}, \bibinfo {author}
  {\bibfnamefont {V.}~\bibnamefont {Magron}},\ and\ \bibinfo {author}
  {\bibfnamefont {A.}~\bibnamefont {Ac\'{\i}n}},\ }\bibfield  {title} {\bibinfo
  {title} {Certifying ground-state properties of many-body systems},\ }\href
  {https://doi.org/10.1103/PhysRevX.14.031006} {\bibfield  {journal} {\bibinfo
  {journal} {Phys. Rev. X}\ }\textbf {\bibinfo {volume} {14}},\ \bibinfo
  {pages} {031006} (\bibinfo {year} {2024})}\BibitemShut {NoStop}%
\bibitem [{\citenamefont {Schouten}\ \emph {et~al.}(2025)\citenamefont
  {Schouten}, \citenamefont {Ewing},\ and\ \citenamefont
  {Mazziotti}}]{Schouten2025}%
  \BibitemOpen
  \bibfield  {author} {\bibinfo {author} {\bibfnamefont {A.~O.}\ \bibnamefont
  {Schouten}}, \bibinfo {author} {\bibfnamefont {S.}~\bibnamefont {Ewing}},\
  and\ \bibinfo {author} {\bibfnamefont {D.~A.}\ \bibnamefont {Mazziotti}},\
  }\href {https://arxiv.org/abs/2504.02861} {\bibinfo {title} {Bootstrapping
  the electronic structure of quantum materials}} (\bibinfo {year} {2025}),\
  \Eprint {https://arxiv.org/abs/2504.02861} {arXiv:2504.02861
  [cond-mat.str-el]} \BibitemShut {NoStop}%
\bibitem [{\citenamefont {Fawzi}\ \emph {et~al.}(2024)\citenamefont {Fawzi},
  \citenamefont {Fawzi},\ and\ \citenamefont {Scalet}}]{Fawzi2024}%
  \BibitemOpen
  \bibfield  {author} {\bibinfo {author} {\bibfnamefont {H.}~\bibnamefont
  {Fawzi}}, \bibinfo {author} {\bibfnamefont {O.}~\bibnamefont {Fawzi}},\ and\
  \bibinfo {author} {\bibfnamefont {S.~O.}\ \bibnamefont {Scalet}},\ }\bibfield
   {title} {\bibinfo {title} {Certified algorithms for equilibrium states of
  local quantum {H}amiltonians},\ }\bibfield  {journal} {\bibinfo  {journal}
  {Nature Communications}\ }\textbf {\bibinfo {volume} {15}},\ \href
  {https://doi.org/10.1038/s41467-024-51592-3} {10.1038/s41467-024-51592-3}
  (\bibinfo {year} {2024})\BibitemShut {NoStop}%
\bibitem [{\citenamefont {Cho}\ \emph {et~al.}(2022)\citenamefont {Cho},
  \citenamefont {Gabai}, \citenamefont {Lin}, \citenamefont {Rodriguez},
  \citenamefont {Sandor},\ and\ \citenamefont {Yin}}]{Cho2022}%
  \BibitemOpen
  \bibfield  {author} {\bibinfo {author} {\bibfnamefont {M.}~\bibnamefont
  {Cho}}, \bibinfo {author} {\bibfnamefont {B.}~\bibnamefont {Gabai}}, \bibinfo
  {author} {\bibfnamefont {Y.-H.}\ \bibnamefont {Lin}}, \bibinfo {author}
  {\bibfnamefont {V.~A.}\ \bibnamefont {Rodriguez}}, \bibinfo {author}
  {\bibfnamefont {J.}~\bibnamefont {Sandor}},\ and\ \bibinfo {author}
  {\bibfnamefont {X.}~\bibnamefont {Yin}},\ }\href@noop {} {\bibinfo {title}
  {{Bootstrapping the Ising Model on the Lattice}}} (\bibinfo {year} {2022}),\
  \Eprint {https://arxiv.org/abs/2206.12538} {arXiv:2206.12538 [hep-th]}
  \BibitemShut {NoStop}%
\bibitem [{\citenamefont {Cho}\ \emph {et~al.}(2025)\citenamefont {Cho},
  \citenamefont {Gabai}, \citenamefont {Sandor},\ and\ \citenamefont
  {Yin}}]{Cho2025}%
  \BibitemOpen
  \bibfield  {author} {\bibinfo {author} {\bibfnamefont {M.}~\bibnamefont
  {Cho}}, \bibinfo {author} {\bibfnamefont {B.}~\bibnamefont {Gabai}}, \bibinfo
  {author} {\bibfnamefont {J.}~\bibnamefont {Sandor}},\ and\ \bibinfo {author}
  {\bibfnamefont {X.}~\bibnamefont {Yin}},\ }\bibfield  {title} {\bibinfo
  {title} {{Thermal bootstrap of matrix quantum mechanics}},\ }\href
  {https://doi.org/10.1007/JHEP04(2025)186} {\bibfield  {journal} {\bibinfo
  {journal} {JHEP}\ }\textbf {\bibinfo {volume} {04}},\ \bibinfo {pages}
  {186}},\ \Eprint {https://arxiv.org/abs/2410.04262} {arXiv:2410.04262
  [hep-th]} \BibitemShut {NoStop}%
\bibitem [{\citenamefont {Cho}\ \emph {et~al.}(2026{\natexlab{a}})\citenamefont
  {Cho}, \citenamefont {Nancarrow}, \citenamefont {Tadi{\'c}}, \citenamefont
  {Xin},\ and\ \citenamefont {Zheng}}]{Cho2026a}%
  \BibitemOpen
  \bibfield  {author} {\bibinfo {author} {\bibfnamefont {M.}~\bibnamefont
  {Cho}}, \bibinfo {author} {\bibfnamefont {C.~O.}\ \bibnamefont {Nancarrow}},
  \bibinfo {author} {\bibfnamefont {P.}~\bibnamefont {Tadi{\'c}}}, \bibinfo
  {author} {\bibfnamefont {Y.}~\bibnamefont {Xin}},\ and\ \bibinfo {author}
  {\bibfnamefont {Z.}~\bibnamefont {Zheng}},\ }\bibfield  {title} {\bibinfo
  {title} {{Coarse-grained bootstrap of quantum many-body systems}},\ }\href
  {https://doi.org/10.1007/JHEP02(2026)222} {\bibfield  {journal} {\bibinfo
  {journal} {JHEP}\ }\textbf {\bibinfo {volume} {02}},\ \bibinfo {pages}
  {222}},\ \Eprint {https://arxiv.org/abs/2412.07837} {arXiv:2412.07837
  [hep-th]} \BibitemShut {NoStop}%
\bibitem [{\citenamefont {Cho}\ \emph {et~al.}(2026{\natexlab{b}})\citenamefont
  {Cho}, \citenamefont {Gabai}, \citenamefont {Lin}, \citenamefont {Yeh},\ and\
  \citenamefont {Zheng}}]{Cho2026b}%
  \BibitemOpen
  \bibfield  {author} {\bibinfo {author} {\bibfnamefont {M.}~\bibnamefont
  {Cho}}, \bibinfo {author} {\bibfnamefont {B.}~\bibnamefont {Gabai}}, \bibinfo
  {author} {\bibfnamefont {H.~W.}\ \bibnamefont {Lin}}, \bibinfo {author}
  {\bibfnamefont {J.}~\bibnamefont {Yeh}},\ and\ \bibinfo {author}
  {\bibfnamefont {Z.}~\bibnamefont {Zheng}},\ }\bibfield  {title} {\bibinfo
  {title} {{Bootstrapping Euclidean two-point correlators}},\ }\href
  {https://doi.org/10.1007/JHEP08(2026)147} {\bibfield  {journal} {\bibinfo
  {journal} {JHEP}\ }\textbf {\bibinfo {volume} {08}},\ \bibinfo {pages}
  {147}},\ \Eprint {https://arxiv.org/abs/2511.08560} {arXiv:2511.08560
  [hep-th]} \BibitemShut {NoStop}%
\bibitem [{\citenamefont {Paul}\ and\ \citenamefont {Refael}(2026)}]{Paul2026}%
  \BibitemOpen
  \bibfield  {author} {\bibinfo {author} {\bibfnamefont {N.}~\bibnamefont
  {Paul}}\ and\ \bibinfo {author} {\bibfnamefont {G.}~\bibnamefont {Refael}},\
  }\href {https://arxiv.org/abs/2605.06784} {\bibinfo {title} {Bootstrapping
  ground state properties of classical frustrated magnets}} (\bibinfo {year}
  {2026}),\ \Eprint {https://arxiv.org/abs/2605.06784} {arXiv:2605.06784
  [cond-mat.stat-mech]} \BibitemShut {NoStop}%
\bibitem [{\citenamefont {Gao}\ \emph {et~al.}(2026)\citenamefont {Gao},
  \citenamefont {Han},\ and\ \citenamefont {Khalaf}}]{Gao2026a}%
  \BibitemOpen
  \bibfield  {author} {\bibinfo {author} {\bibfnamefont {Q.}~\bibnamefont
  {Gao}}, \bibinfo {author} {\bibfnamefont {Z.}~\bibnamefont {Han}},\ and\
  \bibinfo {author} {\bibfnamefont {E.}~\bibnamefont {Khalaf}},\ }\bibfield
  {title} {\bibinfo {title} {Bootstrapping flatband superconductors: Rigorous
  lower bounds on superfluid stiffness},\ }\bibfield  {journal} {\bibinfo
  {journal} {Physical Review Letters}\ }\textbf {\bibinfo {volume} {136}},\
  \href {https://doi.org/10.1103/gw85-5r92} {10.1103/gw85-5r92} (\bibinfo
  {year} {2026})\BibitemShut {NoStop}%
\bibitem [{\citenamefont {Scheer}\ \emph {et~al.}(2025)\citenamefont {Scheer},
  \citenamefont {Chadha}, \citenamefont {Lu},\ and\ \citenamefont
  {Khalaf}}]{Scheer2025}%
  \BibitemOpen
  \bibfield  {author} {\bibinfo {author} {\bibfnamefont {M.~G.}\ \bibnamefont
  {Scheer}}, \bibinfo {author} {\bibfnamefont {N.}~\bibnamefont {Chadha}},
  \bibinfo {author} {\bibfnamefont {D.-C.}\ \bibnamefont {Lu}},\ and\ \bibinfo
  {author} {\bibfnamefont {E.}~\bibnamefont {Khalaf}},\ }\href
  {https://arxiv.org/abs/2511.20860} {\bibinfo {title} {Defect bootstrap: Tight
  ground state bounds in spontaneous symmetry breaking phases}} (\bibinfo
  {year} {2025}),\ \Eprint {https://arxiv.org/abs/2511.20860} {arXiv:2511.20860
  [cond-mat.str-el]} \BibitemShut {NoStop}%
\bibitem [{\citenamefont {Chadha}\ \emph {et~al.}(2026)\citenamefont {Chadha},
  \citenamefont {Scheer},\ and\ \citenamefont {Khalaf}}]{Chadha2026}%
  \BibitemOpen
  \bibfield  {author} {\bibinfo {author} {\bibfnamefont {N.}~\bibnamefont
  {Chadha}}, \bibinfo {author} {\bibfnamefont {M.~G.}\ \bibnamefont {Scheer}},\
  and\ \bibinfo {author} {\bibfnamefont {E.}~\bibnamefont {Khalaf}},\ }\href
  {https://arxiv.org/abs/2606.06584} {\bibinfo {title} {Bootstrap bounds for
  quantum spin systems using string operators}} (\bibinfo {year} {2026}),\
  \Eprint {https://arxiv.org/abs/2606.06584} {arXiv:2606.06584
  [cond-mat.str-el]} \BibitemShut {NoStop}%
\bibitem [{\citenamefont {Önder}\ \emph {et~al.}(2026)\citenamefont {Önder},
  \citenamefont {Scheer}, \citenamefont {Cho},\ and\ \citenamefont
  {Khalaf}}]{Onder2026}%
  \BibitemOpen
  \bibfield  {author} {\bibinfo {author} {\bibfnamefont {Y.}~\bibnamefont
  {Önder}}, \bibinfo {author} {\bibfnamefont {M.~G.}\ \bibnamefont {Scheer}},
  \bibinfo {author} {\bibfnamefont {M.}~\bibnamefont {Cho}},\ and\ \bibinfo
  {author} {\bibfnamefont {E.}~\bibnamefont {Khalaf}},\ }\href@noop {}
  {\bibinfo {title} {Bootstrapping disordered quantum systems}} (\bibinfo
  {year} {2026}),\ \Eprint {https://arxiv.org/abs/arXiv:2609.20916}
  {arXiv:2609.20916} \BibitemShut {NoStop}%
\bibitem [{\citenamefont {Cho}\ \emph {et~al.}(2026{\natexlab{c}})\citenamefont
  {Cho}, \citenamefont {Önder}, \citenamefont {Khalaf},\ and\ \citenamefont
  {Scheer}}]{Cho2026c}%
  \BibitemOpen
  \bibfield  {author} {\bibinfo {author} {\bibfnamefont {M.}~\bibnamefont
  {Cho}}, \bibinfo {author} {\bibfnamefont {Y.}~\bibnamefont {Önder}},
  \bibinfo {author} {\bibfnamefont {E.}~\bibnamefont {Khalaf}},\ and\ \bibinfo
  {author} {\bibfnamefont {M.~G.}\ \bibnamefont {Scheer}},\ }\href@noop {}
  {\bibinfo {title} {Bootstrapping classical systems with quenched disorder}}
  (\bibinfo {year} {2026}{\natexlab{c}}),\ \Eprint
  {https://arxiv.org/abs/arXiv:2609.20927} {arXiv:2609.20927} \BibitemShut
  {NoStop}%
\bibitem [{\citenamefont {Hastings}(2022)}]{Hastings2022}%
  \BibitemOpen
  \bibfield  {author} {\bibinfo {author} {\bibfnamefont {M.~B.}\ \bibnamefont
  {Hastings}},\ }\href@noop {} {\bibinfo {title} {Perturbation theory and the
  sum of squares}} (\bibinfo {year} {2022}),\ \Eprint
  {https://arxiv.org/abs/arXiv:2205.12325} {arXiv:2205.12325} \BibitemShut
  {NoStop}%
\bibitem [{\citenamefont {Schleith}\ \emph {et~al.}(2026)\citenamefont
  {Schleith}, \citenamefont {Soejima},\ and\ \citenamefont
  {Khalaf}}]{Schleith2026}%
  \BibitemOpen
  \bibfield  {author} {\bibinfo {author} {\bibfnamefont {M.-L.}\ \bibnamefont
  {Schleith}}, \bibinfo {author} {\bibfnamefont {T.}~\bibnamefont {Soejima}},\
  and\ \bibinfo {author} {\bibfnamefont {E.}~\bibnamefont {Khalaf}},\
  }\bibfield  {title} {\bibinfo {title} {Anyon dispersion from nonuniform
  magnetic field on the sphere},\ }\href {https://doi.org/10.1103/gzdq-wkxj}
  {\bibfield  {journal} {\bibinfo  {journal} {Phys. Rev. Lett.}\ }\textbf
  {\bibinfo {volume} {137}},\ \bibinfo {pages} {046506} (\bibinfo {year}
  {2026})}\BibitemShut {NoStop}%
\bibitem [{\citenamefont {Gao}\ and\ \citenamefont {Khalaf}(2026)}]{Gao2026b}%
  \BibitemOpen
  \bibfield  {author} {\bibinfo {author} {\bibfnamefont {Q.}~\bibnamefont
  {Gao}}\ and\ \bibinfo {author} {\bibfnamefont {E.}~\bibnamefont {Khalaf}},\
  }\href@noop {} {\bibinfo {title} {Microsopic theory of spin polarons in
  {C}hern ferromagnets}} (\bibinfo {year} {2026}),\ \Eprint
  {https://arxiv.org/abs/arXiv:2605.05292} {arXiv:2605.05292} \BibitemShut
  {NoStop}%
\bibitem [{\citenamefont {Peyrl}\ and\ \citenamefont
  {Parrilo}(2008)}]{Peyrl2008}%
  \BibitemOpen
  \bibfield  {author} {\bibinfo {author} {\bibfnamefont {H.}~\bibnamefont
  {Peyrl}}\ and\ \bibinfo {author} {\bibfnamefont {P.~A.}\ \bibnamefont
  {Parrilo}},\ }\bibfield  {title} {\bibinfo {title} {Computing sum of squares
  decompositions with rational coefficients},\ }\href
  {https://doi.org/10.1016/j.tcs.2008.09.025} {\bibfield  {journal} {\bibinfo
  {journal} {Theoretical Computer Science}\ }\textbf {\bibinfo {volume}
  {409}},\ \bibinfo {pages} {269} (\bibinfo {year} {2008})},\ \bibinfo {note}
  {symbolic-Numerical Computations}\BibitemShut {NoStop}%
\bibitem [{\citenamefont {Kaltofen}\ \emph {et~al.}(2012)\citenamefont
  {Kaltofen}, \citenamefont {Li}, \citenamefont {Yang},\ and\ \citenamefont
  {Zhi}}]{Kaltofen2012}%
  \BibitemOpen
  \bibfield  {author} {\bibinfo {author} {\bibfnamefont {E.~L.}\ \bibnamefont
  {Kaltofen}}, \bibinfo {author} {\bibfnamefont {B.}~\bibnamefont {Li}},
  \bibinfo {author} {\bibfnamefont {Z.}~\bibnamefont {Yang}},\ and\ \bibinfo
  {author} {\bibfnamefont {L.}~\bibnamefont {Zhi}},\ }\bibfield  {title}
  {\bibinfo {title} {Exact certification in global polynomial optimization via
  sums-of-squares of rational functions with rational coefficients},\ }\href
  {https://doi.org/10.1016/j.jsc.2011.08.002} {\bibfield  {journal} {\bibinfo
  {journal} {Journal of Symbolic Computation}\ }\textbf {\bibinfo {volume}
  {47}},\ \bibinfo {pages} {1} (\bibinfo {year} {2012})}\BibitemShut {NoStop}%
\bibitem [{\citenamefont {Naceur}\ \emph {et~al.}(2025)\citenamefont {Naceur},
  \citenamefont {Wang}, \citenamefont {Magron},\ and\ \citenamefont
  {Acín}}]{Naceur2025}%
  \BibitemOpen
  \bibfield  {author} {\bibinfo {author} {\bibfnamefont {Y.}~\bibnamefont
  {Naceur}}, \bibinfo {author} {\bibfnamefont {J.}~\bibnamefont {Wang}},
  \bibinfo {author} {\bibfnamefont {V.}~\bibnamefont {Magron}},\ and\ \bibinfo
  {author} {\bibfnamefont {A.}~\bibnamefont {Acín}},\ }\href@noop {} {\bibinfo
  {title} {Certified bounds on optimization problems in quantum theory}}
  (\bibinfo {year} {2025}),\ \Eprint {https://arxiv.org/abs/arXiv:2512.17713}
  {arXiv:2512.17713} \BibitemShut {NoStop}%
\bibitem [{\citenamefont {Jansson}\ \emph {et~al.}(2008)\citenamefont
  {Jansson}, \citenamefont {Chaykin},\ and\ \citenamefont
  {Keil}}]{Jansson2008}%
  \BibitemOpen
  \bibfield  {author} {\bibinfo {author} {\bibfnamefont {C.}~\bibnamefont
  {Jansson}}, \bibinfo {author} {\bibfnamefont {D.}~\bibnamefont {Chaykin}},\
  and\ \bibinfo {author} {\bibfnamefont {C.}~\bibnamefont {Keil}},\ }\bibfield
  {title} {\bibinfo {title} {Rigorous error bounds for the optimal value in
  semidefinite programming},\ }\href {https://doi.org/10.1137/050622870}
  {\bibfield  {journal} {\bibinfo  {journal} {SIAM Journal on Numerical
  Analysis}\ }\textbf {\bibinfo {volume} {46}},\ \bibinfo {pages} {180}
  (\bibinfo {year} {2008})},\ \Eprint
  {https://arxiv.org/abs/https://doi.org/10.1137/050622870}
  {https://doi.org/10.1137/050622870} \BibitemShut {NoStop}%
\bibitem [{\citenamefont {Robinson}(1975)}]{Robinson1975}%
  \BibitemOpen
  \bibfield  {author} {\bibinfo {author} {\bibfnamefont {S.~M.}\ \bibnamefont
  {Robinson}},\ }\bibfield  {title} {\bibinfo {title} {Stability theory for
  systems of inequalities. {Part I}: Linear systems},\ }\href
  {https://doi.org/10.1137/0712056} {\bibfield  {journal} {\bibinfo  {journal}
  {SIAM J. Numer. Anal.}\ }\textbf {\bibinfo {volume} {12}},\ \bibinfo {pages}
  {754} (\bibinfo {year} {1975})}\BibitemShut {NoStop}%
\bibitem [{\citenamefont {Bonnans}\ and\ \citenamefont
  {Shapiro}(2000)}]{BonnansShapiro2000}%
  \BibitemOpen
  \bibfield  {author} {\bibinfo {author} {\bibfnamefont {J.~F.}\ \bibnamefont
  {Bonnans}}\ and\ \bibinfo {author} {\bibfnamefont {A.}~\bibnamefont
  {Shapiro}},\ }\href {https://doi.org/10.1007/978-1-4612-1394-9} {\emph
  {\bibinfo {title} {Perturbation Analysis of Optimization Problems}}},\
  Springer Series in Operations Research\ (\bibinfo  {publisher} {Springer},\
  \bibinfo {address} {New York},\ \bibinfo {year} {2000})\BibitemShut {NoStop}%
\bibitem [{\citenamefont {Scheer}\ and\ \citenamefont
  {Khalaf}(2026)}]{LaughlinGapCertificate}%
  \BibitemOpen
  \bibfield  {author} {\bibinfo {author} {\bibfnamefont {M.~G.}\ \bibnamefont
  {Scheer}}\ and\ \bibinfo {author} {\bibfnamefont {E.}~\bibnamefont
  {Khalaf}},\ }\href@noop {} {\bibinfo {title} {{LaughlinGapCertificate}}},\
  \bibinfo {howpublished}
  {\url{https://github.com/mgscheer/LaughlinGapCertificate}} (\bibinfo {year}
  {2026}),\ \bibinfo {note} {version 0.1.0}\BibitemShut {NoStop}%
\bibitem [{\citenamefont {ApS}(2026)}]{mosek}%
  \BibitemOpen
  \bibfield  {author} {\bibinfo {author} {\bibfnamefont {M.}~\bibnamefont
  {ApS}},\ }\href {https://docs.mosek.com/latest/juliaapi/index.html} {\emph
  {\bibinfo {title} {MOSEK Optimizer API for Julia 11.2.1}}} (\bibinfo {year}
  {2026})\BibitemShut {NoStop}%
\bibitem [{\citenamefont {Haldane}\ and\ \citenamefont
  {Rezayi}(1985)}]{Haldane1985}%
  \BibitemOpen
  \bibfield  {author} {\bibinfo {author} {\bibfnamefont {F.~D.~M.}\
  \bibnamefont {Haldane}}\ and\ \bibinfo {author} {\bibfnamefont {E.~H.}\
  \bibnamefont {Rezayi}},\ }\bibfield  {title} {\bibinfo {title} {Finite-size
  studies of the incompressible state of the fractionally quantized {H}all
  effect and its excitations},\ }\href
  {https://doi.org/10.1103/PhysRevLett.54.237} {\bibfield  {journal} {\bibinfo
  {journal} {Phys. Rev. Lett.}\ }\textbf {\bibinfo {volume} {54}},\ \bibinfo
  {pages} {237} (\bibinfo {year} {1985})}\BibitemShut {NoStop}%
\bibitem [{\citenamefont {von~zur Gathen}\ and\ \citenamefont
  {Gerhard}(2013)}]{vonzurGathen2013}%
  \BibitemOpen
  \bibfield  {author} {\bibinfo {author} {\bibfnamefont {J.}~\bibnamefont
  {von~zur Gathen}}\ and\ \bibinfo {author} {\bibfnamefont {J.}~\bibnamefont
  {Gerhard}},\ }\href@noop {} {\emph {\bibinfo {title} {Modern Computer
  Algebra}}},\ \bibinfo {edition} {3rd}\ ed.\ (\bibinfo  {publisher} {Cambridge
  University Press},\ \bibinfo {address} {Cambridge},\ \bibinfo {year}
  {2013})\BibitemShut {NoStop}%
\bibitem [{\citenamefont {Davis}\ \emph {et~al.}(2004)\citenamefont {Davis},
  \citenamefont {Gilbert}, \citenamefont {Larimore},\ and\ \citenamefont
  {Ng}}]{Davis2004}%
  \BibitemOpen
  \bibfield  {author} {\bibinfo {author} {\bibfnamefont {T.~A.}\ \bibnamefont
  {Davis}}, \bibinfo {author} {\bibfnamefont {J.~R.}\ \bibnamefont {Gilbert}},
  \bibinfo {author} {\bibfnamefont {S.~I.}\ \bibnamefont {Larimore}},\ and\
  \bibinfo {author} {\bibfnamefont {E.~G.}\ \bibnamefont {Ng}},\ }\bibfield
  {title} {\bibinfo {title} {Algorithm 836: {COLAMD}, a column approximate
  minimum degree ordering algorithm},\ }\href
  {https://doi.org/10.1145/1024074.1024080} {\bibfield  {journal} {\bibinfo
  {journal} {ACM Transactions on Mathematical Software}\ }\textbf {\bibinfo
  {volume} {30}},\ \bibinfo {pages} {377} (\bibinfo {year} {2004})}\BibitemShut
  {NoStop}%
\bibitem [{\citenamefont {Mohammadisiahroudi}\ \emph
  {et~al.}(2025)\citenamefont {Mohammadisiahroudi}, \citenamefont {Augustino},
  \citenamefont {Sampourmahani},\ and\ \citenamefont
  {Terlaky}}]{Mohammadisiahroudi2025}%
  \BibitemOpen
  \bibfield  {author} {\bibinfo {author} {\bibfnamefont {M.}~\bibnamefont
  {Mohammadisiahroudi}}, \bibinfo {author} {\bibfnamefont {B.}~\bibnamefont
  {Augustino}}, \bibinfo {author} {\bibfnamefont {P.}~\bibnamefont
  {Sampourmahani}},\ and\ \bibinfo {author} {\bibfnamefont {T.}~\bibnamefont
  {Terlaky}},\ }\bibfield  {title} {\bibinfo {title} {Quantum computing
  inspired iterative refinement for semidefinite optimization},\ }\bibfield
  {journal} {\bibinfo  {journal} {Mathematical Programming}\ }\href
  {https://doi.org/10.1007/s10107-024-02183-z} {10.1007/s10107-024-02183-z}
  (\bibinfo {year} {2025})\BibitemShut {NoStop}%
\bibitem [{\citenamefont {Dixon}(1982)}]{Dixon1982}%
  \BibitemOpen
  \bibfield  {author} {\bibinfo {author} {\bibfnamefont {J.}~\bibnamefont
  {Dixon}},\ }\bibfield  {title} {\bibinfo {title} {Exact solution of linear
  equations using p-adic expansions},\ }\href {http://eudml.org/doc/132821}
  {\bibfield  {journal} {\bibinfo  {journal} {Numerische Mathematik}\ }\textbf
  {\bibinfo {volume} {40}},\ \bibinfo {pages} {137} (\bibinfo {year}
  {1982})}\BibitemShut {NoStop}%
\bibitem [{\citenamefont {Wang}\ \emph {et~al.}(1982)\citenamefont {Wang},
  \citenamefont {Guy},\ and\ \citenamefont {Davenport}}]{Wang1982}%
  \BibitemOpen
  \bibfield  {author} {\bibinfo {author} {\bibfnamefont {P.~S.}\ \bibnamefont
  {Wang}}, \bibinfo {author} {\bibfnamefont {M.~J.~T.}\ \bibnamefont {Guy}},\
  and\ \bibinfo {author} {\bibfnamefont {J.~H.}\ \bibnamefont {Davenport}},\
  }\bibfield  {title} {\bibinfo {title} {P-adic reconstruction of rational
  numbers},\ }\href {https://doi.org/10.1145/1089292.1089293} {\bibfield
  {journal} {\bibinfo  {journal} {ACM SIGSAM Bulletin}\ }\textbf {\bibinfo
  {volume} {16}},\ \bibinfo {pages} {2} (\bibinfo {year} {1982})}\BibitemShut
  {NoStop}%
\bibitem [{\citenamefont {Wu}\ \emph {et~al.}(2019)\citenamefont {Wu},
  \citenamefont {Lovorn}, \citenamefont {Tutuc}, \citenamefont {Martin},\ and\
  \citenamefont {MacDonald}}]{WuMacdonald}%
  \BibitemOpen
  \bibfield  {author} {\bibinfo {author} {\bibfnamefont {F.}~\bibnamefont
  {Wu}}, \bibinfo {author} {\bibfnamefont {T.}~\bibnamefont {Lovorn}}, \bibinfo
  {author} {\bibfnamefont {E.}~\bibnamefont {Tutuc}}, \bibinfo {author}
  {\bibfnamefont {I.}~\bibnamefont {Martin}},\ and\ \bibinfo {author}
  {\bibfnamefont {A.~H.}\ \bibnamefont {MacDonald}},\ }\bibfield  {title}
  {\bibinfo {title} {Topological {{Insulators}} in {{Twisted Transition Metal
  Dichalcogenide Homobilayers}}},\ }\href
  {https://doi.org/10.1103/PhysRevLett.122.086402} {\bibfield  {journal}
  {\bibinfo  {journal} {Phys. Rev. Lett.}\ }\textbf {\bibinfo {volume} {122}},\
  \bibinfo {pages} {086402} (\bibinfo {year} {2019})}\BibitemShut {NoStop}%
\bibitem [{\citenamefont {Xu}\ \emph {et~al.}(2023)\citenamefont {Xu},
  \citenamefont {Sun}, \citenamefont {Jia}, \citenamefont {Liu}, \citenamefont
  {Xu}, \citenamefont {Li}, \citenamefont {Gu}, \citenamefont {Watanabe},
  \citenamefont {Taniguchi}, \citenamefont {Tong}, \citenamefont {Jia},
  \citenamefont {Shi}, \citenamefont {Jiang}, \citenamefont {Zhang},
  \citenamefont {Liu},\ and\ \citenamefont {Li}}]{xu2023observation}%
  \BibitemOpen
  \bibfield  {author} {\bibinfo {author} {\bibfnamefont {F.}~\bibnamefont
  {Xu}}, \bibinfo {author} {\bibfnamefont {Z.}~\bibnamefont {Sun}}, \bibinfo
  {author} {\bibfnamefont {T.}~\bibnamefont {Jia}}, \bibinfo {author}
  {\bibfnamefont {C.}~\bibnamefont {Liu}}, \bibinfo {author} {\bibfnamefont
  {C.}~\bibnamefont {Xu}}, \bibinfo {author} {\bibfnamefont {C.}~\bibnamefont
  {Li}}, \bibinfo {author} {\bibfnamefont {Y.}~\bibnamefont {Gu}}, \bibinfo
  {author} {\bibfnamefont {K.}~\bibnamefont {Watanabe}}, \bibinfo {author}
  {\bibfnamefont {T.}~\bibnamefont {Taniguchi}}, \bibinfo {author}
  {\bibfnamefont {B.}~\bibnamefont {Tong}}, \bibinfo {author} {\bibfnamefont
  {J.}~\bibnamefont {Jia}}, \bibinfo {author} {\bibfnamefont {Z.}~\bibnamefont
  {Shi}}, \bibinfo {author} {\bibfnamefont {S.}~\bibnamefont {Jiang}}, \bibinfo
  {author} {\bibfnamefont {Y.}~\bibnamefont {Zhang}}, \bibinfo {author}
  {\bibfnamefont {X.}~\bibnamefont {Liu}},\ and\ \bibinfo {author}
  {\bibfnamefont {T.}~\bibnamefont {Li}},\ }\bibfield  {title} {\bibinfo
  {title} {Observation of integer and fractional quantum anomalous {H}all
  effects in twisted bilayer {M}o{T}e$_2$},\ }\href
  {https://doi.org/10.1103/PhysRevX.13.031037} {\bibfield  {journal} {\bibinfo
  {journal} {Phys. Rev. X}\ }\textbf {\bibinfo {volume} {13}},\ \bibinfo
  {pages} {031037} (\bibinfo {year} {2023})}\BibitemShut {NoStop}%
\bibitem [{\citenamefont {Cai}\ \emph {et~al.}(2023)\citenamefont {Cai},
  \citenamefont {Anderson}, \citenamefont {Wang}, \citenamefont {Zhang},
  \citenamefont {Liu}, \citenamefont {Holtzmann}, \citenamefont {Zhang},
  \citenamefont {Fan}, \citenamefont {Taniguchi}, \citenamefont {Watanabe},
  \citenamefont {Ran}, \citenamefont {Cao}, \citenamefont {Fu}, \citenamefont
  {Xiao}, \citenamefont {Yao},\ and\ \citenamefont {Xu}}]{cai2023signatures}%
  \BibitemOpen
  \bibfield  {author} {\bibinfo {author} {\bibfnamefont {J.}~\bibnamefont
  {Cai}}, \bibinfo {author} {\bibfnamefont {E.}~\bibnamefont {Anderson}},
  \bibinfo {author} {\bibfnamefont {C.}~\bibnamefont {Wang}}, \bibinfo {author}
  {\bibfnamefont {X.}~\bibnamefont {Zhang}}, \bibinfo {author} {\bibfnamefont
  {X.}~\bibnamefont {Liu}}, \bibinfo {author} {\bibfnamefont {W.}~\bibnamefont
  {Holtzmann}}, \bibinfo {author} {\bibfnamefont {Y.}~\bibnamefont {Zhang}},
  \bibinfo {author} {\bibfnamefont {F.}~\bibnamefont {Fan}}, \bibinfo {author}
  {\bibfnamefont {T.}~\bibnamefont {Taniguchi}}, \bibinfo {author}
  {\bibfnamefont {K.}~\bibnamefont {Watanabe}}, \bibinfo {author}
  {\bibfnamefont {Y.}~\bibnamefont {Ran}}, \bibinfo {author} {\bibfnamefont
  {T.}~\bibnamefont {Cao}}, \bibinfo {author} {\bibfnamefont {L.}~\bibnamefont
  {Fu}}, \bibinfo {author} {\bibfnamefont {D.}~\bibnamefont {Xiao}}, \bibinfo
  {author} {\bibfnamefont {W.}~\bibnamefont {Yao}},\ and\ \bibinfo {author}
  {\bibfnamefont {X.}~\bibnamefont {Xu}},\ }\bibfield  {title} {\bibinfo
  {title} {Signatures of fractional quantum anomalous {H}all states in twisted
  {M}o{T}e$_2$},\ }\href {https://doi.org/10.1038/s41586-023-06289-w}
  {\bibfield  {journal} {\bibinfo  {journal} {Nature}\ }\textbf {\bibinfo
  {volume} {622}},\ \bibinfo {pages} {63} (\bibinfo {year} {2023})}\BibitemShut
  {NoStop}%
\bibitem [{\citenamefont {Park}\ \emph {et~al.}(2023)\citenamefont {Park},
  \citenamefont {Cai}, \citenamefont {Anderson}, \citenamefont {Zhang},
  \citenamefont {Zhu}, \citenamefont {Liu}, \citenamefont {Wang}, \citenamefont
  {Holtzmann}, \citenamefont {Hu}, \citenamefont {Liu}, \citenamefont
  {Taniguchi}, \citenamefont {Watanabe}, \citenamefont {Chu}, \citenamefont
  {Cao}, \citenamefont {Fu}, \citenamefont {Yao}, \citenamefont {Chang},
  \citenamefont {Cobden}, \citenamefont {Xiao},\ and\ \citenamefont
  {Xu}}]{park2023observation}%
  \BibitemOpen
  \bibfield  {author} {\bibinfo {author} {\bibfnamefont {H.}~\bibnamefont
  {Park}}, \bibinfo {author} {\bibfnamefont {J.}~\bibnamefont {Cai}}, \bibinfo
  {author} {\bibfnamefont {E.}~\bibnamefont {Anderson}}, \bibinfo {author}
  {\bibfnamefont {Y.}~\bibnamefont {Zhang}}, \bibinfo {author} {\bibfnamefont
  {J.}~\bibnamefont {Zhu}}, \bibinfo {author} {\bibfnamefont {X.}~\bibnamefont
  {Liu}}, \bibinfo {author} {\bibfnamefont {C.}~\bibnamefont {Wang}}, \bibinfo
  {author} {\bibfnamefont {W.}~\bibnamefont {Holtzmann}}, \bibinfo {author}
  {\bibfnamefont {C.}~\bibnamefont {Hu}}, \bibinfo {author} {\bibfnamefont
  {Z.}~\bibnamefont {Liu}}, \bibinfo {author} {\bibfnamefont {T.}~\bibnamefont
  {Taniguchi}}, \bibinfo {author} {\bibfnamefont {K.}~\bibnamefont {Watanabe}},
  \bibinfo {author} {\bibfnamefont {J.-H.}\ \bibnamefont {Chu}}, \bibinfo
  {author} {\bibfnamefont {T.}~\bibnamefont {Cao}}, \bibinfo {author}
  {\bibfnamefont {L.}~\bibnamefont {Fu}}, \bibinfo {author} {\bibfnamefont
  {W.}~\bibnamefont {Yao}}, \bibinfo {author} {\bibfnamefont {C.-Z.}\
  \bibnamefont {Chang}}, \bibinfo {author} {\bibfnamefont {D.}~\bibnamefont
  {Cobden}}, \bibinfo {author} {\bibfnamefont {D.}~\bibnamefont {Xiao}},\ and\
  \bibinfo {author} {\bibfnamefont {X.}~\bibnamefont {Xu}},\ }\bibfield
  {title} {\bibinfo {title} {Observation of fractionally quantized anomalous
  {H}all effect},\ }\href {https://doi.org/10.1038/s41586-023-06536-0}
  {\bibfield  {journal} {\bibinfo  {journal} {Nature}\ }\textbf {\bibinfo
  {volume} {622}},\ \bibinfo {pages} {74} (\bibinfo {year} {2023})}\BibitemShut
  {NoStop}%
\bibitem [{\citenamefont {Zeng}\ \emph {et~al.}(2023)\citenamefont {Zeng},
  \citenamefont {Xia}, \citenamefont {Kang}, \citenamefont {Zhu}, \citenamefont
  {Kn{\"u}ppel}, \citenamefont {Vaswani}, \citenamefont {Watanabe},
  \citenamefont {Taniguchi}, \citenamefont {Mak},\ and\ \citenamefont
  {Shan}}]{zeng2023thermodynamic}%
  \BibitemOpen
  \bibfield  {author} {\bibinfo {author} {\bibfnamefont {Y.}~\bibnamefont
  {Zeng}}, \bibinfo {author} {\bibfnamefont {Z.}~\bibnamefont {Xia}}, \bibinfo
  {author} {\bibfnamefont {K.}~\bibnamefont {Kang}}, \bibinfo {author}
  {\bibfnamefont {J.}~\bibnamefont {Zhu}}, \bibinfo {author} {\bibfnamefont
  {P.}~\bibnamefont {Kn{\"u}ppel}}, \bibinfo {author} {\bibfnamefont
  {C.}~\bibnamefont {Vaswani}}, \bibinfo {author} {\bibfnamefont
  {K.}~\bibnamefont {Watanabe}}, \bibinfo {author} {\bibfnamefont
  {T.}~\bibnamefont {Taniguchi}}, \bibinfo {author} {\bibfnamefont {K.~F.}\
  \bibnamefont {Mak}},\ and\ \bibinfo {author} {\bibfnamefont {J.}~\bibnamefont
  {Shan}},\ }\bibfield  {title} {\bibinfo {title} {Thermodynamic evidence of
  fractional {C}hern insulator in moir{\'e} {M}o{T}e$_2$},\ }\href
  {https://doi.org/10.1038/s41586-023-06452-3} {\bibfield  {journal} {\bibinfo
  {journal} {Nature}\ }\textbf {\bibinfo {volume} {622}},\ \bibinfo {pages}
  {69} (\bibinfo {year} {2023})}\BibitemShut {NoStop}%
\bibitem [{\citenamefont {Aharonov}\ and\ \citenamefont
  {Casher}(1979)}]{aharonov1979ground}%
  \BibitemOpen
  \bibfield  {author} {\bibinfo {author} {\bibfnamefont {Y.}~\bibnamefont
  {Aharonov}}\ and\ \bibinfo {author} {\bibfnamefont {A.}~\bibnamefont
  {Casher}},\ }\bibfield  {title} {\bibinfo {title} {Ground state of a
  spin-{$\frac{1}{2}$} charged particle in a two-dimensional magnetic field},\
  }\href {https://doi.org/10.1103/PhysRevA.19.2461} {\bibfield  {journal}
  {\bibinfo  {journal} {Phys. Rev. A}\ }\textbf {\bibinfo {volume} {19}},\
  \bibinfo {pages} {2461} (\bibinfo {year} {1979})}\BibitemShut {NoStop}%
\bibitem [{\citenamefont {Ledwith}\ \emph {et~al.}(2020)\citenamefont
  {Ledwith}, \citenamefont {Tarnopolsky}, \citenamefont {Khalaf},\ and\
  \citenamefont {Vishwanath}}]{ledwithFractionalChernInsulator2020a}%
  \BibitemOpen
  \bibfield  {author} {\bibinfo {author} {\bibfnamefont {P.~J.}\ \bibnamefont
  {Ledwith}}, \bibinfo {author} {\bibfnamefont {G.}~\bibnamefont
  {Tarnopolsky}}, \bibinfo {author} {\bibfnamefont {E.}~\bibnamefont
  {Khalaf}},\ and\ \bibinfo {author} {\bibfnamefont {A.}~\bibnamefont
  {Vishwanath}},\ }\bibfield  {title} {\bibinfo {title} {Fractional {C}hern
  insulator states in twisted bilayer graphene: {A}n analytical approach},\
  }\href {https://doi.org/10.1103/PhysRevResearch.2.023237} {\bibfield
  {journal} {\bibinfo  {journal} {Phys. Rev. Research}\ }\textbf {\bibinfo
  {volume} {2}},\ \bibinfo {pages} {023237} (\bibinfo {year}
  {2020})}\BibitemShut {NoStop}%
\bibitem [{\citenamefont {Wang}\ \emph {et~al.}(2021)\citenamefont {Wang},
  \citenamefont {Cano}, \citenamefont {Millis}, \citenamefont {Liu},\ and\
  \citenamefont {Yang}}]{Wang2021exact}%
  \BibitemOpen
  \bibfield  {author} {\bibinfo {author} {\bibfnamefont {J.}~\bibnamefont
  {Wang}}, \bibinfo {author} {\bibfnamefont {J.}~\bibnamefont {Cano}}, \bibinfo
  {author} {\bibfnamefont {A.~J.}\ \bibnamefont {Millis}}, \bibinfo {author}
  {\bibfnamefont {Z.}~\bibnamefont {Liu}},\ and\ \bibinfo {author}
  {\bibfnamefont {B.}~\bibnamefont {Yang}},\ }\bibfield  {title} {\bibinfo
  {title} {Exact {{Landau Level Description}} of {{Geometry}} and
  {{Interaction}} in a {{Flatband}}},\ }\href
  {https://doi.org/10.1103/PhysRevLett.127.246403} {\bibfield  {journal}
  {\bibinfo  {journal} {Phys. Rev. Lett.}\ }\textbf {\bibinfo {volume} {127}},\
  \bibinfo {pages} {246403} (\bibinfo {year} {2021})}\BibitemShut {NoStop}%
\bibitem [{\citenamefont {Ledwith}\ \emph {et~al.}(2023)\citenamefont
  {Ledwith}, \citenamefont {Vishwanath},\ and\ \citenamefont
  {Parker}}]{ledwith_vortexability_2023}%
  \BibitemOpen
  \bibfield  {author} {\bibinfo {author} {\bibfnamefont {P.~J.}\ \bibnamefont
  {Ledwith}}, \bibinfo {author} {\bibfnamefont {A.}~\bibnamefont
  {Vishwanath}},\ and\ \bibinfo {author} {\bibfnamefont {D.~E.}\ \bibnamefont
  {Parker}},\ }\bibfield  {title} {\bibinfo {title} {Vortexability: {A}
  {Unifying} {Criterion} for {Ideal} {Fractional} {Chern} {Insulators}},\
  }\href {https://doi.org/10.1103/PhysRevB.108.205144} {\bibfield  {journal}
  {\bibinfo  {journal} {Physical Review B}\ }\textbf {\bibinfo {volume}
  {108}},\ \bibinfo {pages} {205144} (\bibinfo {year} {2023})},\ \bibinfo
  {note} {arXiv:2209.15023 [cond-mat]}\BibitemShut {NoStop}%
\bibitem [{\citenamefont {Shi}\ \emph {et~al.}(2024)\citenamefont {Shi},
  \citenamefont {Morales-Dur\'an}, \citenamefont {Khalaf},\ and\ \citenamefont
  {MacDonald}}]{TMDAharonocCasher}%
  \BibitemOpen
  \bibfield  {author} {\bibinfo {author} {\bibfnamefont {J.}~\bibnamefont
  {Shi}}, \bibinfo {author} {\bibfnamefont {N.}~\bibnamefont
  {Morales-Dur\'an}}, \bibinfo {author} {\bibfnamefont {E.}~\bibnamefont
  {Khalaf}},\ and\ \bibinfo {author} {\bibfnamefont {A.~H.}\ \bibnamefont
  {MacDonald}},\ }\bibfield  {title} {\bibinfo {title} {Adiabatic approximation
  and {A}haronov-{C}asher bands in twisted homobilayer transition metal
  dichalcogenides},\ }\href {https://doi.org/10.1103/PhysRevB.110.035130}
  {\bibfield  {journal} {\bibinfo  {journal} {Phys. Rev. B}\ }\textbf {\bibinfo
  {volume} {110}},\ \bibinfo {pages} {035130} (\bibinfo {year}
  {2024})}\BibitemShut {NoStop}%
\bibitem [{\citenamefont
  {{OpenAI}}(2026{\natexlab{a}})}]{OpenAI2026MathRelease}%
  \BibitemOpen
  \bibfield  {author} {\bibinfo {author} {\bibnamefont {{OpenAI}}},\ }\href
  {https://github.com/openai/math} {\bibinfo {title} {{openai/math}:
  mathematical manuscripts and {Lean} formalizations}},\ \bibinfo
  {howpublished} {GitHub repository, released October 6, 2026} (\bibinfo {year}
  {2026}{\natexlab{a}})\BibitemShut {NoStop}%
\bibitem [{\citenamefont
  {{OpenAI}}(2026{\natexlab{b}})}]{OpenAI2026LaughlinGap}%
  \BibitemOpen
  \bibfield  {author} {\bibinfo {author} {\bibnamefont {{OpenAI}}},\ }\href
  {https://github.com/openai/math/blob/main/preprints/A-Fock-space-inequality-and-the-Laughlin-spectral-gap-September-24-2026/A-Fock-space-inequality-and-the-Laughlin-spectral-gap-September-24-2026.pdf}
  {\bibinfo {title} {A {F}ock-space inequality and the {L}aughlin spectral
  gap}},\ \bibinfo {howpublished} {OpenAI Math Release preprint, dated
  September 24, 2026; publicly released October 6, 2026} (\bibinfo {year}
  {2026}{\natexlab{b}}),\ \bibinfo {note} {accessed October 12,
  2026}\BibitemShut {NoStop}%
\bibitem [{\citenamefont {Gao}(2026)}]{Gao2026}%
  \BibitemOpen
  \bibfield  {author} {\bibinfo {author} {\bibfnamefont {Q.}~\bibnamefont
  {Gao}},\ }\href@noop {} {\bibinfo {title} {Improving the lower bound on the
  {L}aughlin gap: An {SDP} search}} (\bibinfo {year} {2026}),\ \Eprint
  {https://arxiv.org/abs/arXiv:2610.12700} {arXiv:2610.12700} \BibitemShut
  {NoStop}%
\bibitem [{\citenamefont {Perelomov}(1977)}]{Perelomov1977}%
  \BibitemOpen
  \bibfield  {author} {\bibinfo {author} {\bibfnamefont {A.~M.}\ \bibnamefont
  {Perelomov}},\ }\bibfield  {title} {\bibinfo {title} {Generalized coherent
  states and some of their applications},\ }\href@noop {} {\bibfield  {journal}
  {\bibinfo  {journal} {Soviet Physics Uspekhi}\ }\textbf {\bibinfo {volume}
  {20}},\ \bibinfo {pages} {703} (\bibinfo {year} {1977})}\BibitemShut
  {NoStop}%
\end{thebibliography}%
\onecolumngrid
\clearpage
\appendix
\crefalias{section}{appendix}
\crefalias{subsection}{appendix}

\section{Review of the Haldane pseudopotentials and Laughlin state on the sphere}\label{app:fqhe-sphere}
Following Ref. \cite{Haldane1983}, we begin by studying the problem of a single quantum particle on the surface of a $2$-sphere of radius $R$ embedded in $\mathbb{R}^3$, at the center of which is placed a magnetic monopole. We denote the magnetic field strength at the surface of the sphere by $B > 0$ so that the magnetic field points radially outwards and the magnetic flux through the sphere is $\Phi = 4\pi R^2 B$. By Dirac quantization, this flux must be an integer multiple $N_\Phi$ of the flux quantum $\Phi_0 = 2\pi\hbar/q_e$ where the electron charge is $-q_e < 0$. We define $s = (R^2 B q_e)/\hbar$ so that $2s = N_\Phi = \Phi/\Phi_0$ is a positive integer, namely the number of flux quanta. The thermodynamic limit is defined by taking $s \to \infty$ and $R \to \infty$ with $R^2/s = \ell_B^2$ fixed, where $\ell_B = \sqrt{\hbar/(B q_e)}$ is the magnetic length. In this limit, the south pole of the sphere converges to a plane with a uniform magnetic field such that the electron charge times magnetic field points in the positive $z$ direction\footnote{We are following here the most commonly used convention which takes the electron charge times magnetic field strength to be negative, i.e. radially inward. This has the benefit of making the connection to the prior literature and also to the plane in the thermodynamic limit particularly clean. However, we note that this choice renders the Chern number of the bundle of wavefunctions to be negative, which makes the math of the Hopf bundle look slightly more complicated. The alternative convention would be to take the electron charge times magnetic field strength to be positive.}.

We can identify the position of a particle on the surface of the sphere by an element of the unit $2$-sphere $S^2$. If there were no magnetic monopole, we could take the single-particle Hilbert space to consist of square-integrable functions from $S^2$ to $\mathbb{C}$. In the presence of the magnetic monopole, the single-particle Hilbert space instead consists of square-integrable sections of a complex line bundle over $S^2$ with structure group $U(1)$ and Chern number $-2s$. To construct this line bundle, we consider an atlas of two charts on $S^2$, namely the northern ($\mathcal{N}$) and southern ($\mathcal{S}$) hemispheres, each enlarged by a small $\varepsilon$ in the polar direction so as to be open sets. More explicitly using spherical coordinates, the northern hemisphere consists of $(\theta, \phi) \in [0, \pi/2 + \varepsilon) \times [0, 2\pi)$ and the southern hemisphere consists of $(\theta, \phi) \in (\pi/2-\varepsilon, \pi] \times [0, 2\pi)$. The two charts overlap on a strip around the equator $(\theta, \phi) \in (\pi/2-\varepsilon, \pi/2+\varepsilon) \times [0, 2\pi)$, on which we supply the transition map (also known as a gauge transformation)
\begin{equation}\label{eq:transition-map}
\Psi_{\mathcal{S}, \mathcal{N}}(\theta, \phi) = e^{-i(2s)\phi}.
\end{equation}
This means that a section of the bundle is determined by its coordinates $\psi_{\mathcal{N}}$ and $\psi_{\mathcal{S}}$ in the two charts, which are related by
\begin{equation}\label{eq:transition-of-sections}
\psi_{\mathcal{N}} = \psi_{\mathcal{S}} \Psi_{\mathcal{S}, \mathcal{N}}
\end{equation}
on the equatorial strip. To see that this bundle has Chern number $-2s$, we consider any connection on the associated principal $U(1)$ bundle with local $u(1)$-valued differential forms $\mathcal{A}_{\mathcal{N}}$ and $\mathcal{A}_{\mathcal{S}}$ on the two charts. By Stokes theorem, the Chern number is the line integral around the equator of $\frac{1}{2\pi i}(\mathcal{A}_{\mathcal{N}} - \mathcal{A}_{\mathcal{S}})$. However, \cref{eq:transition-map} implies
\begin{equation}
\mathcal{A}_{\mathcal{N}} = \mathcal{A}_{\mathcal{S}} + \Psi_{\mathcal{N}, \mathcal{S}} d \Psi_{\mathcal{S}, \mathcal{N}} = \mathcal{A}_\mathcal{S} - i(2s) d\phi
\end{equation}
where we have used $\Psi_{\mathcal{N}, \mathcal{S}} \Psi_{\mathcal{S}, \mathcal{N}} = 1$, and we obtain Chern number $-2s$.

Since it is cumbersome to specify wavefunctions using two charts, we simplify the situation by introducing new coordinates. In the northern hemisphere, we take
\begin{equation}\label{eq:u-N-v-N}
u_{\mathcal{N}} = \cos(\theta/2), \quad v_{\mathcal{N}} = \sin(\theta/2)e^{-i\phi}
\end{equation}
while in the southern hemisphere, we take
\begin{equation}\label{eq:u-S-v-S}
u_{\mathcal{S}} = \cos(\theta/2)e^{i\phi}, \quad v_{\mathcal{S}} = \sin(\theta/2).
\end{equation}
Note that when $\theta = 0$, we have $(u_{\mathcal{N}}, v_{\mathcal{N}}) = (1, 0)$ independent of $\phi$ and similarly when $\theta = \pi$ we have $(u_{\mathcal{S}}, v_{\mathcal{S}}) = (0, 1)$ independent of $\phi$. This means that even though the $(\theta, \phi)$ coordinates are singular at the poles, the new complex coordinates are smooth everywhere within their respective charts. We also note that
\begin{equation}\label{eq:u-v-normalization}
|u_{\mathcal{N}}|^2 + |v_{\mathcal{N}}|^2 = |u_{\mathcal{S}}|^2 + |v_{\mathcal{S}}|^2 = 1.
\end{equation}

Now if $\psi(u,v,\overline{u},\overline{v})$ is a homogeneous polynomial in complex variables $u$, $v$, $\overline{u}$, $\overline{v}$ such that the $u$ and $v$ variables have combined degree $2s + n$ in each monomial while $\overline{u}$ and $\overline{v}$ have combined degree $n$ in each monomial for some integer $n \geq 0$, then the functions $\psi_{\mathcal{N}} = \psi(u_{\mathcal{N}},v_{\mathcal{N}},\overline{u}_{\mathcal{N}},\overline{v}_{\mathcal{N}})$ and $\psi_{\mathcal{S}} = \psi(u_{\mathcal{S}},v_{\mathcal{S}},\overline{u}_{\mathcal{S}},\overline{v}_{\mathcal{S}})$ automatically satisfy \cref{eq:transition-of-sections} on the equatorial strip. Since such a polynomial $\psi$ satisfies
\begin{equation}
\psi(e^{i\alpha} u, e^{i\alpha} v, e^{-i\alpha} \overline{u}, e^{-i\alpha} \overline{v}) = e^{i(2s)\alpha} \psi(u, v, \overline{u}, \overline{v})
\end{equation}
we refer to $\psi$ as an equivariant homogeneous polynomial. Because of \cref{eq:u-v-normalization}, we consider two equivariant homogeneous polynomials to be equivalent if their difference is divisible by $u\overline{u} + v\overline{v} - 1$.

To understand more deeply why this works, we note that $(u,v)$ with $|u|^2 + |v|^2 = 1$ can be thought of as coordinates for the unit $3$-sphere $S^3$, which is a principal $U(1)$ bundle over $S^2$. This bundle is defined by the projection map $\pi_{S^3,S^2} : S^3 \to S^2$ given by
\begin{equation}\label{eq:define-pi-projector}
\pi_{S^3,S^2}(u,v) = (n_x, n_y, n_z) \text{ where } n_\alpha = \begin{pmatrix}
\overline{u} & \overline{v}
\end{pmatrix} \sigma_\alpha^* \begin{pmatrix}
u\\ v
\end{pmatrix}
\end{equation}
and where $\sigma_\alpha$ are the three Pauli matrices. Since $\pi_{S^3,S^2}(e^{i\alpha} u, e^{i\alpha} v) = \pi_{S^3,S^2}(u,v)$, we see that the fiber is indeed $U(1)$. \cref{eq:u-N-v-N,eq:u-S-v-S} are simply local trivializations (also known as gauge choices) for this bundle, and their compositions with $\pi_{S^3, S^2}$ both give
\begin{equation}\label{eq:n-from-theta-phi}
(n_x, n_y, n_z) = (\sin\theta \cos\phi, \sin\theta\sin\phi, \cos\theta)
\end{equation}
which are spherical coordinates for $S^2$. The bundle of $S^3$ over $S^2$ with fiber $U(1)$ is known as the Hopf fibration or Hopf bundle. The key idea is that wavefunctions which are sections of a nontrivial complex line bundle on $S^2$ can be realized as equivariant homogeneous polynomials on $S^3$ composed with Hopf bundle local trivializations.

Finally, we consider the natural representations of $\text{SU}(2)$ symmetry on $S^2$ and $S^3$. $\text{SU}(2)$ acts on $\mathbf{n} = (n_x, n_y, n_z) \in S^2$ by $\mathbf{n} \mapsto \mathcal{R}(U) \mathbf{n}$ where $\mathcal{R} : \text{SU}(2) \to \text{SO}(3)$ is the double covering map. Next, we define $\xi$ to be the column vector $(u, v)$ so the second part of \cref{eq:define-pi-projector} can be written as $\mathbf{n} = \xi^\dagger \boldsymbol{\sigma}^* \xi$. A simple calculation shows
\begin{equation}\label{eq:representation-commutation}
\pi_{S^3, S^2}(U^* \xi) = \mathcal{R}(U) \pi_{S^3, S^2}(\xi),
\end{equation}
so that the corresponding action of $\text{SU}(2)$ on $S^3$ is $\xi \mapsto U^* \xi$.

\subsection{Lowest Landau level}
As explained in Ref. \cite{Haldane1983}, the lowest Landau level (LLL) is the subspace of equivariant homogeneous polynomials that depend only on $u$ and $v$, and not on $\overline{u}$ and $\overline{v}$. We will often call such polynomials LLL wavefunctions. A basis for the LLL is given by
\begin{equation}\label{appeq:define-psi_m}
\psi_m(u,v) = a_m u^{s + m} v^{s - m}
\end{equation}
for some nonzero coefficients $a_m$ which are left unspecified for now, and $m \in \{-s, -(s-1), \dots, s-1, s\}$. The representation $\xi \mapsto U^* \xi$ of $\text{SU}(2)$ on $S^3$ induces a representation on the LLL by
\begin{equation}\label{eq:LLL-representation}
(U \psi)(\xi) = \psi((U^*)^{-1}\xi) = \psi(U^T \xi).
\end{equation}
If we compute the action of $U = e^{-i(\theta_x \sigma_x + \theta_y \sigma_y + \theta_z \sigma_z)/2}$ on $\psi_m$ to first order in $\theta_x$, $\theta_y$, and $\theta_z$, we find that the LLL angular momentum operators take the form
\begin{equation}
L_x = \frac{1}{2}(v\partial_u + u \partial_v),\quad L_y = \frac{i}{2}(v\partial_u - u\partial_v),\quad L_z = \frac{1}{2}(u\partial_u - v\partial_v).
\end{equation}
In particular, we see that $L_z \psi_m = m \psi_m$ so that $m$ is the $z$ component of angular momentum, and the LLL wavefunctions form a spin-$s$ irreducible representation (irrep) of $\text{SU}(2)$. We can also combine $L_x$ and $L_y$ into
\begin{equation}
L_+ = L_x + i L_y = u \partial_v,\quad \quad L_- = L_x - i L_y = v\partial_u.
\end{equation}
These differential expressions for the $L_\alpha$ operators are useful because they allow us to extend their definition to arbitrary polynomials in $u$ and $v$ in a natural way. It is also worth noting that
\begin{equation}
(u\partial_u + v\partial_v - 2s)\psi = 0
\end{equation}
for any LLL wavefunction $\psi$.

Since for any LLL wavefunction $\psi$, the value $\psi(\xi)$ is determined up to a gauge dependent phase by the base point $\mathbf{n} = \xi^\dagger \boldsymbol{\sigma}^* \xi \in S^2$, the product $\psi_1^*(\xi) \psi_2(\xi)$ for any two LLL wavefunctions $\psi_1, \psi_2$ is gauge invariant and determined uniquely by $\mathbf{n}$. We may therefore define an inner product\footnote{We could equivalently have defined $\braket{\psi_1 | \psi_2}$ as an integral over $S^3$ against its symmetric normalized $3$-form $d\xi$. The gauge invariance of $\psi_1^* \psi_2$ implies that the resulting inner product would be the same.}
\begin{equation}\label{eq:inner-product}
\braket{\psi_1 | \psi_2} = \int_{S^2} (\psi_1^* \psi_2)(\mathbf{n})\, d\mathbf{n}
\end{equation}
where $d\mathbf{n} = \frac{1}{4\pi}\sin\theta \,d\theta \wedge d\phi$ is the symmetric normalized $2$-form on $S^2$. Using \cref{eq:representation-commutation} we see that for any $U \in \text{SU}(2)$,
\begin{equation}
\braket{U \psi_1 | U \psi_2} = \int_{S^2} (\psi_1^* \psi_2)(\mathcal{R}(U^{-1})\mathbf{n}) \, d\mathbf{n} = \braket{\psi_1 | \psi_2}
\end{equation}
so that the representation in \cref{eq:LLL-representation} is unitary.

It is convenient to choose the $a_m$ coefficients such that each $\psi_m$ has unit norm. Performing a simple spherical integral, we find that the choice
\begin{equation}\label{appeq:define-a_m}
a_m = \sqrt{(2s+1) \binom{2s}{s+m}}
\end{equation}
ensures $\braket{\psi_m | \psi_m} = 1$. Since $L_z \psi_m = m \psi_m$, we automatically have $\braket{\psi_m | \psi_n} = 0$ for $m \neq n$. As a result, our $\psi_m$ states form an orthonormal basis for the LLL. Furthermore, we have
\begin{equation}\label{eq:Condon-Shortley}
L_\pm \psi_m = \sqrt{s(s+1) - m(m\pm 1)} \psi_{m\pm 1}
\end{equation}
so that the angular momentum operators take the standard Condon--Shortley form.

\subsection{Pseudopotential Hamiltonians and Laughlin state}
The two-particle LLL tensor product Hilbert space consists of homogeneous polynomials in $(u_1, v_1, u_2, v_2) \in S^3 \times S^3$ such that $u_1, v_1$ have combined degree $2s$ in each monomial, and likewise for $u_2, v_2$. The total angular momentum operators are $L^{\text{tot}}_\alpha = L^{(1)}_{\alpha} + L^{(2)}_{\alpha}$ and the standard addition of angular momentum theory applies. By Schur's lemma, any $\text{SU}(2)$ symmetric two-particle interaction must decompose into a direct sum of constant operators on the total spin irreps. For each $0 \leq k \leq 2s$, we define the $k$th Haldane pseudopotential $H^{(2)}_k$ to be the projector onto the spin $2s-k$ irrep, so that any interaction can be expressed as a linear combination of $H^{(2)}_k$ operators. From here on, we assume $s \geq \frac{1}{2}$ so that the first pseudopotential is well defined.

In order to better understand the pseudopotential Hamiltonians, we seek to understand the structure of the total spin irreps. We note that $u_1 v_2 - u_2 v_1$ is annihilated by all $L^{\text{tot}}_\alpha$ operators. It follows that for each $0 \leq n \leq 2s$,
\begin{equation}\label{eq:phi-n}
\tilde{\psi}_n = (u_1 v_2 - u_2 v_1)^n u_1^{2s-n} u_2^{2s-n}
\end{equation}
has $L^{\text{tot}}_z$ eigenvalue $2s - n$ and is annihilated by $L^{\text{tot}}_+$. The spin-$(2s-n)$ irrep is then generated by the polynomials
\begin{equation}\label{eq:2s-n-irrep}
(L^{\text{tot}}_-)^j \tilde{\psi}_n = (u_1 v_2 - u_2 v_1)^n (L^{\text{tot}}_-)^j (u_1^{2s-n} u_2^{2s-n})
\end{equation}
for $0 \leq j \leq 2(2s-n)$. In particular, if a polynomial is divisible by $(u_1 v_2 - u_2 v_1)^n$ then its total spin is at most $2s - n$.

For any $k$, we can consider the $k$th pseudopotential Hamiltonian acting on $N$ particles. This is defined by $H^{(N)}_k = \sum_{p<q} H^{(2,p,q)}_k$ where $H^{(2,p,q)}_k$ is the $k$th pseudopotential Hamiltonian $H^{(2)}_k$ acting on particles $p$ and $q$. The Laughlin state is defined by
\begin{equation}
\psi_{\text{Laughlin}} = \prod_{p<q} (u_p v_q - u_q v_p)^3.
\end{equation}
The total degree for each particle is $3(N-1)$ so this is a valid $N$-particle state when $2s = 3(N-1)$. Since each pair of particles has a factor $(u_p v_q - u_q v_p)^3$ this state has no component in the spin-$(2s-1)$ subspace for any pair of particles and is therefore annihilated by $H^{(N)}_1$. More generally, we consider the Laughlin state with $M$ quasiholes
\begin{equation}
\tilde{\psi}_{\text{Laughlin}} = \psi_{\text{Laughlin}} \prod_{l=1}^M \prod_{p=1}^N (\beta_l u_p - \alpha_l v_p)
\end{equation}
which is a valid $N$-particle state and fully antisymmetric zero mode of $H_1^{(N)}$ for any choice of the $\alpha$ and $\beta$ coefficients, as long as $2s = 3(N-1) + M$. We conclude that the fermionic second quantized form of the first pseudopotential Hamiltonian has zero-energy ground states of particle number $N$ for all integers $0 \leq N \leq \frac{2s}{3} + 1$.

\subsection{Stereographic coordinates and the thermodynamic limit}\label{app:stereographic}
As explained in \cref{app:fqhe-sphere}, in the thermodynamic limit, the south pole of the sphere converges to the plane. In order to make this manifest, we define the stereographic coordinate $z = u/v$. In both the northern and southern hemispheres we have $z = \cot(\theta/2) e^{i\phi}$ by \cref{eq:u-N-v-N,eq:u-S-v-S} and we see that $z$ is a smooth coordinate except at the north pole $u = 1$, $v = 0$. We therefore exclude the north pole and use the southern hemisphere coordinates in \cref{eq:u-S-v-S} everywhere else, so that $v$ is given by
\begin{equation}\label{eq:v-from-z}
v = \frac{1}{\sqrt{1+|z|^2}}.
\end{equation}
Using this notation, the Laughlin wavefunction becomes
\begin{equation}
\psi_{\text{Laughlin}} = \prod_{p < q} (z_p - z_q)^3 \prod_p (1 + |z_p|^2)^{-s}
\end{equation}
where we have used $2s = 3(N-1)$ and \cref{eq:v-from-z}. To second order in $|z|$, we have $(1 + |z_p|^2)^s = e^{s|z_p|^2}$ so that for small $|z|$ we have $(1 + |z_p|^2)^{-s} \sim e^{-s|z_p|^2}$. We recall that the thermodynamic limit is defined by $s = (R/\ell_B)^2$ and $R \to \infty$. If we take $w_p = 2 R z_p$ then in the limit of $R \to \infty$ with $w_p$ fixed, we have $(1 + |z_p|^2)^{-s} \rightarrow e^{-|w_p|^2/(4 \ell_B^2)}$. Up to normalization, we then have
\begin{equation}
\psi_{\text{Laughlin}} \sim \prod_{p < q} (w_p - w_q)^3 \exp\left(-\frac{1}{4\ell_B^2}\sum_p |w_p|^2\right)
\end{equation}
which is the original form of the Laughlin wavefunction on the plane \cite{Laughlin1983}.

\subsection{Coherent state operators}
\label{sec:coherent}
In this section, we will define an alternative coherent state representation of the fermion annihilation operators that makes the locality properties of pseudopotentials in real-space and the relation to the Trugman--Kivelson representation more manifest. Using the stereographic coordinate $z$ introduced in \cref{app:stereographic}, the wavefunctions $\psi_m$ defined by \cref{appeq:define-psi_m,appeq:define-a_m} can be written
\begin{equation}
\psi_m = \sqrt{(2s+1) \binom{2s}{s+m}} (1 + |z|^2)^{-s} z^{s+m}
\end{equation}
where we have used \cref{eq:v-from-z}. We will absorb the non-holomorphic part of this expression along with part of the normalization into the measure. This yields single-particle orbitals
\begin{equation}
\phi_{s,m}(z) = \binom{2s}{s+m}^{1/2} z^{s+m}
\end{equation}
which are holomorphic in $z$ and are orthonormal under the measure
\begin{equation}
d\mu_s(z) = \frac{2s+1}{\pi} \frac{d^2 z}{(1 + |z|^2)^{2(s+1)}}, \qquad d^2z = d \Re z d \Im z.
\end{equation}
We note that using the coordinate $z$ neglects the north pole of the sphere, but since this is a set of measure zero this does not affect integrals. Additionally, we can always change the choice of removed point using an $\text{SU}(2)$ transformation which acts on $z$ as a M\"obius transformation.

Next, we define the second quantized fermionic operators
\begin{equation}
c_{s,z} = \sum_{m=-s}^s \phi_{s,m}(z) c_{s,m}, \qquad \{c_{s,m}, c_{s,m'}\} = 0, \qquad \{c_{s,m}, c_{s,m'}^\dagger \} = \delta_{m,m'}.
\end{equation}
These operators satisfy
\begin{align}
\{c_{s,z}, c_{s,w}^\dagger\} &= \sum_{m=-s}^s \phi_{s,m}(z) \bar \phi_{s,m}(w) = \sum_{m=-s}^s \binom{2s}{s+m} (z \bar w)^{s+m} \nonumber \\
&= (1 + z \bar w)^{2s} = P_s(z, \bar w)
\end{align}
which satisfies the projector identity
\begin{equation}
\int d\mu_s(w) P_s(z, \bar w) c_{s,w} = c_{s,z}.
\end{equation}
We now want to discuss how to construct the two-particle representation $s_1 \otimes s_2$. The basis states are $|s_1,m_1,s_2,m_2 \rangle$ whose position representation is $\langle z_1, z_2|s_1,m_1,s_2,m_2 \rangle = \phi_{s_1,m_1}(z_1) \phi_{s_2,m_2}(z_2)$. This decomposes into a sum of $S = |s_1 - s_2|,\dots,s_1 + s_2$ with
\begin{equation}
|S,M \rangle = \sum_{m_1 + m_2 = M} C^{S,M}_{s_1,m_1,s_2,m_2} | s_1, m_1, s_2, m_2 \rangle.
\end{equation}
We consider the generating function for the CG coefficients \cite{Perelomov1977}
\begin{gather}
\label{CGf}
f_{S,s_1,s_2}(\bar z,z_1, z_2) = \alpha_{S,s_1,s_2} (z_1 - z_2)^{s_1 + s_2 - S} (1 + \bar z z_1)^{s_1 + S - s_2} (1 + \bar z z_2)^{s_2 + S - s_1}  \\
\alpha_{S,s_1,s_2} = \sqrt{\frac{(2S+1)!(2s_1)!(2s_2)!}{(s_1 + s_2 - S)!(s_1 - s_2 + S)!(s_2 - s_1 + S)!(S + s_1 + s_2 + 1)!}}
\end{gather}
which is related to the CG coefficients via
\begin{equation}
f_{S,s_1,s_2}(\bar z, z_1, z_2) = \sum_{m_1=-s_1}^{s_1} \sum_{m_2=-s_2}^{s_2} \sum_{M=-S}^{S} C^{S,M}_{s_1,m_1,s_2,m_2} \phi_{s_1,m_1}(z_1) \phi_{s_2,m_2}(z_2) \phi_{S,M}(\bar z)
\end{equation}
which also gives the inverse relation
\begin{equation}
C^{S,M}_{s_1,m_1,s_2,m_2} = \int d\mu_S(z) d\mu_{s_1}(z_1) d\mu_{s_2}(z_2) \bar \phi_{s_1, m_1}(z_1) \bar \phi_{s_2,m_2}(z_2) \phi_{S,M}(z) f_{S,s_1,s_2}(\bar z, z_1, z_2).
\end{equation}
Note that $f$ satisfies
\begin{equation}
f_{S,s_1,s_2}(\bar z, z_1, z_2) = (-1)^{s_1 + s_2 - S}f_{S,s_2,s_1}(\bar z, z_2, z_1).
\end{equation}

Now consider the 2-body operators defined in the main text
\begin{equation}
Q_{S,M} = \frac{1}{\sqrt{2}} \sum_{m_1, m_2} C^{S,M}_{s_1,m_1,s_2,m_2} c_{s_2, m_2} c_{s_1, m_1} = \int d\mu_S(z) \bar \phi_{S,M}(z) Q_{S,z}
\end{equation}
where we defined
\begin{equation}
\label{Qz}
Q_{S,z} = \sum_{M=-S}^S \phi_{S,M}(z) Q_{S,M} = \frac{1}{\sqrt{2}} \int d\mu_{s_1}(z_1) d\mu_{s_2}(z_2) \bar f_{S,s_1,s_2}(\bar z,z_1, z_2) c_{s_2, z_2} c_{s_1, z_1}.
\end{equation}
For $s_1 = s_2$, this expression is only non-zero for anti-symmetric $f$ which corresponds to odd $s_1 + s_2 - S$ (see \cref{CGf}).

The special case of $s_1 = s_2 = s$ and $S = s_1 + s_2 - 1 = 2s - 1$ corresponds to the first pseudopotential for which $f$ simplifies to
\begin{equation}
f_{2s - 1,s, s}(\bar z, z_1, z_2) = \sqrt{s} (z_1 - z_2) (1 + \bar z z_1)^{2s - 1} (1 + \bar z z_2)^{2s - 1} = \frac{1}{2\sqrt{s}}[P_{s}(z_2, \bar z) \bar \partial_z P_{s}(z_1,\bar z) - P_{s}(z_1, \bar z) \bar \partial_z P_{s}(z_2,\bar z)].
\end{equation}
Substituting in \cref{Qz}, we find
\begin{equation}
Q_{2s-1,z} \propto \int d\mu_{s}(z_1) d\mu_{s}(z_2) P_{s}(z, \bar z_2) \partial_z P_{s}(z, \bar z_1) c_{s, z_2} c_{s, z_1} \propto c_{s, z} \partial_z c_{s, z}.
\end{equation}

To relate this to the first quantized Trugman--Kivelson representation, we write the second quantized Laughlin state
\begin{equation}
|\psi_{\rm Laughlin} \rangle \propto \int d\mu_s(z_1) \dots d\mu_s(z_N) \prod_{i<j} (z_i - z_j)^3 c_{s,z_1}^\dagger \dots c_{s,z_N}^\dagger |0 \rangle
\end{equation}
and note that
\begin{align}
Q_{2s-1,z} |\psi_{\rm Laughlin} \rangle &\propto \int d\mu_s(z_1) \dots d\mu_s(z_N) P_s(z, \bar z_1) \partial_z P_s(z,\bar z_2)\prod_{i<j} (z_i - z_j)^3 c_{s,z_3}^\dagger \dots c_{s,z_N}^\dagger |0 \rangle \nonumber \\
&= \int d\mu_s(z_3) \dots d\mu_s(z_N) \, \partial_{z_2} \Big[ (z_1 - z_2)^3 \prod_{i=3}^N (z_1 - z_i)^3 (z_2 - z_i)^3 \Big]_{z_1 = z_2 = z} \prod_{3 \leq i<j \leq N} (z_i - z_j)^3 c_{s,z_3}^\dagger \dots c_{s,z_N}^\dagger |0 \rangle \nonumber \\
&= 0.
\end{align}

We can also derive a version of the $H^2$ identity that does not rely on ${\rm SU}(2)$ symmetry. Consider the expression (here and below, $\int_{z_1, z_2, \dots}$ denotes integration with $d\mu_s$ over each single-particle coordinate)
\begin{align}
F_{z,w} &= Q_{2s-1,z}^\dagger [Q_{2s-1,z}, Q_{2s-1,w}^\dagger] Q_{2s-1,w} \nonumber \\
&= \frac{1}{2} \int_{z_1, z_2, w_1, w_2} \bar f_{2s-1,s,s}(\bar z,z_1, z_2) f_{2s-1,s,s}(\bar w, w_1, w_2) Q_{2s-1,z}^\dagger [c_{s,z_2} c_{s,z_1}, c_{s,w_1}^\dagger c_{s,w_2}^\dagger] Q_{2s-1,w} \nonumber \\
&=  \int_{z_1, z_2, w_1, w_2} \bar f_{2s-1,s,s}(\bar z,z_1, z_2) f_{2s-1,s,s}(\bar w,w_1, w_2) \nonumber \\
&\qquad \times \{Q_{2s-1,z}^\dagger c_{s,z_1} c_{s,w_1}^\dagger Q_{2s-1,w} P_s(z_2, \bar w_2) - Q_{2s-1,z}^\dagger c_{s,w_2}^\dagger c_{s,z_2} Q_{2s-1,w} P_s(z_1, \bar w_1) \} \nonumber \\
&= \int_{z_1, z_2, w_1, w_2} \bar f_{2s-1,s,s}(\bar z,z_1, z_2) f_{2s-1,s,s}(\bar w,w_1, w_2) \nonumber \\
&\qquad \times \{2Q_{2s-1,z}^\dagger c_{s,z_1} c_{s,w_1}^\dagger Q_{2s-1,w} P_s(z_2, \bar w_2) - Q_{2s-1,z}^\dagger Q_{2s-1,w} P_s(z_1, \bar w_1) P_s(z_2, \bar w_2) \}
\nonumber \\
&= 2\int_{z_1, z_2, w_1} \bar f_{2s-1,s,s}(\bar z,z_1, z_2) f_{2s-1,s,s}(\bar w,w_1, z_2) Q_{2s-1,z}^\dagger c_{s,z_1} c_{s,w_1}^\dagger Q_{2s-1,w} \nonumber \\
&\qquad - \int_{z_1, z_2} \bar f_{2s-1,s,s}(\bar z,z_1, z_2) f_{2s-1,s,s}(\bar w,z_1, z_2) Q_{2s-1,z}^\dagger Q_{2s-1,w}
\end{align}
We now note that
\begin{equation}
\int_{z_1, z_2} \bar f_{2s-1,s,s}(\bar z,z_1, z_2) f_{2s-1,s,s}(\bar w,z_1, z_2) = \sum_{M}  \phi_{2s-1,M}(z) \bar \phi_{2s-1,M}(w) = P_{2s-1}(z, \bar w).
\end{equation}
We can then write
\begin{equation}
\int d\mu_{2s-1}(z) d\mu_{2s-1}(w) F_{z,w} = 2\int d\mu_s(z)  R_{s,z}^\dagger R_{s,z} - H
\end{equation}
where $H = \int d\mu_{2s-1}(z) Q_{2s-1,z}^\dagger Q_{2s-1,z}$ and $R_{s,z}$ is
\begin{equation}
\label{Rz}
R_{s,z} = \int d\mu_{2s-1}(w) d\mu_s(w_1) f_{2s-1,s,s}(\bar w, w_1, z) c_{s,w_1}^\dagger Q_{2s-1,w}.
\end{equation}
Note that $R$ is associated with angular momentum $s$. This leads to the identity
\begin{align}
\label{H2}
H^2 &= \int d\mu_{2s-1}(z) d\mu_{2s-1}(w) Q_{2s-1,z}^\dagger Q_{2s-1,z} Q_{2s-1,w}^\dagger Q_{2s-1,w} \nonumber \\
&= \int d\mu_{2s-1}(z) Q_{2s-1,z}^\dagger H Q_{2s-1,z} + 2\int d\mu_s(z) R_{s,z}^\dagger R_{s,z} - H.
\end{align}
To relate this to the identity in the main text, let us define
\begin{align}
\Gamma_{J,j} &= \sum_{M,m} [C^{J,j}_{S,M,s,m}]^* (-1)^{s-m} c^\dagger_{s,-m} Q_{S,M} \nonumber \\
&= \int d\mu_J(z) d\mu_S(z_1) d\mu_s(z_2) \bar \phi_{J,j}(z) \bar f_{J,S,s}(\bar z,z_1, z_2)  \sum_{M,m} (-1)^{s-m} c^\dagger_{s,-m} Q_{S,M} \phi_{S,M}(z_1) \phi_{s,m}(z_2).
\end{align}
We now notice that
\begin{equation}
(-1)^{s-m} \phi_{s,m}(-1/\bar z) = (-1)^{2s} \binom{2s}{s+m}^{1/2} (\bar z)^{-s-m} = (-1)^{2s} (\bar z)^{-2s} \binom{2s}{s-m}^{1/2} (\bar z)^{s-m} = (-1)^{2s} (\bar z)^{-2s} \phi_{s,-m}(\bar z)
\end{equation}
and
\begin{equation}
d\mu_s(-1/\bar z) = \frac{2s+1}{\pi} \frac{d^2 (1/\bar z)}{(1 + |1/z|^2)^{2(s+1)}} = \frac{2s+1}{\pi} \frac{|z|^{4(s+1)}}{|z|^4} \frac{d^2 z}{(1 + |z|^2)^{2(s+1)}} = |z|^{4s} d\mu_s(z).
\end{equation}
We can now change the integration variable $z_2 \mapsto -1/\bar z_2$ to simplify
\begin{align}
\sum_m \int d\mu_s(z_2) \bar f_{J,S,s}(\bar z,z_1, z_2) (-1)^{s-m} \phi_{s,m}(z_2) c_{s,-m}^\dagger &= (-1)^{2s} \sum_m \int d\mu_s(z_2) \bar f_{J,S,s}(\bar z,z_1, -1/\bar z_2) z_2^{2s} \phi_{s,-m}(\bar z_2) c_{s,-m}^\dagger \nonumber \\
&= (-1)^{2s} \int d\mu_s(z_2) \bar f_{J,S,s}(\bar z,z_1, -1/\bar z_2) z_2^{2s} c_{s,z_2}^\dagger \nonumber \\
&= (-1)^{2s} \int d\mu_s(z_2) f_{J,S,s}( z,\bar z_1, -1/z_2) z_2^{2s} c_{s,z_2}^\dagger.
\end{align}
where we used $\bar f(z, z_1, z_2) = f(\bar z,\bar z_1, \bar z_2)$. Then $\Gamma$ becomes
\begin{equation}
\Gamma_{J,j} = (-1)^{2s} \int d\mu_J(z) d\mu_S(z_1) d\mu_s(z_2) \bar \phi_{J,j}(z) f_{J,S,s}(z, \bar z_1, -1/z_2) z_2^{2s}  c^\dagger_{s,z_2} Q_{S,z_1}.
\end{equation}
This can be simplified using the identity
\begin{equation}
z_2^{2j_2} f_{J,j_1,j_2}( z, \bar z_1,-1/z_2) = \sqrt{\frac{2J+1}{2j_1+1}} f_{j_1,j_2,J}(\bar z_1, z_2, z).
\end{equation}
Then we can define
\begin{equation}
\Gamma_z = \sum_j \Gamma_{s,j} \phi_{s,j}(z) = (-1)^{2s} \sqrt{\frac{2s+1}{4s-1}} \int d\mu_{2s-1}(z_1) d\mu_s(z_2) f_{2s-1,s,s}(\bar z_1, z_2, z)  c^\dagger_{s,z_2} Q_{2s-1,z_1}.
\end{equation}
Comparing with \cref{Rz}, we see that  $\Gamma_z = (-1)^{2s} \sqrt{\frac{2s+1}{4s-1}} R_{s,z}$. Hence, $\int d\mu_s(z) R_{s,z}^\dagger R_{s,z} = \frac{4s-1}{2s+1} \sum_j R_{s,j}^\dagger R_{s,j} = (4s-1) \mathcal S(R_{s,s_0}^\dagger R_{s,s_0})$ reproducing \cref{eq:H-squared-second-term-intermediate} of the main text. 

\section{Properties of the Clebsch--Gordan tensor}\label{app:Clebsch-Gordan}
We review here some of the useful properties of the Clebsch--Gordan tensor $C^{J,M}_{j_1,m_1,j_2,m_2}$. First, the values $J$, $j_1$, $j_2$, $m_1$, $m_2$, and $M$ are all integers or half-integers. $J$, $j_1$, and $j_2$ are nonnegative and $J$ lies in the range from $|j_1-j_2|$ to $j_1 + j_2$ with integer steps. $M$ lies in the range from $-J$ to $J$ with integer steps, $m_1$ lies in the range from $-j_1$ to $j_1$ with integer steps, and $m_2$ lies in the range from $-j_2$ to $j_2$ with integer steps. We work in the Condon--Shortley basis $|j,m\rangle$ of the spin-$j$ irrep, defined by
\begin{equation}
    L_z |j,m\rangle = m |j,m\rangle, \qquad L_\pm |j,m\rangle = \sqrt{j(j+1) - m(m\pm 1)}\, |j,m\pm 1\rangle,
\end{equation}
and denote the corresponding unitary matrix representation of ${\rm SU}(2)$ by
\begin{equation}
    \label{Ddef}
    D_{n,m}(j,g) = \langle j,n|U(g)|j,m\rangle, \qquad U(g)|j,m\rangle = \sum_n D_{n,m}(j,g)\,|j,n\rangle.
\end{equation}
The Clebsch--Gordan coefficients are the expansion coefficients
\begin{equation}
    C^{J,M}_{j_1,m_1,j_2,m_2} = \langle j_1,m_1; j_2,m_2|J,M\rangle,
\end{equation}
where $|J,M\rangle$ is the Condon--Shortley basis of the spin-$J$ irrep in $j_1 \otimes j_2$, with the phase of each irrep fixed by $C^{J,J}_{j_1,j_1,j_2,J-j_1} > 0$. With this convention the Clebsch--Gordan coefficients are real, and they intertwine the representations,
\begin{equation}
    \label{CGintertwine}
    \sum_{m_1,m_2} D_{n_1,m_1}(j_1,g)\, D_{n_2,m_2}(j_2,g)\, C^{J,M}_{j_1,m_1,j_2,m_2} = \sum_N D_{N,M}(J,g)\, C^{J,N}_{j_1,n_1,j_2,n_2}
\end{equation}
for all $g \in {\rm SU}(2)$.

 The unitarity identity is
\begin{equation}
\sum_{m_1,m_2} C^{J,M}_{j_1,m_1,j_2,m_2} C^{J',M'}_{j_1,m_1,j_2,m_2} = \delta_{J,J'}\delta_{M,M'}
\end{equation}
for fixed $j_1$, $j_2$. The completeness identity is
\begin{equation}
\sum_{J,M} C^{J,M}_{j_1,m_1,j_2,m_2} C^{J,M}_{j_1,m_1',j_2,m_2'} = \delta_{m_1,m_1'}\delta_{m_2,m_2'}
\end{equation}
for fixed $j_1$, $j_2$. The exchange identity is
\begin{equation}
C^{J,M}_{j_2,m_2,j_1,m_1} = (-1)^{j_1+j_2-J} C^{J,M}_{j_1,m_1,j_2,m_2}.
\end{equation}
Finally, the Racah formula gives an explicit expression for the Clebsch--Gordan tensor, namely
\begin{equation}\label{appeq:Racah-formula}
C^{J,M}_{j_1,m_1,j_2,m_2} = \delta_{M,m_1+m_2} \sqrt{a_{J,j_1,j_2} b_{J,M} c_{j_1,m_1,j_2,m_2}} \sum_k \frac{(-1)^k}{d_{J,j_1,j_2,m_1,m_2,k}}
\end{equation}
where
\begin{align}
a_{J,j_1,j_2} &= \frac{(2J+1)(J+j_1-j_2)!(J-j_1+j_2)!(j_1+j_2-J)!}{(j_1+j_2+J+1)!}\\
b_{J,M} &= (J+M)!(J-M)!\\
c_{j_1,m_1,j_2,m_2} &= (j_1-m_1)!(j_1+m_1)!(j_2-m_2)!(j_2+m_2)!\\
d_{J,j_1,j_2,m_1,m_2,k} &= k!(j_1 + j_2 - J - k)! (j_1 - m_1 - k)!(j_2+m_2-k)!(J-j_2+m_1+k)!(J-j_1-m_2+k)!\label{appeq:Racah-delta-term}
\end{align}
The summation in \cref{appeq:Racah-formula} includes all $k$ for which each factorial in \cref{appeq:Racah-delta-term} has a nonnegative argument.

\section{Supporting derivations}
\subsection{Clebsch--Gordan summation identity}\label{app:CG-summation-identity}
We now prove the identity in \cref{eq:CG-sum-identity}. By construction, $C^{2s-1, \cdot}_{s,\cdot,s,\cdot}$ is an intertwiner satisfying
\begin{equation}
\sum_{m_1, m_2} D_{n_1, m_1}(s,g) D_{n_2,m_2} (s,g) C^{2s-1,M}_{s,m_1,s,m_2} = \sum_N D_{N,M}(2s-1,g) C^{2s-1,N}_{s,n_1, s,n_2}
\end{equation}
while $\tilde{C}^{J, \cdot}_{2s-1, \cdot, s, \cdot}$ is an intertwiner satisfying
\begin{equation}
\sum_{M,m} D_{N,M}(2s-1,g) D^*_{n,m}(s,g) \tilde{C}^{J,j}_{2s-1,M,s,m} = \sum_k D_{k,j}(J,g) \tilde{C}^{J,k}_{2s-1,N,s,n}.
\end{equation}
Composing these relations, we see that
\begin{equation}
B^{J,j}_{l,m,n} = \sum_M C^{2s-1,M}_{s,l,s,m} \tilde{C}^{J,j}_{2s-1,M,s,n}
\end{equation}
is an intertwiner satisfying
\begin{equation}
\sum_{l,m,n} D_{l',l}(s,g) D_{m',m}(s,g) D^*_{n',n}(s,g) B^{J,j}_{l,m,n} = \sum_k D_{k,j}(J,g) B^{J,k}_{l',m',n'}.
\end{equation}
We now look at
\begin{equation}
\begin{split}
\tilde{B}^J_{l,j} &= \sum_m B^{J,j}_{l,m,m}\\
&= \sum_{m,n} B^{J,j}_{l,m,n} (D^\dagger D)_{n,m}(s,g)\\
&= \sum_{m,n,n'} B^{J,j}_{l,m,n} D_{n',m}(s,g) D^*_{n',n}(s,g).
\end{split}
\end{equation}
This satisfies
\begin{equation}
\begin{split}
\sum_l D_{l',l}(s,g) \tilde{B}^J_{l,j} &= \sum_{l,m,n,n'} D_{l',l}(s,g) D_{n',m}(s,g) D^*_{n',n}(s,g) B^{J,j}_{l,m,n}\\
&= \sum_{k,n'} D_{k,j}(J,g) B^{J,k}_{l',n',n'}\\
&= \sum_k \tilde{B}^J_{l',k} D_{k,j}(J,g)
\end{split}
\end{equation}
so that $\tilde{B}^J$ is an intertwiner between irreps $D(s,g)$ and $D(J,g)$. Schur's lemma then implies $\tilde{B}^J_{l,j} = \lambda \delta_{J,s} \delta_{j,l}$ for some constant $\lambda$. We now compute
\begin{equation}
\begin{split}
(2s+1)\lambda^2 &= \sum_{J,l,j} (\tilde{B}^J_{l,j})^2\\
&= \sum_l \sum_{M,m} \sum_{N,n} C^{2s-1,M}_{s,l,s,m} C^{2s-1,N}_{s,l,s,n} \sum_{J,j} \tilde{C}^{J,j}_{2s-1,M,s,m} \tilde{C}^{J,j}_{2s-1,N,s,n}\\
&= \sum_{M,l,m} \left(C^{2s-1,M}_{s,l,s,m}\right)^2\\
&= 4s-1
\end{split}
\end{equation}
by unitarity. This completes the proof of \cref{eq:CG-sum-identity}.

\subsection{Lower bound for $\mathcal{S}(\Omega_2)$}\label{app:analytical-bound}
We now derive \cref{eq:bound-equation-Omega-2,eq:define-omega_2,eq:define-chi_M}. We first note that for any operators $X$ and $Y$,
\begin{equation}
(X\pm Y)^\dagger (X \pm Y) \succeq 0.
\end{equation}
Rearranging this inequality yields
\begin{equation}\label{appeq:positivity-inequality}
x(X^\dagger Y + Y^\dagger X) \succeq -|x|(X^\dagger X + Y^\dagger Y)
\end{equation}
for any real $x$. Since all terms in \cref{eq:Omega-2-PSD} are PSD, we have $\mathcal{S}(\Omega_2) \succeq 2\mathcal{S}(\tilde{R}^\dagger_1 \tilde{R}_2 + \tilde{R}^\dagger_2 \tilde{R}_1)$. Additionally, \cref{eq:define-tilde-C,eq:H-symmetrized,eq:define-tilde-R-1,eq:define-tilde-R-2,appeq:positivity-inequality} imply
\begin{align}
2\mathcal{S}(\tilde{R}^\dagger_1 \tilde{R}_2 + \tilde{R}^\dagger_2 \tilde{R}_1) &= 2\sum_{M=1}^r \sum_{N=r+1}^{2s} \tilde{C}^{s,-s}_{2s-1,-2s+M,s,-s+M} \tilde{C}^{s,-s}_{2s-1,-2s+N,s,-s+N}\mathcal{S}(\tilde{Q}^\dagger_M \tilde{c}_M \tilde{c}^\dagger_N \tilde{Q}_N + \tilde{Q}^\dagger_N \tilde{c}_N \tilde{c}^\dagger_M \tilde{Q}_M)\\
&\succeq -2\sum_{M=1}^r \sum_{N=r+1}^{2s}|C^{s,-s}_{2s-1,-2s+M,s,s-M} C^{s,-s}_{2s-1,-2s+N,s,s-N}|\mathcal{S}(\tilde{Q}^\dagger_M \tilde{c}_M \tilde{c}^\dagger_M \tilde{Q}_M + \tilde{Q}^\dagger_N \tilde{c}_N \tilde{c}^\dagger_N \tilde{Q}_N)\\
&\succeq -2\sum_{M=1}^r \sum_{N=r+1}^{2s}|C^{s,-s}_{2s-1,-2s+M,s,s-M} C^{s,-s}_{2s-1,-2s+N,s,s-N}|\mathcal{S}(\tilde{Q}^\dagger_M \tilde{Q}_M + \tilde{Q}^\dagger_N \tilde{Q}_N)\\
&= -\omega_2 \mathcal{S}(\tilde{Q}^\dagger_M \tilde{Q}_M)
\end{align}
for any $M$. The value of $\omega_2$ is given by
\begin{align}
\omega_2 &= 4\sum_{M=1}^r \sum_{N=r+1}^{2s}|C^{s,-s}_{2s-1,-2s+M,s,s-M} C^{s,-s}_{2s-1,-2s+N,s,s-N}|\\
&= 4 \chi_r (\chi_{2s} - \chi_r)\label{appeq:define-omega_2}
\end{align}
where
\begin{equation}\label{appeq:define-chi_M}
\chi_M = \sum_{N=1}^M |C^{s,-s}_{2s-1,-2s+N,s,s-N}|.
\end{equation}

\subsection{$\tilde{Q}_M$ operators in terms of $f^\dagger_m$ and $f_m$}\label{app:tilde-Q-f_m}
We now express $\tilde{Q}_M$ in terms of $f^\dagger_m$ and $f_m$. We begin by computing the CG coefficient appearing in \cref{eq:define-tilde-Q}. \cref{appeq:Racah-formula} implies
\begin{equation}
C^{2s-1,-2s+M}_{s,-s+m,s,-s+M-m} = (2m-M)\sqrt{\frac{4s-1}{M(4s-M)}}\sqrt{\frac{\binom{4s-M}{2s-m}\binom{M}{m}}{\binom{4s}{2s}}}.
\end{equation}
In terms of $t = \frac{1}{2s}$, we have
\begin{align}
\frac{\binom{4s-M}{2s-m}}{\binom{4s}{2s}} &= \frac{(4s-M)!}{(4s)!}\frac{(2s)!}{(2s-m)!}\frac{(2s)!}{(2s-M+m)!}\\
&= \frac{((2s) \cdots (2s-m + 1)) ((2s) \cdots (2s - M + m + 1))}{(4s) \cdots (4s-M + 1)}\\
&= \frac{\prod_{j=0}^{m-1} (1 - tj) \prod_{k=0}^{M-m-1} (1-tk)}{\prod_{l=0}^{M-1} (2 - lt)}\\
&= \frac{\xi_m \xi_{M-m}}{m! (M-m)!\prod_{l=0}^{M-1} (2 - lt)},\label{appeq:binomial-ratio-2}
\end{align}
where we have used \cref{eq:define-xi_m}. We then have
\begin{align}
C^{2s-1,-2s+M}_{s,-s+m,s,-s+M-m} &= \frac{1}{\sqrt{2}} \frac{2m-M}{m! (M-m)!} \sqrt{\frac{(M-1)! \xi_m \xi_{M-m}}{\prod_{l=2}^M (2 - lt)}}\\
C^{2s-1,-2s+M}_{s,-s+m,s,-s+M-m} \tilde{c}_{M-m} \tilde{c}_m &= \frac{1}{\sqrt{2}}\frac{2m-M}{m! (M-m)!} \sqrt{\frac{(M-1)!}{\prod_{l=2}^M (2 - lt)}} f_{M-m} f_m
\end{align}
where we have used \cref{eq:define-f}. Finally, \cref{eq:define-tilde-Q} implies
\begin{align}
\tilde{Q}_M &= \frac{1}{\sqrt{2}} \sum_m C^{2s-1,-2s+M}_{s,-s+m,s,-s+M-m} \tilde{c}_{M-m} \tilde{c}_m\\
&= q_M \sqrt{\frac{(M-1)!}{\eta_M}}\label{appeq:Q-in-q}\\
q_M &= \frac{1}{2}\sum_m \frac{2m-M}{m! (M-m)!} f_{M-m} f_m\\
\eta_M &= \prod_{l=2}^M (2 - lt).\label{appeq:define-eta_M}
\end{align}

\subsection{$\mathcal{S}(2\tilde{R}^\dagger_1 \tilde{R}_1)$ in terms of $q^\dagger_m$, $q_m$, $f^\dagger_m$, and $f_m$}\label{app:tilde-R_1-f_m}
We now express $\mathcal{S}(2\tilde{R}^\dagger_1 \tilde{R}_1)$ in terms of $f^\dagger_m$ and $f_m$. We begin by computing the CG coefficient appearing in \cref{eq:define-tilde-R-1}. \cref{eq:define-tilde-C,appeq:Racah-formula} imply
\begin{equation}
\tilde{C}^{s,-s}_{2s-1,-2s+M,s,-s+M} = \sqrt{\frac{(2s+1)M}{4s-M}}\sqrt{\frac{\binom{4s-M}{2s}}{\binom{4s}{2s}}}.
\end{equation}
\cref{appeq:binomial-ratio-2} allows us to write
\begin{align}
\tilde{C}^{s,-s}_{2s-1,-2s+M,s,-s+M} &= \sqrt{\frac{1}{(M-1)!} \frac{(1+t)\xi_M}{\prod_{l=0}^M (2 - lt)}}\label{appeq:tilde-C-expression}\\
\tilde{C}^{s,-s}_{2s-1,-2s+M,s,-s+M} \tilde{Q}^\dagger_M \tilde{c}_M &= \frac{1}{\eta_M} \sqrt{\frac{1+t}{2(2-t)}} q^\dagger_M f_M
\end{align}
where we have used \cref{eq:define-f,appeq:Q-in-q,appeq:define-eta_M}. \cref{eq:define-tilde-R-1,eq:define-xi_m,eq:f-anticommutator-delta,appeq:Q-in-q} then give
\begin{align}
\tilde{R}^\dagger_1 &= \sqrt{\frac{1+t}{2(2-t)}} \sum_{M=1}^r \frac{1}{\eta_M} q^\dagger_M f_M\\
\tilde{R}^\dagger_1 \tilde{R}_1 &= \frac{1+t}{2(2-t)} \sum_{M,N=1}^r \frac{1}{\eta_M\eta_N} q^\dagger_M f_M f^\dagger_N q_N\\
&= \frac{1+t}{2(2-t)} \sum_{M=1}^r \frac{\xi_M}{(M-1)!\eta_M} \tilde{Q}^\dagger_M \tilde{Q}_M - \frac{1+t}{2(2-t)} \sum_{M,N=1}^r \frac{1}{\eta_M\eta_N} q^\dagger_M f^\dagger_N f_M q_N.
\end{align}
By \cref{eq:H-symmetrized}, we find
\begin{equation}
\mathcal{S}(2\tilde{R}^\dagger_1 \tilde{R}_1) = (2-\delta) \mathcal{S}(\tilde{Q}^\dagger_M \tilde{Q}_M) - \mathcal{S}(W)
\end{equation}
for any $M$, where
\begin{align}
\delta &= 2 - \frac{1+t}{2-t} \sum_{M=1}^r \frac{\xi_M}{(M-1)!\eta_M}\label{appeq:define-delta}\\
W &= \frac{1+t}{2-t} \sum_{M,N=1}^r \frac{1}{\eta_M\eta_N} q^\dagger_M f^\dagger_N f_M q_N.
\end{align}

\subsection{$\omega_2$ and $\delta$ in the thermodynamic limit}\label{app:omega-2-delta-thermodynamic}
We now compute $\tilde{\omega}_2 = \lim_{s\to\infty} \omega_2$ where $\omega_2$ is given in \cref{appeq:define-omega_2,appeq:define-chi_M}. We take $t \to 0$ in \cref{appeq:tilde-C-expression} and use \cref{eq:define-tilde-C,eq:define-xi_m} to find
\begin{equation}
\lim_{s\to\infty} |C^{s,-s}_{2s-1,-2s+M,s,s-M}| = \sqrt{\frac{M}{2^{M+1}}}.
\end{equation}
\cref{appeq:define-omega_2,appeq:define-chi_M} along with a standard dominated convergence argument imply
\begin{equation}\label{appeq:tilde-omega_2}
\tilde{\omega}_2 = 4 \tilde{\chi}_r (\tilde{\chi}_\infty - \tilde{\chi}_r)
\end{equation}
where
\begin{align}
\tilde{\chi}_M &= \lim_{s\to\infty} \chi_M = \sum_{N=1}^M \sqrt{\frac{N}{2^{N+1}}}\\
\tilde{\chi}_\infty &= \lim_{M\to\infty} \tilde{\chi}_M = \sum_{N=1}^\infty \sqrt{\frac{N}{2^{N+1}}}.
\end{align}

In order to lower and upper bound $\tilde{\chi}_\infty$, we define
\begin{align}
g_{0,M}(x) &= \sum_{N=0}^M x^N = \frac{x^{M+1} - 1}{x-1}\\
g_{0,\infty}(x) &= \lim_{M\to\infty} g_{0,M}(x) = \frac{1}{1-x}
\end{align}
for $|x| < 1$. Taking the derivative of $g_{0,M}(x)$ and multiplying by $x$ gives
\begin{align}
g_{1,M}(x) &= \sum_{N=0}^M N x^N = \frac{M x^{M+2} - (M+1) x^{M+1} + x}{(x-1)^2}\label{appeq:g-1-M}\\
g_{1,\infty}(x) &= \lim_{M\to\infty} g_{1,M}(x) = \frac{x}{(1-x)^2}.
\end{align}
Using $x_0 = \frac{1}{\sqrt{2}}$, for any $N_0 > 1$ we have
\begin{align}
\tilde{\chi}_\infty &< \tilde{\chi}_{N_0} + \sum_{N=N_0+1}^\infty N x_0^{N+1}\\
&= \tilde{\chi}_{N_0} + x_0(g_{1,\infty}(x_0) - g_{1,N_0}(x_0))\\
&= \tilde{\chi}_{N_0} + \frac{x_0^{N_0+2}(N_0 + 1 - N_0 x_0)}{(1-x_0)^2}
\end{align}
and similarly
\begin{align}
\tilde{\chi}_\infty &> \tilde{\chi}_{N_0} + \sum_{N=N_0+1}^\infty x_0^{N+1}\\
&= \tilde{\chi}_{N_0} + x_0(g_{0,\infty}(x_0) - g_{0,N_0}(x_0))\\
&= \tilde{\chi}_{N_0} + \frac{x_0^{N_0+2}}{1-x_0}.
\end{align}
Finally, for any $N_0 \geq r$ we have
\begin{equation}\label{eq:tilde-omega_2-bounded}
4 \tilde{\chi}_r \left(\frac{x_0^{N_0+2}}{1-x_0} + \sum_{N=r+1}^{N_0} \sqrt{\frac{N}{2^{N+1}}}\right) < \tilde{\omega}_2 < 4 \tilde{\chi}_r \left(\frac{x_0^{N_0+2}(N_0 + 1 - N_0 x_0)}{(1-x_0)^2} + \sum_{N=r+1}^{N_0} \sqrt{\frac{N}{2^{N+1}}}\right).
\end{equation}
See \cref{fig:bounds}(a) for a plot of $\tilde{\omega}_2$. For our exactly certified spectral gap bound, we evaluate the upper bound in \cref{eq:tilde-omega_2-bounded} using rigorous interval arithmetic with $N_0 = 200$ and $r = 22$.

Next, we compute $\tilde{\delta} = \lim_{s\to\infty} \delta$ where $\delta$ is given in \cref{appeq:define-delta}. Taking $t \to 0$ in \cref{appeq:define-delta} and using \cref{appeq:define-eta_M,eq:define-xi_m,appeq:g-1-M} gives
\begin{align}
\tilde{\delta} &= 2 - \sum_{M=1}^r \frac{M}{2^M}\\
&= 2 - g_{1,r}(1/2)\\
&= \frac{r+2}{2^r}.
\end{align}

\end{document}